\documentclass[a4paper,12pt,twoside]{report}
\usepackage{dissertacao}
\allowdisplaybreaks
\newcommand{\titleinfoMain}{Heavy-Quarkonium Potential with Input \\[2mm]
                                          from Lattice Gauge Theory}%
\newcommand{\titleinfoLowerCase}{Heavy-Quarkonium Potential with Input \\[2mm]
                                          from Lattice Gauge Theory}
\newcommand{\titleinfo}{\bf Heavy-quarkonium potential with input
                                          from lattice gauge theory}
\newcommand{\titleinfoPort}{\bf Potencial de quarks pesados com input
                                          de teorias de gauge na rede}                                         
\newcommand{\authorinfo}{Willian Matioli Serenone}
\newcommand{\bibauthorinfo}{WILLIAN, M. S.}
\newcommand{\vect}[1]{\boldsymbol{#1}}
\let\ophat\hat
\renewcommand{\hat}[1]{\boldsymbol{\ophat{#1}}}

\usetikzlibrary{patterns,decorations.markings,decorations.pathreplacing}
\usepgfplotslibrary{units,external}
\begin{document}

\thispagestyle{empty}
\begin{center}
\setcounter{page}{-1}
{\LARGE UNIVERSIDADE DE SÃO PAULO\\[2mm]
  INSTITUTO DE FÍSICA DE SÃO CARLOS\\}

\vspace*{4cm}

{\Large \authorinfo \\}

\vspace*{8.5cm}

{\LARGE \titleinfoMain \\}

\vspace*{5cm}
São Carlos\\
\number\year
\end{center}

\newpage\
\thispagestyle{empty}
\newpage
\thispagestyle{empty}

\baselineskip 30pt
\thispagestyle{empty}

\setcounter{footnote}{1}

\begin{center}
{ \large \sc  \authorinfo
\\}
\end{center}

\vspace*{6cm}

\begin{center} {\LARGE \bfseries \titleinfoLowerCase}
\end{center}

\vspace*{1.5cm}

\begin{flushright}
\begin{minipage}{11cm}

\baselineskip 18pt

\vspace{1.5cm}

Dissertation presented to the Graduate Program in Physics at
the Instituto de Física de São Carlos, Universidade de São Paulo
to obtain the degree of Master of Science.
\vspace{0.5cm}

Concentration area: Fundamental Physics \\
Advisor: Prof.\ Dr.\ Tereza Cristina da Rocha Mendes\\ 
\end{minipage}

\vspace*{3cm}
\centerline{Corrected Version}
\centerline{\small(Original version avaiable at the Unity that lodges the Program)}
\end{flushright}

\baselineskip 17pt

\vspace{0.5cm}
\begin{center}
{\bf São Carlos}\\
{\bf \number\year}
\end{center}

\baselineskip 23.5pt

\includepdf[pages=1]{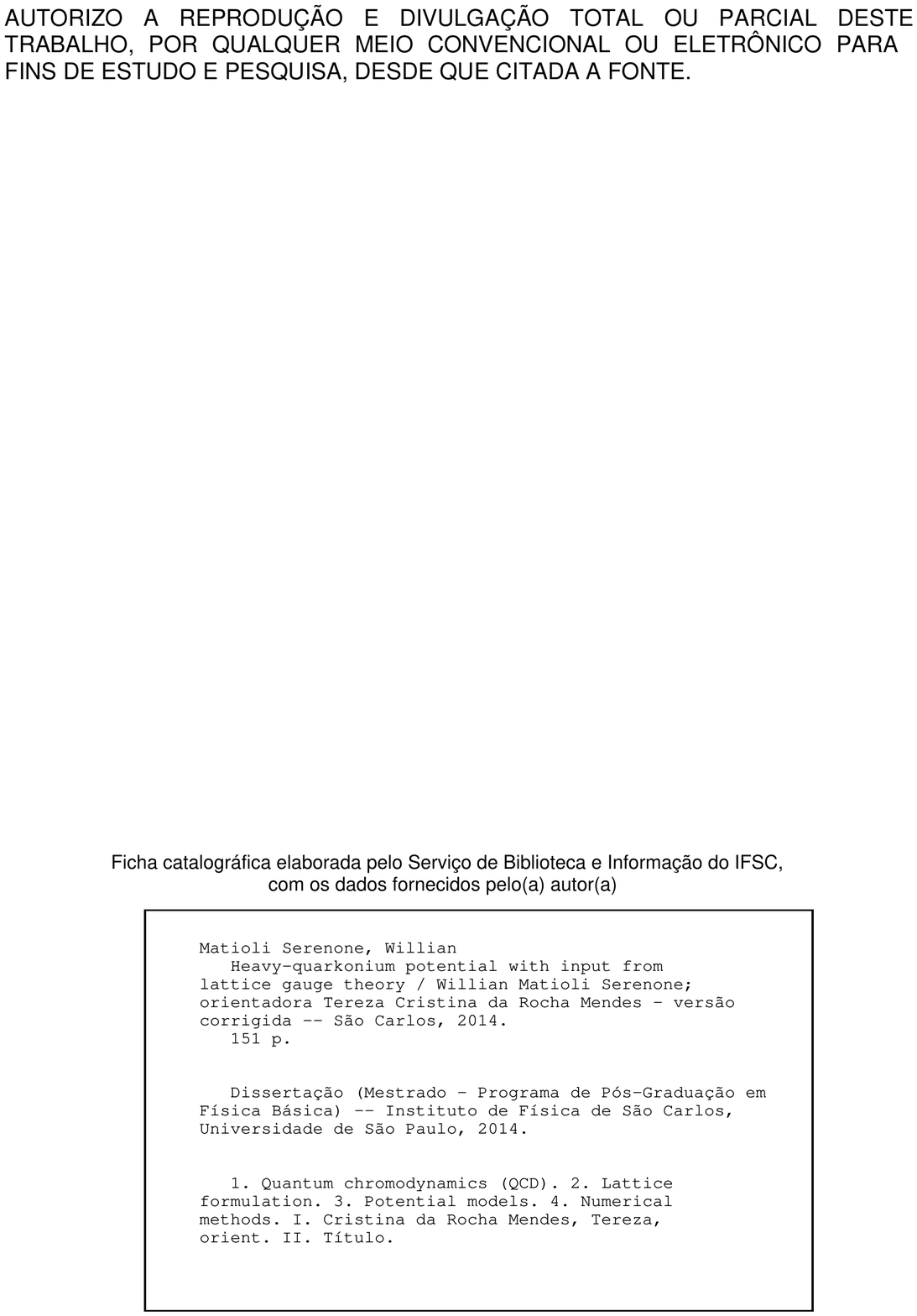}
\thispagestyle{empty}
\begin{center}
	THIS PAGE WAS INTENTIONALLY LEFT BLANK
\end{center}
\newpage\
\thispagestyle{empty}
\newpage
\thispagestyle{empty}

\vspace*{22cm}

\begin{minipage}{15cm}
\begin{flushright}
  \emph{To my parents,\\[2mm] who taught me the value of knowledge}
\end{flushright}
\end{minipage}
\newpage\
\thispagestyle{empty}
\newpage
\thispagestyle{empty}
\vspace{-0.1cm}
\begin{center}
{\LARGE \textsc{\textbf{ACKNOWLEDGEMENTS}}}
\end{center}

\vspace{0.3cm}
\noindent I thank my parents, who always supported me with words of encouragement 
and valuable advice, not only during this project, but throughout my academic 
life. Thanks to them, several times I gave up the idea of giving up. 
\vspace{0.05cm}

\noindent I would like to thank especially my supervisor Tereza Mendes for the guidance and support offered,
as well for her huge patience with the silly mistakes of a student who is starting his 
academic life. Undoubtedly, I could not have finished this project without her guidance, advice
and corrections. I am grateful for her believing in my potential and presenting me to wonderful 
opportunities that I could never imagine I would have and that proved essential for my formation.

\vspace{0.05cm}
\noindent One of these opportunities was the DESY Summer Student Proggramme of 2012,
at DESY-Zeuthen in Germany. During those two months in the mid-2012 I learned key concepts 
for this dissertation and I am really grateful for the opportunity, as well as for the work 
and dedication of Dr.\ Karl Jansen who was my supervisor at the time. I thank the 
collaboration of Aleksandra S\l{}apik during this time, who helped me at the developed project
and became a friend.

\vspace{0.05cm}
\noindent I would like to thank Cesar Uliana, Prof.~Dr.~Attilio Cucchieri and Dr.~Beno\^it Blossier,
for taking the time to read through this work, giving me useful suggestions to the text and collaborating 
to its precision.

\vspace{0.05cm}
\noindent I also thank the staff of the the Instituto de Física de São 
Carlos (IFSC) for helping me countless times with the paperwork necessary to carry out my 
research. I am grateful to the professors of IFSC for increasing my knowledge in several 
aspects of physics. 

\vspace{0.05cm}
\noindent This research was funded by a FAPESP (São Paulo Research Foundation) fellowship, grant \#2012/04811-0. Initial funding from CNPq is also acknowledged.
I especially thank CRInt/IFSC for financial support, allowing my participation in the DESY Summer Student Programme 
mentioned above and in the Lattice 2013 Conference.

\newpage\
\thispagestyle{empty}
\newpage
\thispagestyle{empty}
\begin{minipage}{15cm}
\begin{flushright}
  \emph{All things being consider'd, it seems probable to me, that God in 
  the Beginning form'd Matter in solid, massy, hard, impenetrable, moveable Particles of 
  such Sizes and Figures, and with such other Properties, and in such Proportion to Space,
  as most conduced to the End for which he form'd them; and that these primitive Particles,
  being solids, are incomparably harder than any porous Bodies compounded of them; even so 
  very hard as never to wear or break in pieces; no ordinary Power being able to divide what
  God Himself made one in the first Creation.}
  ---Isaac Newton, Opticks
\end{flushright}
\end{minipage}
\vfill

\begin{minipage}{15cm}
\begin{flushright}
\emph{Music happens to be a case of Artificial Quantization\ldots As a result of the effective quantization, all
significant descriptions of musical phenomena (\ldots) end up being expressed as dimensionless
ratios between integers, \ldots, henceforth known as pythagoreanisms. Pythagoras is known to have conjectured that it should be possible to express
the whole of physical science as pythagoreanisms\ldots  Considering
that in truth the world is quantized\ldots we have to concede
that Pythagoras' guess was a real hit.}
---Yuval Ne'eman, Symmetry and "Magic" Numbers or
From the Pythagoreans to Eugene Wigner\\Proceedings of the Wigner Centennial Conference
\end{flushright}
\end{minipage}
\vfill

\begin{minipage}{15cm}
\begin{flushright}
  \emph{The fact that, at least indirectly, one can actually see a single elementary particle --- 
  in a cloud chamber, say, or a bubble chamber --- supports the view that the smallest units 
  of matter are real physical objects, existing in the same sense that stones or flowers do.}
  ---Werner Heisenberg, The Physicist's Conception of Nature
\end{flushright}
\end{minipage}
\newpage\
\thispagestyle{empty}
\newpage\
\thispagestyle{empty}
\vspace*{-42pt}
\begin{abstract}
  \singlespace
	\vspace*{3pt}
  \noindent \bibauthorinfo \, {\bf \titleinfo}. \number\year. \pageref{LastPage} p. Dissertation (Master of Science) - Instituto de Física de São Carlos, Universidade de São Paulo, São Carlos, \number\year. 

  \vspace{0.5cm}
  \onehalfspacing
  \noindent In this dissertation we study potential models incorporating
	a nonperturbative propagator obtained from lattice simulations of a pure gauge theory.
	Initially we review general aspects of gauge theories,
the principles of the lattice formulation of quantum chromodynamics
(QCD) and some properties of heavy quarkonia, i.e.\ bound states of a heavy 
quark and its antiquark. As an illustration of Monte Carlo simulations of
lattice models, we present applications in the case of the harmonic 
oscillator and SU(2) gauge theory.
We then study the effect of using a gluon propagator
from lattice simulations of pure SU(2) theory as an input in a potential model for the 
description of quarkonium, in the case of bottomonium and charmonium.
We use, in both cases, a numerical approach to evaluate masses of quarkonium states.
The resulting spectra are compared to calculations using
the Coulomb plus linear (or Cornell) potential.

Potential models. Numerical methods.
\end{abstract}
\newpage\
\thispagestyle{empty}
\newpage
\thispagestyle{empty}
\selectlanguage{brazil}

\begin{abstract}
  \singlespace
  \noindent \bibauthorinfo \, {\bf \titleinfoPort}. \number\year. \pageref{LastPage} p. Dissertação (Mestrado em Ciências) -  Instituto de Física de São Carlos, Universidade de São Paulo, São Carlos, \number\year. 

  \vspace{0.5cm}
  \onehalfspacing
  \noindent Nesta dissertação estudamos modelos de potenciais com a incorporação de 
	um propagador não-perturbativo obtido através de simulações de rede para uma teoria de gauge pura. 
	Inicialmente fazemos uma revisão de aspectos gerais de teorias de gauge,
dos príncipios da formulação de rede da cromodinâmica quântica (QCD) e de algumas 
propriedades de quarkônios pesados, i.e.\ estados ligados de um quark pesado 
e seu antiquark. Como um exemplo de simulações de Monte Carlo de modelos de rede, 
apresentamos aplicações nos casos do oscilador harmônico e teorias de gauge SU(2).
Passamos então ao estudo do efeito de usar um propagador de glúon
de simulações na rede como input em um modelo de potencial para a descrição do quarkônio,
no caso do botômomio e do charmônio. Nós usamos, em ambos os casos, uma abordagem numérica 
para calcular as massas dos estados de quarkônio.
Os espectros resultantes são comparados com cálculos usando 
o potencial de ``Coulomb mais linear'' (ou potencial Cornell).

  \noindent Palavras-chave: Cromodinâmica quântica (QCD). Formulação na rede. Modelos de potencial. Métodos numéricos.
\end{abstract}
\newpage\
\thispagestyle{empty}
\newpage
\thispagestyle{empty}

\selectlanguage{english}

\clearpage
\listoffigures
\thispagestyle{empty}

\fancyhead[LE,RO]{}
\clearpage
\listoftables
\thispagestyle{empty}

\newpage\
\thispagestyle{empty}
\newpage
\thispagestyle{empty}

\fancyhf{}
\fancyhead[LE,RO]{}
\fancyhead[LO]{}
\fancyhead[RE]{}
\renewcommand{\headrulewidth}{0pt}
\renewcommand{\footrulewidth}{0pt}
\tocloftpagestyle{limpa}
\tableofcontents
\thispagestyle{empty}

\fancyhf{}
\fancyhead[LE,RO]{\bfseries\thepage}
\fancyhead[LO]{\bfseries\rightmark}
\fancyhead[RE]{\bfseries\leftmark}
\renewcommand{\headrulewidth}{0.5pt}
\renewcommand{\footrulewidth}{0pt}

\cleardoublepage
\chapter{Introduction}
\label{sec:introduction}
\thispagestyle{capitulo}

The study of bound states has been important throughout the history of
particle physics. In the beginning of the 20th century,
the need to understand the atom structure was one of the challenges that led
to quantum mechanics. When physicists started to probe the
structure of matter at even smaller scales, they found out that the atomic nucleus
was a bound state as well. Then, the
need to understand how protons and neutrons bind together to form the nucleus 
led to the proposal of the strong force, later to be described by
quantum chromodynamics (QCD).
According to QCD, a quantum field theory with 
local gauge symmetry, the proton itself is a bound state of \emph{quarks}.
The study of bound states due to the strong force can be implemented
directly from QCD, in a first principles approach, using the lattice formulation of gauge theory.
It is interesting to note that the quark model, proposed in 1964, became
widely accepted only in 1974, after the discovery
of the $J/\psi$ particle, which is a bound state of charm and 
anticharm quarks \cite{riordan1987hunting}. In this dissertation, we consider bound states of
a heavy quark and its antiquark in the nonrelativistic approximation,
using input from lattice-gauge-theory calculations. 

Gauge theories are the foundation of the \emph{Standard Model} of particle physics. Today the model
includes six flavors of quarks, carrying electric charge $+2/3$ (quarks $u$, $c$, $t$) 
or $-1/3$ (quarks $d$, $s$, $b$). There are six leptons as well: $e^-$, $\mu^-$, $\tau^-$ and their corresponding
neutrinos $\nu_{e^-}$, $\nu_{\mu^-}$, $\nu_{\tau^-}$. These are the fermions of the standard model.
The electromagnetic force is described in a unified way with the weak force, which is responsible for the decay 
of quarks and leptons. The bosons associated with the electroweak force are the photon (carrier
of the electromagnetic force) and the $W^\pm$, $Z^0$ particles, associated with the weak interaction. Since in the unified theory 
the $W^\pm$ and $Z^0$ bosons are massless,
conflicting with the experimental evidence, a new boson is introduced: the Higgs 
boson. The interaction of the particles with the Higgs
field explains the origin of their masses (although no prediction is made for the mass values, except 
for the mediating bosons $W^{\pm}$, $Z$). 
The existence of the Higgs particle was confirmed in 2012. The bosons associated with QCD 
comprise eight color-charged gluons, the force carriers of the strong interaction, responsible for ``gluing'' the
quarks. Gravity is not a part of the standard model, although the existence of
the graviton, which would be the carrier boson of the gravitational force, is speculated.
All the three forces of the standard model are described by gauge theories. Nevertheless,
for a long time, the importance of gauge theories was neglected. During the
first half of the 20th century, physicists regarded the gauge symmetry 
of electromagnetism as an accident. This ``accident'' was the only thing
keeping gauge invariance alive \cite{moriyasu1983elementary}. The
importance of gauge symmetries was only recognized later, in the 1950's,
when quantum field theory was being
developed.
Even then, the description of the strong force by a gauge theory 
remained a challenge for about two decades.

In the early 1930's, quantum mechanics was well developed.
The first steps toward a quantum field theory were being made,
following the theoretical prediction of antiparticles by Dirac.
This prediction was confirmed in 1932, with the discovery of the positron in cosmic rays.
It was understood in that period that the electromagnetic force was mediated 
by the exchange of photons.
Physicists of the time knew as well that an atom was made up of electrons,
protons and neutrons, with the neutron also being discovered in 1932. However, 
this picture had unsolved issues. One of these issues was the beta decay, which was unexplained
and required the conversion of a neutron into a proton. The process seemed to
violate conservation of energy. This led to the proposal by Pauli in 1931
of another, much lighter, neutral particle. The idea was used by Fermi in 1934 
to propose a new interaction, later
called the \emph{weak force}, which was responsible for the creation of the missing
particle. This particle was named neutrino some years later \cite{Ne'eman1996particle}.

Another issue was: what could hold dozens of positively charged particles bound
in a small region of space such as the atomic nucleus? And why was the neutron 
there and did not wander outside the nucleus? The 
solution was to postulate the existence of an unknown \emph{strong force}
binding the nucleus together. Also, this force should be of short range,
confined to the atomic nucleus. Under the picture of forces being 
mediated by the exchange of particles, Yukawa proposed in 1935 the existence 
of a \emph{massive} particle that would be the strong-force equivalent of the photon 
in electromagnetism. Experimental confirmation came only in 1947
when Lattes, Muirhead, Occhialini and Powell discovered the $\pi$-meson \cite{lattes1947}. 
In the following years, several other strongly interacting particles, or hadrons, were discovered.
The abundance of\, ``elementary'' particles made
physicists wonder if all of them were truly elementary, or if they were different
bound states of a handful of particles. Indeed, in 1964 Gell-Mann and Zweig independently 
proposed the existence of particles that were more elementary and composed the proton, neutron, pion and
the other recently discovered hadrons. Gell-Mann named them quarks.
In the late 1960s, the observation 
of high-energy electron scattering in a proton showed that the proton behavior was indeed
that of a bound state of three point-like particles \cite{Ne'eman1996particle}.

Despite this, the quark model was not well accepted until the so-called ``November revolution'' in 1974,
with the discovery of the $J/\psi$ meson. A compelling reason not to believe in quarks was that 
they were never observed in isolation. Also, the model predicted that quarks were
1/2-spin particles. But, for some hadrons such as the $\Delta^{+ +}$ particle,
three quarks would have to be in the same quantum state, thus conflicting with Pauli's exclusion principle.
The solution came from the 
hypothesis that these particles had
a new quantum number --- a new charge, the \emph{color charge} --- which could be of 
three different types: red, green or blue. The fact that they had different
quantum numbers allowed them to occupy the same 
state without violating Pauli's principle.
These new charges would be the source of the strong interaction. 
The force binds the quarks inside hadrons in such a way that
the bound states are all colorless (neutral, or white). More precisely, 
the three quarks inside a baryon 
are of different colors, resulting in a colorless state, 
or the quark and antiquark in a meson carry a color and its
respective anticolor. As a consequence, the color charge cannot
be observed. It is confined inside the hadrons.

The theory of the strong force is called \emph{quantum chromodynamics}, 
due to the analogy with color combinations. Since quantum 
electrodynamics (QED) had yielded excellent results and was unified
with the weak force by Weinberg and Salam
in 1967-1968, it was natural to suppose that QCD would be a gauge
theory as well. However, the conventional treatment presented problems
in the QCD case, since the theory was predicted to have a diverging 
coupling constant at physically relevant short distances.

The issue was solved in 1973 by Gross, Politzer and Wilczek \cite{PhysRevLett.30.1343,PhysRevLett.30.1346}, 
who showed that the theory's coupling tends to zero in the limit of high energy (short distances),
in accord with the deep-inelastic-scattering experiments mentioned above, which showed that 
the quarks behave as free particles at high energies. This phenomenon is called asymptotic
freedom. Conversely, this behavior suggested that the coupling constant could be strong at low 
energies (long distances), which would account for the confinement of quarks. 
It is important to note that the methods used to study QED cannot
be applied in the low-energy regime of QCD, since
they rely mainly on the fact that 
the QED coupling parameter, the fine-structure constant, is much smaller 
than one, making the theory suitable for perturbation theory. In fact,
asymptotic freedom tells us that perturbation theory is only applicable to QCD at 
high energies. In other words, the nature of the QCD coupling, 
which is small at high energies and large in the low-energy regime, 
determines that QCD bound states, such as the proton, cannot be described perturbatively.

A method to perform non-perturbative computations in QCD came in 1974 
when Wilson showed how to discretize the theory on a lattice, retaining its gauge symmetry exactly \cite{Wils74}. 
This enabled him to perform a strong-coupling expansion 
to calculate path-integral averages, using a similar method to the high-temperature 
expansion of classical statistical mechanics. In particular, 
Wilson could prove confinement of static quarks by a linear potential 
in the strong-coupling limit, i.e.\ when the lattice parameter $\beta$ is small.
In order to extend this result to the physical limit,
however, it is necessary to consider large values of the lattice parameter,
which was not possible using the strong-coupling expansion.
The lattice formulation was later used in
computational simulations to obtain nonperturbative expectation 
values of QCD observables at physical values of the coupling. 
It was then possible to show that, in the 
limiting case of infinitely massive quarks, 
the interaction potential between a quark and its respective antiquark 
rises linearly \cite{gattringer2009quantum}.

Bound states composed of a quark-antiquark pair are called quarkonia.
The consideration of the limit of heavy quarks (charm, bottom) in QCD
offers the opportunity of a more direct approach to several fundamental aspects 
of the theory, such as probing perturbative effects on the gluon fields \cite{Shifman:1995dn}. 
Also, the study of 
heavy quarkonia may be done in a quite precise way
by nonrelativistic potential models \cite{Lucha:1991vn}.
This is possible because
in this case the binding energy becomes much smaller than the
quark mass, justifying the nonrelativistic approximation. In this sense, 
these states may be viewed as the positronium of QCD, since we can try to 
obtain the system's several observed states using Schrödinger's equation. The method allows the 
study of all the energy spectra, while enabling an explicit physical interpretation
of the interactions between the particles in the bound state. 
The potential should be modeled with respect to the physical characteristics 
of the QCD running coupling, described above.
Generally, the potential is a sum of two terms: The first one, obtained perturbatively, comes
from the quark-antiquark interaction in the approximation of 
one-gluon exchange (OGE). It can be related to elastic scattering inside the meson
[see \cite[Chap.\ 6]{perkins2000introduction}].
The second one is a linearly-rising confining
potential, inspired by lattice calculations. The resulting potential
is called ``Coulomb-plus-linear potential'' or
Cornell potential
\begin{equation}
V(r)\;=\; -\frac{4}{3}\frac{\alpha_s}{r} \,+\, F_0\,r\,,
\end{equation}
where $\alpha_s$ and $F_0$ are suitable constants.
A list of common potentials is given in Ref.\ \cite{Lucha:1991vn}.
The binding energy of the quark-antiquark pair may be obtained 
by a numerical integration of the Schrödinger equation, 
yielding the mass spectrum of the system. 
Alternatively, the contribution to the potential coming from the OGE term
cited above may be obtained directly from the theory's gluon propagator.
This is what we propose in our study, using data for this propagator obtained from lattice simulations
of the theory [see e.g.\ \cite{Cucchieri:2007rg}].

In Chapter \ref{sec:Overview_Gauge_Field} of this dissertation we review the path-integral formalism in
gauge theories, both in the Abelian and the non-Abelian case. Also,
we discuss the quantization of gauge theories using path integrals and
comment on general features of QCD. We proceed in Chapter \ref{sec:LQCD_and_MCS}
to discuss the discretization of QCD on a lattice, which enabled
the strong-coupling expansion mentioned above and justifies the linearly rising potential that we will be
using later. The lattice discretization is well suited to computer simulations of the theory, 
using \emph{Monte Carlo} methods. As an application, we developed a program to simulate 
the case of pure $SU(2)$ gauge theory (i.e.\ without dynamical fermions)
and present some of our results. Chapter \ref{sec:Pot_Model_Heavy_Quark} details the
calculation of the OGE contribution to the potential as mentioned above and describes the
numerical method we use to find the eigenenergies in the Schrödinger equation. Our results, 
incorporating the nonperturbative gluon propagator into the OGE term of the potential,
are presented and discussed in Chapter \ref{sec:results}. Our conclusions
are drawn in Chapter \ref{sec:conclusion}. Appendix \ref{sec:notation} summarizes
the used notation, while Appendix \ref{sec:group_theory_review} reviews some aspects of
group theory.

\chapter{Overview of Gauge Field Theories}
\label{sec:Overview_Gauge_Field}
\thispagestyle{capitulo}

\epigraph{``\textit{Don't turn your back. Don't look away. And don't blink. Good Luck.}''}{The Tenth Doctor\\Doctor Who: Blink}

In this chapter we review some basic aspects of gauge theory. Firstly, we introduce the Feynman
path-integral formalism, showing that it is equivalent to the operator formalism of
nonrelativistic quantum mechanics \cite{feynman2012quantum}. It is possible to show that the
path-integral formalism leads to Hamilton's principle in the classical limit. The same procedure applies to quantum field
theories \cite{zee2010quantum}. We start by briefly reviewing the case of electromagnetism, emphasizing its Abelian 
symmetry. Then we build on top of it a gauge theory with non-Abelian symmetry \cite{moriyasu1983elementary,creutz1983quarks}. We also discuss the inclusion of fermions.
Finally, we perform the quantization of the classical gauge theory through the 
path-integral formalism. This provided a framework to study quantum field theory by
perturbation theory, based on an expansion in Feynman diagrams, which was remarkably successful for
QED. This treatment is applicable to QCD only in the high-energy regime, where the property
of asymptotic freedom holds.

We note that, due to the way in which it is defined, the path integral also provides a 
natural connection of quantum field theories with statistical mechanics \cite{Creutz:1980gp}.
In fact, it is one of the main ingredients of the lattice formulation of gauge theories 
\cite{creutz1983quarks}, which allows the nonperturbative 
investigation of QCD using statistical mechanical methods, such as the strong-coupling expansion and
Monte Carlo simulations. This will be instrumental in Chapter \ref{sec:LQCD_and_MCS}.
\section{Feynman's Path Integral}
\label{sec:path_integral}

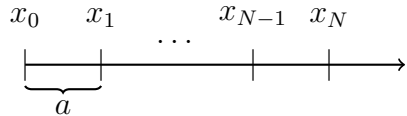
\begin{figure}[ht]
\centering
\caption{The process of discretizing the time. In our case $x_0 \equiv x_I$, $x_N \equiv x_F$.}
\label{fig:discretized_time}
\begin{tikzpicture}
 \node[label=above:{$x_0$}]     (xi)   at (0,0)   {|};
 \node[label=above:{$x_1$}]     (x1)   at (1,0)   {|};
 \node[label=above:{$\dots$}]   (dots) at (2,0)   {};
 \node[label=above:{$x_{N-1}$}] (xn-1) at (3,0)   {|};
 \node[label=above:{$x_N$}]     (xf)   at (4,0)   {|};
 \draw[thick,->] (0,0) -- (5,0);
 \draw [thick,decorate,decoration={brace,mirror},yshift=-8](0,0) -- (1,0) node[below,yshift=-1] at (0.5,0) {$a$};
\end{tikzpicture}
\sourcebytheauthour
\end{figure}

Our aim here is to derive the Feynman path integral from the usual operator formulation used in nonrelativistic 
quantum mechanics \cite{zee2010quantum,feynman2012quantum}. For simplicity, let us consider a one-dimensional 
system. We start by calculating 
the probability amplitude of a particle leaving a position $x_I$ to arrive at a position $x_F$ under the action of a 
potential $V(x)$ during a time $T$. 
The probability is given\footnote{We will adopt natural units for all our calculations, see Appendix \ref{sec:notation}.} by
$\langle x_F|e^{-i H T} |x_I\rangle$. We divide the time $T$ in 
$N$ steps of size $a$, so that $N a = T$ (see Fig.\ \ref{fig:discretized_time}). This will result in
\begin{equation}
 \langle x_F|e^{-i H T} |x_I\rangle \;=\; \langle x_F| \underbrace{e^{- i H a} e^{- i H a} \ldots e^{- i H a}}_\text{$N$ factors} | x_I \rangle\,.
\end{equation}
  
We then insert the completeness relation $\,\int{dx\, |x\rangle \langle x|} = 1\,$ between each two consecutive exponentials, to get
\begin{align}
\langle x_F|e^{-i H T} |x_I\rangle \;=\; \left(\prod_{i=1}^{N-1} \int dx_i\right)\langle x_F| 
         e^{- i H a} |x_{N-1}\rangle \langle x_{N-1}| e^{- i H a} | x_{N-2} \rangle \ldots  \quad \quad \hskip 6mm \nonumber \\ 
				\langle x_2 | e^{- i H a} |x_1\rangle \langle x_1|e^{- i H a}| x_I \rangle\,.
\end{align}
 
The interpretation of the above expression is that we 
compute the probability amplitude for the particle
leaving a position $x_i$ at time $t_i$ to reach the position $x_{i+1}$ 
at time $t_{i+1}$, for all the $N$ time slices. The next step is to multiply 
these amplitudes to obtain the probability amplitude for the particle leaving $x_I$ to arrive at $x_F$
in a time $T$ \emph{along a specific path}. We repeat this for all possible paths and sum the 
corresponding amplitudes to obtain the total probability amplitude 
$\langle x_F|e^{-i H T} |x_I\rangle$.

Let us proceed then to compute $\langle x_{i+1}| e^{- i H a} | x_i \rangle$. 
The Hamiltonian for a particle of mass $m$ is written as $H = K(p) + V(x)$,
corresponding to kinetic and potential terms respectively.
Since we plan to take the limit $a \rightarrow 0$ later,
we may approximate
\begin{equation}
e^{ -iaK(p)-iaV(x) }\; =\; e^{ -iaK(p) }\,e^{ -iaV(x) }\,e^{ \frac { a^2 }{ 2 } [K,V] }\dots \;\approx\; e^{ -iaK(p) }\,e^{ -iaV(x) }\,,
\label{eq:Zassenhaus_formula_approx}
\end{equation}
where we retained only the leading contribution in the Zassenhaus 
formula [see e.g.\ Eq.\ 1.2 of Ref.\ \cite{Casas20122386}].
Next, we use Eq.\ \ref{eq:Zassenhaus_formula_approx} and the completeness relation in the 
momentum basis $\,\int{dp\, |p\rangle \langle p|} = 2 \pi\,$ to write $\langle x_{i+1}| e^{- i H a} | x_i \rangle$ as
\begin{align}
  \label{eq:sum_over_paths}
  \langle x_{i+1}| e^{- i H a} | x_i \rangle \; & \approx\; \langle x_{i+1}|  e^{ -iaK(p) }\,e^{ -iaV(x) } | x_i \rangle \; \nonumber \\[1mm]
	&=\; e^{- i a V(x_{i})} \int \frac{dp}{2 \pi} \;\langle x_{i+1}| e^{- i  a \frac{p^2}{2m}}|p\rangle \langle p| x_i \rangle\,. 
\end{align}
We then use the relation $\langle x| p\rangle = e^{i p x}$ to obtain a Gaussian integral. We get
\begin{equation}
  \langle x_{i+1}| e^{- i H a} | x_i \rangle \;=\; \frac{e^{- i a V(x_{i})}}{2 \pi} \int dp\,
   \exp \left[-i a \frac{p^2}{2 m} + i p (x_{i+1}-x_i)\right]\,,
\end{equation}
which can be easily evaluated by completing squares
\begin{equation}
 \label{eq:time_slice_amplitude}
  \langle x_{ i+1 }|e^{ -i H a }|x_{ i }\rangle \;=\; \left( \frac { -im }{ 2\pi a }  \right) ^{ 1/2 }\exp\left\{ ia\left[ \frac { m }{ 2 } \left( \frac { x_{ i+1 }-x_{ i } }{ a }  \right) ^{ 2 }-V(x_{ i }) \right]  \right\} \,. 
\end{equation}
 
Now we have the amplitude for a particle leaving position $x_i$ to arrive at position $x_{i+1}$ in a time $a$.
We can insert Eq.\ \ref{eq:time_slice_amplitude} into Eq.\ \ref{eq:sum_over_paths} to obtain the total probability 
amplitude
\begin{align}
 \label{eq:discretized_path_integral}
 \langle x_{ F }|e^{ -iHT }|x_{ I }\rangle \;=\; \hskip 1cm & \nonumber \\[2mm] 
\left( \frac { -im }{ 2\pi a }  \right) ^{ \frac { N-1 }{ 2 }  } & \left( \prod _{ i=1 }^{ N-1 } \int  dx_{ i } \right) \exp\left\{ ia\sum _{ j=0 }^{ N-1 } \left[ \frac { m }{ 2 } \left( \frac { x_{ j+1 }-x_{ j } }{ a }  \right) ^{ 2 }-V(x_{ j }) \right]  \right\} \,.
\end{align}
 
But time is not a discrete quantity. This means that we must consider now the continuum limit, making $a$ progressively smaller 
(and, conversely, $N$ progressively larger). In this limit we define
\begin{align}
\left( \frac { -i m }{ 2\pi a }  \right) ^{ \frac{N-1}{2} } \left(\prod_{i=1}^{N-1} \int dx_i\right) & \;\to\; \int \mathcal{D}[x(t)], \nonumber \\[2mm] 
\frac{x_{j+1}-x_j}{a} & \;\to\; \dot{x},  \nonumber \\[2mm]
a \sum_{j=0}^{N-1} & \;\to\; \int dt,
\label{eq:path_integral_continuum_limit}
\end{align}
to obtain the final expression for the probability amplitude
\begin{equation}
 \langle x_F|e^{-i H T} |x_I\rangle  \;=\; \int{ \mathcal{D}[x(t)]\, \exp \left\{ i \int{ dt\, \left[\frac{m \dot{x}^2}{2} - V(x)\right]}\right\}}\,.
\end{equation}

We can see that the integral in the exponential is just the definition of the classical action. Finally, we rewrite the probability 
amplitude as
\begin{equation}
 \label{eq:Path_Integral}
 \langle x_F|e^{-i H T} |x_I \rangle  \; \equiv \; \int{ \mathcal{D}[x(t)]\, e^{i S[x(t)]}}\,.
\end{equation}

The inverse process, where we start from the path integral formulation and show its
equivalence with the Schrödinger equation, is shown e.g.\ in Ref.\ \cite{shankar1994principles}.

Let us notice that this formulation keeps a strong connection with classical 
physics. In fact, we are using the classical action to perform the 
calculations. It is also possible to show, as a limiting case,
that one recovers Hamilton's principle (i.e.\ $\delta S = 0$) in the
classical regime \cite{feynman2012quantum}. To see this, 
we need to reintroduce Planck's constant $\hbar$ into
Eq.\ \ref{eq:Path_Integral}
\begin{equation}
 \label{eq:Path_Integral2}
 \langle x_F|e^{-i H T/\hbar} |x_I \rangle  \;=\; \int{ \mathcal{D}[x(t)]\, e^{i S[x(t)]/\hbar}}\,.
\end{equation}
Now we treat $\hbar$ as a parameter, for which the limit $\hbar \rightarrow 0$
can be taken. Considering neighboring paths, the action $S[x(t)]$
varies slightly. However, in units of $\hbar$ this variation is enormous,
so that the phase $ e^{i S[x(t)]/\hbar}$ will be a rapidly varying (periodic)
function.
As a consequence, the contributions of these paths will cancel each other in the sum
and the corresponding probability amplitude will be zero. 
The only nonzero contribution comes from a path corresponding to an extremum
of the phase, i.e.\ the one for which the phase is stationary. This is 
the same as requesting that the action be extremized and therefore we recover
Hamilton's principle. An alternative to this qualitative argument 
is described in Ref.\ \cite{zee2010quantum}.

\section{Connection with Statistical Mechanics}
\label{sec:QM_Conn_w_SM}

The path integral has an analytic solution only in some simple cases. Also,
in the form in Eq.\ \ref{eq:Path_Integral}, it is not efficient to evaluate it 
numerically. Our aim here will be to relate the path integral to the 
partition function in statistical mechanics, which will enable us to 
use statistical mechanical techniques to address the problem. 
In particular, as will be discussed in Chap.\ \ref{sec:LQCD_and_MCS}, 
we may use Monte Carlo simulations 
to evaluate the discretized version of the path integral.
 
We transform from the Minkowski time $t$ to the Euclidean time $\tau = i t$. The
names come from the fact that, when we do this transformation, the Minkowski metric
is replaced by the Euclidean metric (see Appendix \ref{sec:notation}). This procedure
is known as Wick rotation \cite{PhysRev.96.1124}. We get
\begin{align}
 Z_{FI} &\;\equiv\; \langle x_F | e^{-i H T} |x_I\rangle \nonumber \\[1mm]
        &\;=\; \int{ \mathcal{D}[x(\tau)]\, \exp \left\{ \int{ d\tau\, \left[-\frac{m}{2} \left(\frac{dx}{d\tau}\right)^2 - V(x)\right]}\right\}}
         \;\equiv\; \int{ \mathcal{D}[x(\tau)]\, e^{-S[x(\tau)]}}\,,
 \label{eq:Z_FI}
\end{align}
where $S[x(\tau)]$ is the Euclidean action. Notice that, in Euclidean time, the Lagrangian assumes the expression of the Hamiltonian in
Minkowski time. Also, the path integral is now extremely similar to the partition function in statistical mechanics. This means that we 
can use statistical mechanical methods to evaluate it and to obtain other observable quantities such as the ground-state
energy of the system.
 
Before proceeding, we define the average of an operator $\ophat{A}$ as
\begin{equation}
 \label{eq:average_definition}
 \langle \ophat{A} \rangle \; \equiv \; \frac{\Tr \left(e^{-H T} \ophat{A}\right)}{Z} \; = \; \frac{\int \langle x| e^{-H T} \ophat{A} |x\rangle  \ dx}{Z}\,,
\end{equation}
where $Z$ is defined by [see Ref.\ \cite{Creutz:1980gp}]
\begin{equation}
 \label{eq:partition_func_definition}
 Z \;\equiv\; \Tr Z_{FI} \;=\; \int dx\, \langle x| e^{-H T} |x \rangle \;=\; \int{dx_I\, dx_F\, Z_{FI}\, \delta(x_F - x_I)}\,, 
\end{equation}
with $Z_{FI}$ given in Eq.\ \ref{eq:Z_FI}.

We remark that $\langle \ophat{A} \rangle$ stands for a statistical average, while $\langle x|  \ophat{A} |x\rangle$ is the quantum mechanical
expectation value. Note also that $Z$ is just $Z_{FI}$ with $x_F = x_I$ and integrated over all possible values for initial (or final)
points. Thus it is a path integral. By the same reasoning, the average $\langle \ophat{A} \rangle$ in Eq.\ \ref{eq:average_definition} 
is a path integral as well. We can write it as
\begin{equation}
 \label{eq:average_definition_path_int}
 \langle \ophat{A} \rangle = \frac{\int{ \mathcal{D}[x(\tau)]\, A[x(\tau)]\,e^{-S[x(\tau)]}}}{\int{ \mathcal{D}[x(\tau)]\, e^{-S[x(\tau)]}}}\,,
\end{equation}
where the Euclidean action $S[x(\tau)]$ is defined in Eq.\ \ref{eq:Z_FI}.
Note that we slightly changed our definition of path integral here, to include the end points of the paths.
This means that the product defined for discrete time steps in Eq.\ \ref{eq:path_integral_continuum_limit} now runs from $i = 0$ 
until $N$, where $x_0 = x_I $, $x_N = x_F$ and $x_I = x_F$. We will use this notation from now on.

If our system has $M$ ($M$ may be infinity) discrete energy levels, we can insert the completeness relation $\sum_{n=0}^{M-1} | 
n \rangle \langle n | = 1$ into Eq.\ \ref{eq:average_definition}, obtaining
\begin{equation}
 \langle \ophat{A} \rangle \;=\; \frac{1}{Z} \int{\sum_{n=0}^{M-1} \langle x| e^{-H T} | n \rangle \langle n | \ophat{A} |x\rangle  \ dx} \;=
\; \frac { 1 }{ Z } \sum_{n=0}^{M-1} e^{ -E_{ n }T }\langle n|\ophat { A } |n\rangle\,,
\label{eq:expected_value_op_A}
\end{equation}
where $n = 0$ is the ground state. Also, we interchanged the sum and the integral, and used the completeness relation $\int{dx\, | x \rangle \langle x |} = 1$.

Taking the limit $T \rightarrow \infty$, we find
\begin{equation}
 \label{eq:op_average_T_large}
\langle \ophat{A} \rangle \;\approx\; \langle 0 | \ophat{A} | 0 \rangle\,.
\end{equation}
Note that the absence of the term $e^{-E_0 T}$ comes from the fact that $Z \approx e^{-E_0 T}$ in this limit,
leading to a cancellation. This can be easily verified by setting $\ophat{A} = \dblone$ in Eq.\ \ref{eq:expected_value_op_A}.

In this way, we can isolate the ground energy level by studying the behavior of the Hamiltonian at large $T$, which
means evaluating the path integral in Eq.\ \ref{eq:average_definition_path_int} in this limit. For our calculations we 
make use of the virial theorem\cite{shankar1994principles}
\begin{equation}
 \label{eq:Virial_Theorem}
 \langle m \dot{x}^2 \rangle \;=\; \langle x V'(x) \rangle\,,
\end{equation}
which allows us to write Eq.\ \ref{eq:average_definition_path_int} as
\begin{equation}
 E_0 \;=\; \langle 0 | H | 0 \rangle \;=\; \lim_{T \rightarrow \infty}\langle \ophat{H} \rangle \;=\; \frac{\int{ \mathcal{D}[x(t)] \left[x V'(x)/2+V(x) \right]}\,e^{-S[x(t)]}}{\int{\mathcal{D}[x(t)]\,e^{-S[x(t)]}}}.
\end{equation}

For the calculation of the first excited state, we introduce the time-ordered (in Euclidean time)
connected $N$-point function\footnote{By time-ordered
we mean that we will not be able to commute the time-dependent operators evaluated at different times. The times are ordered
from higher to lower values of $\tau$, respectively from left to right, i.e.\ $\tau_1 > \tau_2 > \dots > \tau_n$.}
\begin{equation}
 \label{eq:gamma_n_c}
 \Gamma^{(N)}_c \;\equiv\; \left. \left[ \left(\prod_{i=1}^n \frac{\delta}{\delta J(\tau_i)} \right) \ln Z(J)\right] \right|_{J=0} \,,
\end{equation}
where $J \equiv J(\tau)$ is a function of the Euclidean time and represents an external source added
to the partition function [see e.g.\ Ref.\ \cite{binney1992theory}], i.e.\ $Z(J)$ generalizes $Z$ in Eq.\ \ref{eq:partition_func_definition} to
\begin{equation}
 Z(J)= \Tr \left\{ \exp \left[ -H T + \int{x(\tau') J(\tau') d\tau'}\right] \right\}\,.
 \label{eq:Z_J}
\end{equation}
Notice that in Eq.\ \ref{eq:gamma_n_c} we mean a functional derivative.
In particular, $\delta J(\tau)/ \delta J(\tau_i) = \delta(\tau-\tau_i)$. Also, by $J = 0$
we mean to set $J(\tau)$ to zero at all times.
 
If we choose $N = 2$ and $\tau_1=\tau$, $\tau_2=0$ to evaluate Eq.\ \ref{eq:gamma_n_c}, we obtain
\begin{equation}
\Gamma^{(2)}_c(\tau) \;=\; \frac{\Tr\left[ e^{-H T} x(\tau) x(0)\right]}{Z}-\frac{\Tr\left[ e^{-H T} x(\tau) \right]}{Z} \frac{\Tr\left[ e^{-H T} x(0) \right]}{Z}\,,
\end{equation}
which gives, using Eq.\ \ref{eq:average_definition},
\begin{equation}
 \Gamma^{(2)}_c(\tau) \;=\; \langle x(\tau)\,x(0) \rangle -\langle x(\tau) \rangle\langle x(0) \rangle\,.
 \label{eq:2-point_connect_corr_func-def}
\end{equation}

We then consider the limit $T \rightarrow \infty$, which allows us to use Eq.\ \ref{eq:op_average_T_large}, and rewrite
the expression above as
\begin{equation}
 \Gamma^{(2)}_c(\tau) \;=\; \langle 0| x(\tau)\,x(0)  |0\rangle - \langle0| x(\tau)|0 \rangle \langle0| x(0) |0\rangle \,.
\end{equation}
By inserting the completeness relation $\sum_{n=0} | n \rangle \langle n| = 1$ between $x(\tau)$ and $x(0)$, we find
\begin{equation}
 \Gamma^{(2)}_c(\tau) \;=\; \sum_{n\neq 0} \langle 0| x(\tau) | n \rangle \langle n| x(0) |0\rangle\,.
\end{equation}
 
We must highlight that the operators $x(\tau)$ and $x(0)$ are in the Heisenberg representation. This means that
 there is a time dependence in the above equation. We can extract it by transforming to the Schrödinger 
representation through the relation $x(\tau)_H = e^{H \tau} \,x_S\, e^{-H \tau}$. We arrive then at
\begin{equation}
 \label{eq:gamma_n_c_large_time}
 \Gamma^{(2)}_c(\tau) = \sum_{n\neq 0} \left|\langle 0| x | n \rangle \right|^2 e^{-(E_n-E_0)\tau}\,.
\end{equation}

If we consider the limit of large $\tau$, we actually manage to isolate the difference between the ground state and the first excited
state
\begin{equation}
  \Gamma^{(2)}_c(\tau) \;\approx\; \left|\langle 0| x | 1 \rangle \right|^2 e^{-(E_1-E_0)\tau}\,.
	\label{eq:gamma_n_c_E1_isolated}
\end{equation}

This approach is of central importance in numerical simulations, which are done based on 
the discretized version of Eq.\ \ref{eq:Z_FI}, namely
\begin{equation}
 \label{eq:discretized_Z_FI}
 Z_{FI} \;=\;
\left( \frac { -im }{ 2\pi a }  \right) ^{ \frac { N-1 }{ 2 }  } \left( \prod _{ i=1 }^{ N-1 } \int  dx_{ i } \right) \exp\left\{ -a\sum _{ j=0 }^{ N-1 } \left[ \frac { m }{ 2 } \left( \frac { x_{ j+1 }-x_{ j } }{ a }  \right) ^{ 2 } + V(x_{ j }) \right]  \right\} \,.
\end{equation}
In fact, all the above expressions for $\Gamma^{(2)}_c(\tau)$ carry over to discrete times,
i.e.\ by taking $x(\tau) \to x_i $ and $J(\tau) \to J_i$.
The two-point function is obtained directly from the thermalized $x_i$ configurations generated in a Monte Carlo simulation.

There are then two possibilities for determining $E_1$ from the computation of $ \Gamma^{(2)}_c(\tau)$. The first one consists in plotting the results for $\Gamma^{(2)}_c$ as 
a function of $\tau$ and then fitting to these points the function Eq.\ \ref{eq:gamma_n_c_E1_isolated}. This will work better
if we consider the larger values of $\tau$. The other possibility is to plot
the quantity $\,\ln\left[\Gamma^{(2)}_c (\tau+\Delta \tau)/\Gamma^{(2)}_c (\tau)\right]$. This will remove the constant
in front of the exponential in Eq.\ \ref{eq:gamma_n_c_E1_isolated} and then we consider the expression
\begin{equation}
 \label{eq:Calc_E1}
 E_1 - E_0 \;=\; -\frac{1}{\Delta \tau} \ln \left[ \frac{\Gamma^{(2)}_c(\tau+\Delta \tau)}{\Gamma^{(2)}_c(\tau)}\right] \,. 
\end{equation}

This second method is more reliable since we look for a plateau for the quantity on the RHS of the above equation
at large values of $\tau$ and read off $E_1 - E_0$ directly.
\section{Abelian Gauge Fields}
\label{sec:Gauge_Fields}
In this section 
we will briefly review the case of electromagnetism, emphasizing
its Abelian symmetry \cite{moriyasu1983elementary} . 

We postulate the existence of the vector potential $A_{\mu}$ and then define the field-strength tensor as
\begin{equation}
  \label{eq:field_tensor}
  F_{\mu \nu} \;=\; \partial_{\mu}A_{\nu} - \partial_{\nu}A_{\mu}\,.
\end{equation}

We adopt the usual covariant notation, with $\mu = 0,1,2,3$ and Einstein sum convention for repeated indices. 
Details of the used notation can be found in Appendix \ref{sec:notation}.

The theory's Lagrangian density is given by
\begin{equation}
 \label{eq:EM_lagrangian}
 \mathcal{L} \;=\; -\frac{1}{4} F_{\mu \nu} F^{\mu \nu} - j^{\mu} A_{\mu}\,.
\end{equation}
It is tempting to identify $j^{\mu}$ with the electromagnetic-charge current density.
However, we will refrain from doing this for a while. For now,
it will be treated as an external current, as was done in Eq.\ \ref{eq:Z_J},
which may be useful in computing correlation functions of $A_\mu(x)$. Later, in Section \ref{sec:Fermions},
we will be able to give physical meaning to it.

We want to substitute Eq.\ \ref{eq:EM_lagrangian} into the Euler-Lagrange 
equations
\begin{equation}
 \frac{\delta \mathcal{L}}{\delta A_{\nu}} \,-\, 
\partial_{\mu}\left[\frac{\delta \mathcal{L}}{\delta 
(\partial_{\mu} A_{\nu})} \right] \;=\; 0\,.
 \label{eq:Euler_Lagrange}
\end{equation}
Noting that
\begin{align}
\frac{\delta (j^{\mu} A_{\mu})}{\delta A_{\nu}} & \;=\; j^{\nu}\,, \\[2mm]
\frac{\delta F_{\lambda \sigma}}{\delta (\partial_{\mu} A_{\nu})} & \;=\; \delta_\lambda^\mu \delta_\sigma^\nu-\delta_\sigma^\mu \delta_\lambda^\nu\,,
\end{align}
we obtain 
\begin{align}
j^{\nu}-\frac{1}{4} \partial_{\mu} \left[\left(\delta_{\lambda}^{\mu}\delta_{\sigma}^{\nu}-\delta_{\sigma}^{\mu}\delta_{\lambda}^{\nu} \right) F^{\lambda \sigma}
                    +F_{\lambda \sigma  }\left(\delta^{\lambda \mu}  \delta^{\sigma \nu}  -\delta^{\sigma \mu}  \delta^{\lambda \nu} \right)  \right] \;=\; & \nonumber\\[2mm]
j^{\nu}-\frac{1}{4} \partial_{\mu}\left[ \left( F^{\mu \nu}-F^{\nu \mu} \right) + \left( F^{\mu \nu}-F^{\nu \mu} \right)  \right] \;=\; & \nonumber\\[2mm]
j^{\nu}-\frac{1}{2} \partial_{\mu}\left[ \left( F^{\mu \nu}+F^{\mu \nu} \right) \right] \;=\; & 0 \,,
\end{align}
where, in the last step, we used the fact that the field-strength 
tensor is antisymmetric with respect to its indices, which can be 
easily seen from its definition in Eq.\ \ref{eq:field_tensor}.
We thus have the equations of motion
\begin{equation}
\partial_{\mu}F^{\mu \nu} \;=\; j^{\nu}\,,
\label{eq:EM_motion_equation}
\end{equation}
which is the covariant form of the two Maxwell equations with sources, i.e.\
the Gauss law and the Maxwell-Ampère law. It remains to prove that $F_{\mu \nu}$
will obey the Gauss law for the magnetic field and Faraday's law.
Let us notice that
\begin{align}
\partial_\rho F_{\mu \nu} + \partial_\nu F_{\rho \mu} & \;=\; 
        \partial_\rho \partial_\mu A_\nu - \partial_\rho \partial_\nu A_\mu 
      + \partial_\nu \partial_\rho A_\mu - \partial_\nu \partial_\mu A_\rho \nonumber \\
& \;=\; \partial_\mu \partial_\rho A_\nu - \partial_\mu \partial_\nu A_\rho\
  \;=\;-\partial_\mu F_{\nu \rho}\,. 
\label{eq:deduction_Bianchini}
\end{align}
This leads us to the identity [see e.g.\ Ref.\ \cite{moriyasu1983elementary}]
\begin{equation}
\partial_\rho F_{\mu \nu} + \partial_\nu F_{\rho \mu} + \partial_\mu F_{\nu \rho} \;=\; 0\,,
\label{eq:Bianchi_identity}
\end{equation}
which accounts for the two missing laws.

Note that, if we differentiate Eq.\ \ref{eq:EM_motion_equation}
with respect to $\partial_\nu$, we obtain an antisymmetric expression 
on the LHS. The RHS, however, is not altered by the exchange of indices,
which leads us to the continuity equation $\,\partial_{\mu} j^{\mu} = 0\,$.

We are interested in understanding the symmetry of the theory. By this we mean a transformation that will keep the equations
of motion in Eq.\ \ref{eq:EM_motion_equation} invariant.
This transformation is called a gauge  transformation and is written as
\begin{equation}
 \label{eq:gauge_transform}
 A_{\mu} \;\to\; A_{\mu}' \;=\; A_{\mu} \,+\, \partial_{\mu} \Lambda(x)\,,
\end{equation}
where $\Lambda(x)$ is an arbitrary function of the space coordinates.
We say that the symmetry group of electromagnetism is the $U(1)$ group. We can explicitly see this by representing a given
group element $U$ of $U(1)$ by
\begin{equation}
 U \;=\; e^{i g_0 \Lambda(x)}\,,
\label{eq:group_element}
\end{equation}
where $g_0$ is a coupling constant, which in the present case can be identified with the elementary electric charge.
Writing $U$ in this way effectively associates a group element to each point in space-time. Using this representation,
we may write the gauge transformation as
\begin{equation}
 \label{eq:generalized_gauge_transform}
 A_{\mu} \;\to\; A_{\mu}' \;=\; U A_\mu U^{-1}
                                - \frac{i}{g_0}(\partial_\mu U)U^{-1}\,.
\end{equation}

Eq.\ \ref{eq:generalized_gauge_transform} seems to be an unnecessary complication if compared to Eq.\ \ref{eq:gauge_transform}. 
However, we will see that this formulation will allow us to easily use electromagnetism as a basis for a theory with 
$SU(N)$ gauge symmetry.

Let us use Eq.\ \ref{eq:gauge_transform} to see how the field-strength tensor $F_{\mu \nu}$ transforms under a gauge transformation. We get
\begin{equation}
 F_{\mu \nu} \;\to\; F_{\mu \nu}' \;=\; \partial_{\mu}A_{\nu}' \,-\, \partial_{\nu}A_{\mu}' \;=\; 
 \partial_{\mu}(A_{\nu} \,+\, \partial_{\nu} \Lambda) \,-\, \partial_{\nu}(A_{\mu} \,+\, \partial_{\mu} \Lambda) \;=\; F_{\mu \nu}\,.
\end{equation}

In this way we determine that $F_{\mu \nu}$ is invariant under a gauge 
transformation. Since $j^\mu$ is an external current, we have the freedom to ``define'' its behavior
under a gauge transformation. In particular, we may assume 
that $j^\mu$ will transform in such a way that the term
involving it will be kept invariant. 
When interpreting $j^\mu$ as physical current in Section \ref{sec:Fermions}, we 
will be able to see that the term $j^\mu A_\mu$ is not invariant\footnote{Note that if we choose $\,{j^\mu}'=j^\mu$, 
the term involving $j^\mu$ will transform as $j^\mu A_\mu \to j^\mu A_\mu + j^\mu \partial_\mu \Lambda$. Nevertheless,
this additional term does not affect the equations of motion (since the additional term does not depend on $A_\mu$) and, for now, 
no harm is done by imposing $j^\mu A_\mu$ to be gauge-invariant.}\label{footnote:current_abelian}. Indeed, the extra term
that appears is needed to keep the full Lagrangian (involving the gauge fields and fermions) invariant.
Since $F_{\mu \nu}$ is invariant under gauge transformation, we get that the Lagrangian will be invariant
as well
\begin{equation}
 \mathcal{L} \;\to\; \mathcal{L}' \;=\; -F_{\mu \nu} F^{\mu \nu} \,-\, j^{\mu} A_{\mu}\,.
\end{equation}

Also, notice that, since the gauge transformation does not affect $F_{\mu \nu}$, Eq.\ \ref{eq:Bianchi_identity}
will be invariant as well.
\section{Non-Abelian Theory}
\label{sec:Non-Abelian_Theory}
We proceed to build a gauge theory with non-Abelian symmetry \cite{creutz1983quarks}. 
We perform the same calculations as above but considering a non-Abelian gauge theory 
with $SU(N)$ symmetry. As above, we will write the group elements as
\begin{equation}
 U \;=\; e^{i g_0 \Lambda^a(x) \lambda^a}\,,
 \label{eq:group_element_SU_N}
\end{equation}
where $\lambda^a$ are the group generators (see Appendix \ref{sec:group_theory_review})
and $g_0$ is the bare coupling constant of the theory. The index $a$ is called 
a color index and runs from 1 to $N^2-1$ (see Appendix \ref{sec:notation}). The generators obey the relation
\begin{equation}
 [\lambda^a,\lambda^b] \;=\; i f^{a b c} \lambda^c\,,
 \label{eq:gen_commutators}
\end{equation}
where the braces stands for the commutator.
We adopt the normalization
\begin{equation}
 \label{eq:gen_normalization}
 \Tr(\lambda^a \lambda^b) \;=\; \frac{1}{2} \delta^{a b}\,.
\end{equation}

One important point to notice is that when we allow $U$ to be a group element of $SU(N)$
the vector potential becomes a non-commuting object as well, which can be 
represented in matrix form. 
We can decompose it as a linear combination of the $SU(N)$ generators $\lambda^a$
\begin{equation}
 A_{\mu} \;=\; A^{a}_{\mu} \,\lambda^{a}\,.
 \label{eq:matrix_potential}
\end{equation}

We can invert this relation by multiplying both sides of the above equation
by $\lambda^{b}$. Then we take the 
trace and use Eq.\ \ref{eq:gen_normalization} on the RHS. After 
rearranging the result, we obtain
\begin{equation}
 \label{eq:vector_potential}
 A^{a}_{\mu} \;=\; 2 \Tr (\lambda^{a} A_{\mu})\,.
\end{equation}

Note that the coefficient $A^{a}_{\mu}$ is a scalar.
This is the analogue of the vector potential 
in electromagnetism.

We use the definition of gauge transformation in Eq.\ \ref{eq:generalized_gauge_transform}
\begin{equation}
A_\mu \;\to\; {A_\mu}' \;=\; U A_\mu U^{-1} - \frac{i}{g_0}(\partial_\mu U)\,U^{-1}
\label{eq:generalized_gauge_transform_2}
\end{equation}
and proceed to analyze 
the behavior of the field-strength tensor under gauge transformations. 
Let us try to use the Abelian-case definition of $F_{\mu \nu}$ in Eq.\ \ref{eq:field_tensor}.
Noticing the property
\begin{equation}
 \partial_{\mu} U^{-1} \;=\; -U^{-1}\,(\partial_{\mu} U)\, U^{-1}\,,
 \label{eq:property_diff_group_element}
\end{equation}
which can be easily proven from $\,\partial_{\mu}( U U^{-1}) = 0$, we obtain
\begin{align}
F_{\mu \nu}' \;={}&\; \partial_{\mu}\left[U A_{\nu}U^{-1} - \frac{i}{g_0} (\partial_{\nu}U)U^{-1} \right] 
             \,-\,  \partial_{\nu}\left[U A_{\mu}U^{-1} - \frac{i}{g_0} (\partial_{\mu}U)U^{-1} \right]\nonumber \\[3mm]
             \;={}&\; U\left(\partial_{\mu}A_{\nu} - \partial_{\nu}A_{\mu}\right)U^{-1}
             \,+\, \big[\left(\partial_{\mu}U\right)A_{\nu} - \left(\partial_{\nu}U\right)A_{\mu}\big]U^{-1} \nonumber \\
          &\!\!+\, U\left[A_{\nu}\left(\partial_{\mu}U^{-1} \right) - A_{\mu}\left(\partial_{\nu}U^{-1}\right)\right]
             \,-\,\frac{i}{g_0}\left[ \left(\partial_{\nu}U\right)(\partial_{\mu}U^{-1})
						       -  \left(\partial_{\mu}U\right) (\partial_{\nu}U^{-1}) \right] \nonumber \\[2mm]
              \;={}&\; U F_{\mu \nu}\, U^{-1}
              \,+\, \left[(\partial_{\mu}U)(U^{-1} U) A_{\nu} - (\partial_{\nu}U) (U^{-1} U) A_{\mu}\right]U^{-1} \nonumber \\
          & \!\!-\, U\left[A_{\nu}\,U^{-1} (\partial_{\mu}U)\,U^{-1} - A_{\mu}\,U^{-1} (\partial_{\nu}U)\,U^{-1}\right] \nonumber \\
          & \!\!+\, \frac{i}{g_0}\left[(\partial_{\nu}U)\,U^{-1}(\partial_{\mu}U)\,U^{-1} - (\partial_{\mu}U)\,U^{-1}(\partial_{\nu}U)\,U^{-1} \right]  \nonumber \\[2mm]
              \;={}&\; U^{-1} F_{\mu \nu}\,U \,-\, \left[U A_{\nu}\,U^{-1},\,(\partial_{\mu}U)\,U^{-1}\right]
							\,-\,  \left[(\partial_{\nu}U)\,U^{-1},\,U A_{\mu}\, U^{-1} \right] \nonumber \\
          & \!\!+\, \frac{i}{g_0} \left[(\partial_{\nu}U)\,U^{-1},\,(\partial_{\mu}U)\,U^{-1}\right]\,,
\label{eq:wrong_tensor_transform}
\end{align}
where we introduced the commutator in the last step.

Such a transformation will induce extra terms with dependence on $A_{\mu}$, 
changing the final equations of motion. We need thus to redefine our
procedure, but we wish to keep our definition of a gauge transformation in
Eq.\ \ref{eq:generalized_gauge_transform_2}. Note that if we calculate 
$[A_{\mu}',A_{\nu}']$ we obtain
\begin{align}
\left[ A_{\mu}',A_{\nu}' \right] \;=\;& \left[U A_{\mu}\,U^{-1} - \frac{i}{g_0} (\partial_{\mu}U)\, U^{-1},\;U A_{\nu}\,U^{-1}-\frac{i}{g_0} (\partial_{\nu}U)\, U^{-1} \right] \nonumber \\[2mm]
\;=\;& \left[ U A_{ \mu  }\, U^{ -1 },\,U A_{ \nu  }\, U^{ -1 } \right]
+ \frac { 1 }{ g_{ 0 }^{ 2 } } \left[ (\partial _{ \nu  }U)\, U^{ -1 },\, (\partial_{ \mu  }U)\, U^{ -1 } \right]  \nonumber \\
& +\, \frac { i }{ g_0 } \left[ U A_{ \nu  }\,U^{ -1 },\,(\partial _{ \mu  }U)\,U^{ -1 } \right]
  + \frac { i }{ g_0 } \left[ (\partial _{ \nu  }U)\,U^{ -1 },\,U A_{ \mu  }\,U^{ -1 } \right]\,.
\end{align}

In the above expression, we notice the appearance of several terms that are in common with Eq.\ \ref{eq:wrong_tensor_transform}.
Taking this into account, let us rewrite the transformation of $F_{\mu \nu}$ as defined in Eq.\ \ref{eq:field_tensor}
\begin{equation}
 F_{\mu \nu} \;\to\; F_{\mu \nu}' \;=\; U F_{\mu \nu}\,U^{-1} \,-\,
i g_0 [U A_{\mu}\,U^{-1},U A_{\nu}\,U^{-1}] \,+\, i g_0\, [A_{\mu}',A_{\nu}'] \,,
\label{eq:wrong_tensor_transform_short}
\end{equation}
or, equivalently,
\begin{equation}
F_{\mu \nu}' \,-\,  i g_0\, [A_{\mu}',A_{\nu}'] \;=\; 
U \Big(F_{\mu \nu} -i g_0\, [A_{\mu},A_{\nu}]\Big)\,U^{-1}\,.
\end{equation}

This hints at a natural definition of the field-strength tensor in 
the non-Abelian case as
\begin{equation}
  F_{\mu \nu} \;=\; \partial_{\mu}A_{\nu} - \partial_{\nu}A_{\mu} - i g_0\, [A_{\mu},A_{\nu}]\,.
  \label{eq:generalized_field_tensor}
\end{equation}

Using this definition and recalling the transformation in
Eq.\ \ref{eq:generalized_gauge_transform}, the result is that the
field-strength tensor $F_{\mu \nu}$ behaves under gauge transformations as
\begin{equation}
 F_{\mu \nu} \;\to\; F_{\mu \nu}' \;=\; U F_{\mu \nu}U^{-1}\,.
\end{equation}
We see that, contrary to what happened in the Abelian case, the field-strength
tensor is not gauge-invariant.

Note that $F_{\mu \nu}$ has a matrix nature as well and can be decomposed in
terms of the generators just as was done with the vector potential
\begin{equation}
 F_{\mu \nu} \;=\; F_{\mu \nu}^c \lambda^c\,,
\end{equation}
with $F_{\mu \nu}^c$ given by
\begin{equation}
F_{\mu \nu}^c \;=\; 2 \Tr(\lambda^c F_{\mu \nu}) \;=\; \partial_{\mu}A_{\nu}^c - \partial_{\nu}A_{\mu}^c \,+\, g_0 f^{a b c} A_{\mu}^a\,A_{\nu}^b\,,
\label{eq:non-abelian_tensor}
\end{equation}
where we used Eqs. \ref{eq:generalized_field_tensor}, \ref{eq:gen_normalization} and \ref{eq:gen_commutators}.

It is natural to extend this notation to the current $j^{\mu}$, i.e.\
\begin{equation}
 j^{\mu} \;=\; j^{\mu,\,a} \lambda^{a}\,, \qquad j^{\mu,\,a} \;=\; 2 \Tr ( \lambda^{a} j^{\mu})\,.
\end{equation}

As was done for the Abelian theory, we will treat $j^{\mu,\,a}$ for now as an external current, which can be
useful for calculating correlation functions of the field $A_\mu^a$. It will gain a physical
interpretation as the current of color-charged fermions in Section \ref{sec:Fermions}. 

Now, having the Lagrangian density of electromagnetism in Eq.\ \ref{eq:EM_lagrangian} as motivation,
we write the Lagrangian density for the non-Abelian theory as
\begin{equation}
 \label{eq:non-abelian_lagrangian}
 \mathcal{L} \;=\; -\frac{1}{4} F_{\mu \nu}^a\, F^{\mu \nu,\,a} - j^{\mu,\,a} A_\mu^a\,.
\end{equation}

The above expression can be written in matrix notation by inserting a Kronecker delta in each term.
\begin{align}
 \mathcal{L} & \;=\; -\frac{1}{4}\, F_{\mu \nu}^{a}\, \delta^{a b} F^{\mu \nu,\,b} \,-\, 
                j^{\mu,\,a} \delta^{a b} A^{b}_{\mu} \nonumber \\[2mm] 
             & \;=\; -\frac{1}{2}\, F_{\mu \nu}^{a}\, F^{\mu \nu,\,b} \Tr(\lambda^{a} \lambda^{b}) \,-\, 
              2\, j^{\mu,\,a} A^{b}_{\mu} \Tr(\lambda^{a} \lambda^{b}) \nonumber \\[2mm]
             & \;=\; -\frac{1}{2} \Tr(F_{\mu \nu} F^{\mu \nu}) \,-\, 2\Tr(j^{\mu} A_{\mu})\,,
\end{align}
where we used the normalization in Eq.\ \ref{eq:gen_normalization}.

We remark that $j^\mu = j^{\mu,\,a} \lambda^a$ is an external current and therefore we may choose its transformation to keep
the terms involving it gauge-invariant. In any case, as in the Abelian theory, 
we will be able to interpret $j^\mu$ physically in Section \ref{sec:Fermions}, obtaining that 
$\Tr(j^\mu A_\mu)$ is not invariant under gauge transformations
\footnote{Note that if we choose ${j^\mu}'= U j^\mu U^{-1}$, then the term containing the current will transform as 
$\Tr(j^\mu A_\mu) \to \Tr(j^\mu A_\mu) - \frac{i}{g_0}\Tr(j^\mu U^{-1} \partial_\mu U)$. Notice that, again,
this extra term does not affect the equations of motion (see footnote in p.\ \pageref{footnote:current_abelian})and, for the moment, there is no impact in choosing
$\Tr(j^\mu A_\mu)$ to be invariant under gauge transformations.}. As happened before, it turns out that this 
extra term it is needed to keep
the full Lagrangian invariant under gauge transformation.

In this way, the Lagrangian is left invariant under gauge transformations
\begin{align}
\mathcal{L}' \;=\;& -\frac{1}{2} \Tr \left(F_{\mu \nu}' {F^{\mu \nu}}' \right) - 2\Tr \left({j^{\mu}}' A_{\mu}' \right) \nonumber\\[2mm]
             \;=\;& -\frac{1}{2} \Tr \left(U F_{\mu \nu} U^{-1} U F^{\mu \nu} U^{-1} \right)
						  - 2\Tr \left(j^{\mu} A_{\mu} \right) \nonumber\\[2mm]
						 \;=\;& -\frac{1}{2} \Tr \left(F_{\mu \nu} F^{\mu \nu} \right) - 2\Tr \left(j^{\mu} A_{\mu} \right) \;=\; \mathcal{L}\,.
\label{eq:lagrangian_gauge_field_transform}
\end{align}

Finally, we may derive the equations of motion. The Euler-Lagrange equations are 
\begin{equation}
 \frac{\delta \mathcal{L}}{\delta A_{\nu}^a} - \partial_{\mu}\frac{\delta \mathcal{L}}{\delta (\partial_{\mu} A_{\nu}^a)} \;=\; 0\,.
 \label{eq:non-abelian_euler-lagrange}
\end{equation}

Let us substitute Eq.\ \ref{eq:non-abelian_lagrangian} into Eq.\ \ref{eq:non-abelian_euler-lagrange} and use
\begin{align}
\frac{\delta{\left(j^{\rho,\,d}A_{\rho}^{d} \right)}}{\delta A_\nu^c} &\;=\; j^{\nu,\,c}\\[2mm]
\frac{\delta{F_{\rho \tau}^{d}}}{\delta A_{\nu}^{c}} &\;=\;  g_0 f^{a b d}
      \left(\delta_\rho^\nu \delta^{a c}A_{\tau}^{b} + \delta_\tau^\nu \delta^{b c}A_{\rho}^{a} \right) \;=\; 
			 g_0 f^{c b d}A_\tau^b \delta^\nu_\rho + g_0 f^{a c d} A_\rho^a \delta^\nu_\tau \\[2mm]
\frac{\delta{ \left(j^{\rho,\,d}A_{\rho}^{d} \right) } }{ \delta(\partial_{\mu}A_{\nu}^{c})} &\;=\; 0 \\[2mm]
\frac{\delta{F_{\rho \tau}^d} }{\delta(\partial_{\mu}A_{\nu}^{c})} &\;=\;
      \delta^\mu_\rho \delta^\nu_\tau \delta^{c d}-\delta^\nu_\rho \delta^\mu_\tau \delta^{c d}
\end{align}
along with its similar versions for $F^{\mu \nu,\,c}$, which can be  obtained through contractions
with the metric tensor (see Appendix \ref{sec:notation}), to obtain
\begin{align}
j^{\nu,\,c} \;= {} &\; -\frac{1}{4} \left( g_0\, f^{c b d}A_\tau^b \delta^\nu_\rho + g_0\, f^{a c d} A_\rho^a \delta^\nu_\tau \right)F^{\rho \tau,\,d}
                   -\frac{1}{4} \left( g_0\, f^{c b d}A^{\tau,\,b} \delta^{\rho \nu} + g_0\, f^{a c d} A^{\rho,\,a} \delta^{ \tau \nu} \right) F_{\rho \tau}^d \nonumber \\[1mm]
                 & +\frac{1}{4}\, \partial_{\mu}\Big[
								    \left( \delta^\mu_\rho \delta^\nu_\tau \delta^{c d}-\delta^\nu_\rho \delta^\mu_\tau \delta^{c d} \right)F^{\rho \tau,\,d} + 
								    \left( \delta^{\rho \mu} \delta^{\tau \nu} \delta^{c d}-\delta^{\rho \nu} \delta^{\tau \mu} \delta^{c d} \right)F_{\rho \tau}^d
								\Big] \nonumber \\[2mm]
            \;= {} &\; -\frac{1}{4} \left( g_0\, f^{c b d}A_\tau^b\, F^{\nu \tau,\,d} + g_0\, f^{a c d} A_\rho^a\, F^{\rho \nu,\,d} \right)
                   -\frac{1}{4} \left( g_0\, f^{c b d}A^{\tau,\,b}\, F^{\nu,\,d}_{\;\,\tau}  +  g_0\, f^{a c d} A^{\rho,\,a}\, F_\rho^{\;\,\nu,\,d} \right) \nonumber \\[1mm]
                  &+\frac{2}{4}\, \partial_{\mu}\left[\left( F_{\mu \nu}^c - F_{\nu \mu}^c \right)\right] \nonumber \\[2mm]
            \;= {} &\; -\frac{1}{2} \left( g_0\, f^{a c d}A_\rho^a\, F^{\rho \nu,\,d} + g_0\, f^{a c d} A_\rho^a\, F^{\rho \nu,\,d} \right) 
									+ \frac{2}{2}\, \partial_{\mu}F^{\mu \nu,\,c} \nonumber \\[2mm]
            \;= {} &\; \partial_{\mu}F^{\mu \nu,\, c}\, +\, g_0\, f^{c a d} A_{\mu}^{a}\, F^{\mu \nu,\,d}
            \;=\;      \partial_{\mu}F^{\mu \nu,\, c}\, +\, g_0\, f^{a b c} A_{\mu}^{a}\, F^{\mu \nu,\,b}\,,
						\label{eq:non-abelian_EoM}
\end{align}
which is the desired equation of motion. Here we used the cyclic property of the structure constant derived in Appendix \ref{sec:group_theory_review}
and in Ref.\ \cite{Agostinho}. We also used that $f^{a b c}$ is (in particular) antisymmetric in its first two indices, i.e.\ $f^{a b c} = - f^{b a c}$.
This can be easily seen from Eq.\ \ref{eq:gen_normalization}.

If we compare the result obtained in Eq.\ \ref{eq:non-abelian_EoM} with Eq.\ \ref{eq:EM_motion_equation}, we notice
a new term. This term arises from the fact that the usual derivative $\partial_\mu$ does not take into account that the
fields at different points do not commute. A derivative that takes this effect into account is called \emph{covariant derivative}.
Its form can be obtained if we notice that, when we calculated $\,\partial_\mu F^{\mu \nu}$ in the Abelian case, we were in fact
computing a divergence, which is the difference in flux between infinitesimal positions. When we perform this computation
for the non-Abelian theory, we see that the covariant derivative of a generic color-charged field $B^a$ can be defined as
\begin{equation}
D_\mu B^a \;\equiv\; \partial_\mu B^a+g_0 f^{a b c} A_\mu^b B^c\,.
\label{eq:covariant_diff_def}
\end{equation}

Using the matrix notation, the definition of the covariant derivative is 
\begin{equation}
D_\mu B \equiv \partial_\mu B - i g_0 [A_\mu,B]\,.
\label{eq:covariant_diff_def-matrix_notation}
\end{equation}
Also, note that if $B$ transforms as $B \to U B U^{-1}$, then the covariant derivative has the property
\begin{align}
(D_\mu B)' \;&=\; \partial_\mu(U B U^{-1}) -i g_0 \left[U A_\mu U^{-1} - \frac{i}{g_0}(\partial_\mu U) U^{-1}, U B U^{-1}\right] \nonumber \\
                   \;&=\; U\left(\partial_\mu B -i g_0 \left[ A_\mu, B \right]\right) U^{-1} + [(\partial_\mu U)U^{-1},U B U^{-1}] - [(\partial_\mu U)U^{-1},U B U^{-1}] \nonumber \\
									 \;&=\; U D_\mu BU^{-1}\,.
\label{eq:gauge_transf_cov_diff}
\end{align}
For details see Ref.\ \cite[Chap. 5]{moriyasu1983elementary}.

Therefore we may write the equations of motion as
\begin{equation}
 \partial_{\mu}F^{\mu \nu,\,a} \,+\, g_0\, f^{a b c}A_\mu^b\, F^{\mu \nu,\,c} \;=\; D_\mu F^{\mu \nu,\,a} \;=\; j^{\nu,\,a}\,.
 \label{eq:eq_motion_non-abelian}
\end{equation}

If we apply $\partial_\nu$ to the above equation, we find that the current $j^{\nu,\,a}$ is not necessarily conserved.
This will seem reasonable when we interpret $j^{\nu,\,a}$ as a fermionic current in Section \ref{sec:Fermions}. 
Also, let us note that the term $\,A_{\mu}^a A_{\nu}^b\,$ in Eq.\ \ref{eq:non-abelian_tensor} represents
the field interacting with itself and thus it must have color charge. 
This motivates us to rewrite Eq.\ \ref{eq:eq_motion_non-abelian} as
\begin{equation}
 \partial_{\mu}F^{\mu \nu,\,a} \;=\; j^{\nu,\,a} - g_0 f^{a b c}A_{\mu}^{b}F^{\mu \nu,\,c}\ \;\equiv\; J^{\nu,\,a}\,.
\end{equation}

We expect that  $J^{\nu,\,a}$ will be the full color current. Using the same reasoning as for the Abelian theory,
i.e.\ noticing that the LHS of the above equation is antisymmetric while the RHS is not,
we obtain the continuity equation
\begin{equation}
\partial_\nu J^{\nu,\,a} \;=\; 0\,.
\end{equation}

It remains now to show that the field $F_{\mu \nu}$ obeys a similar identity to Eq.\ \ref{eq:Bianchi_identity}.
Our attempt will be to just replace the usual derivative by the covariant derivative. 
With this, we obtain
\begin{align}
& D_\rho F_{\mu \nu} + D_\nu F_{\rho \mu} + D_\mu F_{\nu \rho} \;=\; \nonumber \\
& - i g_0 [\partial_\rho A_\mu,A_\nu] - i g_0 [A_\mu,\partial_\rho A_\nu]
  - i g_0 [\partial_\nu A_\rho,A_\mu] - i g_0 [A_\rho,\partial_\nu A_\mu] \nonumber \\
& - i g_0 [\partial_\mu A_\nu,A_\rho] - i g_0 [A_\nu,\partial_\mu A_\rho] 
  - i g_0 [A_\rho,\partial_\mu A_\nu] + i g_0 [A_\rho,\partial_\nu A_\mu] \nonumber \\
& - i g_0 [A_\nu,\partial_\rho A_\mu] + i g_0 [A_\nu,\partial_\mu A_\rho]
  - i g_0 [A_\mu,\partial_\nu A_\rho] + i g_0 [A_\mu,\partial_\rho A_\nu] \;=\; 0\,,
\label{eq:bianchi_identity_not_abelian}
\end{align}
where we used Eq.\ \ref{eq:deduction_Bianchini} and Jacobi's identity (see Appendix \ref{sec:group_theory_review}).
Also, using the property in Eq.\ \ref{eq:gauge_transf_cov_diff}, we discover that the LHS of this identity transforms in the same way as $F_{\mu \nu}$.
Moreover, if we left multiply the transformed identity by $U^{-1}$ and right multiply by $U$, we cancel out the outer matrices
and recover its original form.
\section{Fermions}
\label{sec:Fermions}

As was done for the gauge fields, we will first discuss the Abelian case.
We start by considering the Dirac equation
\begin{equation}
\left(i\slashed{\partial} - m\right) \psi \;=\; 0\,,
\label{eq:Dirac_equation}
\end{equation}
which can be obtained from the Lagrangian density
\begin{equation}
\mathcal{L}_\psi \;=\; \bar{\psi} \left(i\slashed{\partial} - m\right) \psi \,
\label{eq:fermionic_lagrangian_no_gauge_field}
\end{equation}
if we treat $\bar{\psi}$ and $\psi$ as independent dynamical variables. 
Here $\psi$ is the Dirac 
spinor, $\bar{\psi} = \psi^\dagger \gamma^0$, $\slashed{\partial} = \gamma^\mu \partial_\mu$ and
$\gamma^\mu$ are the $4 \times 4$ Dirac matrices. Details of the 
used notation can be found in Appendix \ref{sec:notation}.

In fact, using the Euler-Lagrange equations
\begin{equation}
                     \frac{\delta \mathcal{L}_\psi}{\delta \bar{\psi}} \;=\; 
\partial_{\mu}\left[ \frac{\delta \mathcal{L}_\psi}{\delta(\partial_{\mu} \bar{\psi})} \right]\,,
\label{eq:Euler_Lagrange_Fermions}
\end{equation}
and considering that there is no term $\partial_\mu \bar{\psi}$ in the Lagrangian density in
Eq.\ \ref{eq:fermionic_lagrangian_no_gauge_field},
the RHS of Eq.\ \ref{eq:Euler_Lagrange_Fermions} is zero and we get the Dirac equation
\begin{equation}
\frac{\delta \mathcal{L}_\psi}{\delta \bar{\psi}} \;=\; \left(i\slashed{\partial} - m\right) \psi \;=\; 0\,. 
\end{equation}

In addition, we will impose that the Lagrangian for the field theory have \emph{local gauge symmetry}, i.e.\ the transformation
\begin{align}
 \psi &\;\to\; e^{i g_0 \Lambda(x)} \psi \;=\; U\psi
 \label{eq:spinor_gauge_transform}  \\
 \bar{\psi} &\;\to\; \bar{\psi}\, e^{ -i g_0 \Lambda(x)} \;=\; \bar{\psi}\,U^{-1}\,
 \label{eq:spinor_dagger_gauge_transform}  
\end{align}
(see Eq.\ \ref{eq:group_element}) should keep the Lagrangian invariant.

However, when we perform this transformation in Eq.\ref{eq:fermionic_lagrangian_no_gauge_field}, the Lagrangian becomes
\begin{equation}
 {\mathcal{L}_\psi} '\;=\; i \bar{\psi}\, \slashed{\partial}\psi 
                         - g_0 \,\bar{\psi} \psi\, \slashed{\partial}\Lambda(x) - m \bar{\psi}\psi\,
\end{equation}
and therefore it is not gauge invariant. This result comes from the fact that the derivative $\partial_\mu$ introduces
into the Lagrangian fields that depend on positions $x$ and $\,x+\epsilon e_\mu$, i.e.\ the Lagrangian
is not local\footnote{Here $\epsilon$ is a small number and $e_\mu$ is a versor pointing in the direction $\mu$.}. Since we have a local symmetry,
it could have been expected that the Lagrangian would not be invariant.

The way to fix this is
to introduce a vector field that will connect the spinors at different positions, by making them transform
with the same phase [for details, see Ref.\ \cite{peskin1995introduction}]. We can verify that the Lagrangian
\begin{equation}
 \mathcal{L}_\psi \;=\; \bar{\psi} \left(i\slashed{\partial} + g_0\slashed{A} - m\right) \psi
 \label{eq:fermion_interacting_lagrangian}
\end{equation}
is invariant under local gauge transformations, where $A_\mu$ is the vector potential and transforms as
in Eq.\ \ref{eq:generalized_gauge_transform}
\begin{equation}
 A_{\mu} \;\to\; A_{\mu}' \;=\; A_{\mu} \,+\, \partial_{\mu} \Lambda(x) \;=\; U A_\mu U^{-1}
                                - \frac{i}{g_0}(\partial_\mu U)U^{-1} \,,
\label{eq:vec_pot_gauge_transform_ferm_sec}
\end{equation}
Indeed, under the transformations in Eqs.\ \ref{eq:spinor_gauge_transform}, \ref{eq:spinor_gauge_transform} and
 \ref{eq:vec_pot_gauge_transform_ferm_sec}, the Lagrangian in Eq.\ \ref{eq:fermion_interacting_lagrangian} becomes
\begin{equation}
 {\mathcal{L}_\psi}' \;=\; i \bar{\psi}\, \slashed{\partial}\psi 
                         - g_0\, \bar{\psi}\, (\slashed{\partial}\Lambda) \psi
                         + g_0\, \bar{\psi}\, \slashed{A}\psi 
                         + g_0\, \bar{\psi}\, (\slashed{\partial} \Lambda)\psi
                         - m \bar{\psi}\psi\, \;=\; \mathcal{L}\,.
\end{equation}

Although the vector field $A_\mu$ was introduced in the fermionic Lagrangian due to
a mathematical necessity, it has physical effects. We can see this by rearranging the Lagrangian in Eq.\
\ref{eq:fermion_interacting_lagrangian} to get
\begin{equation}
 \mathcal{L}_\psi \;=\; \bar{\psi}(i\slashed{\partial}-m)\psi + g_0 \bar{\psi}\gamma^\mu \psi A_\mu \;\equiv\;
                        \bar{\psi}(i\slashed{\partial}-m)\psi - j^\mu A_\mu\,,
\end{equation}
where we defined the current\footnote{Here $j_{\mu}$ is the electromagnetic-charge current density.
However, in the case of a pure gauge theory (fermion 
fields $\psi = 0$), it can be treated as an external current, 
as was done in Eq.\ \ref{eq:Z_J}. As in that case, this is useful 
when computing correlation functions, as explained below Eq.\ \ref{eq:EM_lagrangian}.} $j^\mu$ as
\begin{equation}
 j^\mu \;\equiv\; -g_0\, \bar{\psi}\,\gamma^\mu\psi\,.
\end{equation}
Notice that the term $j^\mu A_\mu$ is the same that appears in the Abelian-theory Lagrangian in Eq.\ \ref{eq:EM_lagrangian}.
The difference is that previously it was an ``artificial'' term that could lead to extra contributions
in the Lagrangian when performing a gauge transformation (see footnote on page \pageref{footnote:current_abelian}). Here it appears naturally and it is needed
to keep the Lagrangian invariant. We emphasize that if we are to consider the full theory (gauge fields plus
fermions) this term cannot be counted twice, i.e.\ the full Lagrangian becomes
\begin{equation}
 \mathcal{L} \;=\; \bar{\psi}(i\slashed{\partial}-m)\psi - j^\mu A_\mu - \frac{1}{4}F_{\mu \nu}F^{\mu \nu}\,.
\end{equation}

This Lagrangian density reproduces all the equations of motion previously deduced for the gauge fields. However,
Dirac's equation is modified and becomes
\begin{equation}
\left(i\slashed{\partial} + g_0 \slashed{A} - m\right) \psi \;=\; 0\,.
\end{equation}

This can be easily extended to the non-Abelian theory. In this case, the Dirac spinor
gains a column vector $c$ accompanying it\footnote{We adopt the notation used in Ref.\ \cite{griffiths2008introduction}
rather than the usual choice  of giving the wave function an extra index. This choice will simplify calculations
in Chapter \ref{sec:Pot_Model_Heavy_Quark}.}. The gauge transformation defined in Eq.\ \ref{eq:spinor_gauge_transform}
will act on this vector $c$ (notice that $U$ is now a matrix). The interpretation of $c$ is that 
the fermion is now a vector in an internal (color) space. What the gauge transformation does is to rotate 
this vector, making it point in another direction. And since we imposed gauge symmetry, this rotation 
does not cause physical consequences. In this way the fermionic Lagrangian density for the non-Abelian theory is
\begin{align}
\mathcal{L} \;&=\; \,\bar{\psi} c^\dagger \left(i\slashed{\partial} + g_0 \slashed{A} - m\right) c \psi 
            \;=\; \,\bar{\psi} c^\dagger \left(i\slashed{\partial} - m\right)c \psi \,+\, g_0\, \,\bar{\psi} c^\dagger\, \gamma^\mu \lambda^a \,c \psi \,A_\mu^a \nonumber \\
						\;&\equiv\; \,\bar{\psi} c^\dagger \left(i\slashed{\partial} - m\right)c \psi - j^{\mu,\,a} A_\mu^a\,,
\label{eq:fermionic_lagrangian_not_Abelian}
\end{align}
where we used Eq.\ \ref{eq:matrix_potential} and defined the fermionic current as 
\begin{equation}
j^{\mu,\,a}\;\equiv\; -\,g_0\, \bar{\psi}\, c^\dagger\, \gamma^\mu\, \lambda^a\, c \psi\,.
\end{equation}

Notice that the current appears again naturally and as a necessity to keep the Lagrangian
invariant under gauge transformations. It is completely analogous to the non-Abelian case. In this
way we obtain that the full Lagrangian density is
\begin{equation}
\mathcal{L} \;=\; \bar{\psi} c^\dagger (i\slashed{\partial}+g_0 \slashed{A}-m)\,c \psi - \frac{1}{2} \Tr \left(F_{\mu \nu} F^{\mu \nu}\right)\,.
\label{eq:Full_Lagrangian_non_abelian}
\end{equation}
The Dirac equation keeps its structure, the only addition being the vector $c$
\begin{equation}
(i\slashed{\partial}+g_0 \slashed{A}-m)c \psi \;=\; 0\,.
\label{eq:dirac_equation_non_Abelian}
\end{equation}

The equations of motion for the gauge field are not altered, since we did not introduce any new
term involving them in the Lagrangian density.

Since the vector potential was introduced to compensate phase differences due to the gauge symmetry,
it is natural to define the covariant derivative as
\begin{equation}
 D_\mu \;=\; \partial_\mu - i g_0 A_\mu\,.
 \label{eq:cov_diff_fermions}
\end{equation}

This definition seems to conflict with the definition in Eq.\ \ref{eq:covariant_diff_def-matrix_notation}
for the covariant derivative applied to a field given by a linear combination of the group generators.
However, it is possible to show that Eq.\ \ref{eq:covariant_diff_def-matrix_notation} can be derived
from Eq.\ \ref{eq:cov_diff_fermions}. Considering this definition, let us calculate $D_\mu (B c \psi)$,
where, as in Section \ref{sec:Non-Abelian_Theory}, $B$ is a quantity that transforms as $B \to U B U^{-1}$.
We will impose that the covariant derivative obey the Leibniz rule and get \cite{pokorski2000gauge}
\begin{equation}
 D_\mu (B c \psi) \;=\; (D_\mu B)c \psi + B D_\mu (c \psi) \;=\;  (D_\mu B)c \psi + B \partial_\mu (c \psi) - i g_0 B A_\mu c \psi\,.
\end{equation}
Notice that since $B c \psi$ behaves as a spinor, we have as well from Eq.\ \ref{eq:cov_diff_fermions}
\begin{equation}
 D_\mu (B c \psi) \;=\; (\partial_\mu B)c \psi + B \partial_\mu (c \psi) - i g_0 A_\mu B c \psi\,.
\end{equation}
From these two equations we obtain
\begin{align}
 (D_\mu B) c \psi \;&=\; (\partial_\mu B)c \psi + B \partial_\mu (c \psi) - i g_0 A_\mu B c \psi
                         -B \partial_\mu (c \psi) + i g_0 B A_\mu c \psi \nonumber\\
                  \;&=\; (\partial_\mu B)c \psi - i g_0 A_\mu B c \psi + i g_0 B A_\mu c \psi\,.
\end{align}

We obtain then that the covariant derivative acting on Lie-algebra elements
(see Appendix \ref{sec:group_theory_review}), such as the test field $B$, behaves as 
in Eq.\ \ref{eq:covariant_diff_def-matrix_notation}
\begin{equation}
D_\mu B \;=\; \partial_\mu B - i g_0 [A_\mu,B]\,.
\end{equation}

The expression in Eq.\ \ref{eq:Full_Lagrangian_non_abelian} is called the Yang-Mills Lagrangian, since 
it was Yang and Mills who first developed the idea in 1954, although
not aiming at the description of the electroweak or strong forces as it is used nowadays.

We finish this section remarking that we have a freedom to choose the sign in Eq.\ \ref{eq:group_element_SU_N}.
This would affect the sign of several of our definitions, however without impact on physical observables. 
Also, some textbooks choose to incorporate $g_0$ into $A_\mu$, rescaling it. This will change all equations involving $A_\mu$ 
by factors of $g_0$, in order to keep the physical results.
Details of several such transformations for the theory can be found in Ref.\ \cite{santosqcd}.

\section{Euclidean Action for Gauge Theories}

Lastly, we will define the Euclidean action for a pure gauge theory\footnote{In this dissertation, we will not
deal with simulations of fermions in the lattice. Therefore we do not need the fermionic portion of the 
Euclidean action.}. To obtain it, we extend the Wick rotation 
to be applied to any four-vector. In covariant notation (see Appendix \ref{sec:notation}), the rotation applied 
to a position vector is given by
\begin{equation}
 x^0 \;=\; -i x_4^{\rm E}\,, 
 \label{eq:wick_rotation_contra_vec}
\end{equation}
where $\,x^0 \equiv t$ and $\,x_4^{\rm E} \equiv \tau$. From now on, we will use a superscript ${\rm E}$ to indicate quantities in the Euclidean case.

Notice that, with the Euclidean metric, there is no difference between a covariant or contravariant vector 
and therefore we will always use a lower index. We postulate that all contravariant vectors will transform in
the same way. For a covariant vector, let us take as prototype $\partial_\mu$. When performing the Wick rotation
in Eq.\ \ref{eq:wick_rotation_contra_vec}, we obtain
\begin{equation}
\partial_0 \;\equiv\; \frac{\partial}{\partial x^0} \,\;=\;\, i \frac{\partial}{\partial x_4^{\rm E}} \;\equiv\; i \partial_4^{\rm E}\,.
\end{equation}
This means that $A_0 = i A_0^{\rm E} $ and therefore $F_{\mu \nu}$ will transform as
\begin{equation}
F_{0 i} \;=\; i F_{4 i}^{\rm E}\,,
\end{equation}
while the remaining components remain unchanged. The next step is to apply the Wick rotation
to $F_{\mu \nu} F^{\mu \nu}$. The expression
\begin{equation}
 F_{\mu \nu}F^{\mu \nu} \;=\; -F_{0 i} F_{0 i} - F_{i 0} F_{i 0} + F_{i j} F_{i j}
\end{equation}
is modified to
\begin{equation}
 F_{4 i}^{\rm E} F_{4 i}^{\rm E} \,+\, F_{i 4}^{\rm E} F_{i 4}^{\rm E} \,+\, F_{i j}^{\rm E} F_{i j}^{\rm E} \;=\; F_{\mu \nu}^{\rm E}F_{\mu \nu}^{\rm E}   \,.
\end{equation}

This allows us to obtain the Wick rotation for the Lagrangian term involving the gauge fields. We need 
just to notice that
\begin{equation}
\mathcal{L} \;=\; -\frac{1}{2} \Tr F_{\mu \nu} F^{\mu \nu} 
\end{equation}
will become
\begin{equation}
 -\frac{1}{2} \Tr F_{\mu \nu}^{\rm E} F_{\mu \nu}^{\rm E} \;\equiv\; -\mathcal{L}_{\rm E}  \,.
\end{equation}

Also, we are able to see that $\,d^4x = dx^0\,d^3x\,$ turns into $\,-i dx_4^{\rm E} \,d^3x^{\rm E} = -i d^4x^{\rm E}$
and we obtain that the expression
\begin{equation}
iS \;=\; i\int{d^4x}\, \mathcal{L}
\end{equation}
becomes
\begin{equation}
 - \int{d^4x^{\rm E}\, \mathcal{L}_{\rm E}} \;\equiv\; - S_{\rm E}  \,,
\end{equation}
where we defined the Euclidean action
\begin{equation}
 S_{\rm E} \;\equiv\; \int{d^4x^{\rm E}\, \frac{1}{2} \Tr F_{\mu \nu} F^{\mu \nu}}\,.
  \label{eq:full_euclidean_action}
\end{equation}

In Chapter \ref{sec:LQCD_and_MCS} it is assumed that the Wick rotation was already performed and when we 
refer to the system's action, we will be referring to Eq.\ \ref{eq:full_euclidean_action}. Notice that this
is the Euclidean analogue of the integral\footnote{In our work we will not consider field sources 
and therefore $j^{\mu,\,a} = 0$.} of Eq.\ \ref{eq:non-abelian_lagrangian}.

\section{Quantum Gauge Field Theory}
\label{sec:QGFT}
In this section we discuss the quantization of gauge theories, directly for the non-Abelian case, since 
the same procedure applies to the Abelian case in a straightforward way [for details see Ref.\ \cite{moriyasu1983elementary}].

Let us consider a fermion in a region of space without any gauge fields ($A_\mu = 0$). Its wave function obeys
\begin{equation}
 \left(i\slashed{\partial} - m\right)c_0 \psi_0 \;=\; 0\,.
 \label{eq:free_fermion}
\end{equation}

We now turn on the fields and make the Ansatz $\,c \psi = f(x) c_0 \psi_0$, where $f(x)$ is a matrix in color space,
dependent on the space time coordinates. We have then, using Eq.\ \ref{eq:dirac_equation_non_Abelian},
\begin{equation}
 i f(x) \,c_0\, \slashed{\partial}\psi_0 \,-\, m f(x) \,c_0\, \psi_0 \,+\, i\, [\partial_\mu f(x)]\, c_0 \gamma^\mu \psi_0 \,+\, g_0 A_\mu(x) f(x) \, c_0 \gamma^\mu \psi_0 \;=\; 0\,.
\end{equation}

Now using Eq.\ \ref{eq:free_fermion} we get
\begin{equation}
 \gamma^\mu \psi_0 \big[i \partial_\mu f(x) + g_0 A_\mu f(x)\big] c_0 \;=\; 0\,,
\end{equation}
which implies
\begin{equation}
f(x) \;=\; C \exp\left(\int{i g_0 A_\mu\, dx_\mu}\right)\,.
\end{equation}

This lead us to the conclusion that a particle interacting with a gauge
field gains a phase factor in its wave function
\begin{equation}
 \label{eq:phase_equation}
 c \psi \;\to\; \exp\left(\int_P i g_0 A_\mu\, dx_\mu\right)c \psi \;\equiv\; U(P) c \psi\,.
\end{equation}

Here we already chose the normalization constant in front of the exponential 
to be unity, due to the requirement of normalization of the wave function.
Notice that $A_\mu$ is now a matrix,
which will be different at each point in space. Matrices at different points do not commute with each other. Therefore, there is the need to
order the sequence in which the integral will be performed in Eq.\ \ref{eq:phase_equation}. The most natural way is 
to place the matrices associated with the beginning of the path to the left.

When we perform a gauge transformation, $U(P)$ will only be sensitive to the matrix elements at the end points of the path $P$.
\begin{equation}
 U(P) \;\to\; V(x_{\mu,\text{i}}) U(P) V^{-1}(x_{\mu,\text{f}})
\label{eq:group_gauge_transformation}
\end{equation}

This opens the possibility of defining a gauge-invariant quantity called the Wilson loop, given by
\begin{equation}
 W(C) = \Tr[U(C)]\,,
 \label{eq:def_continuum_Wilson_loop}
\end{equation}
where $C$ is a closed path.

This approach will be very important when we discretize the theory in Chap.\ \ref{sec:LQCD_and_MCS}.
\vskip3mm

So far the treatment we gave to the theory is a classical one. The Euler-Lagrange equations are obtained
applying the Hamilton principle, i.e.\ that the field configuration will be the one that extremizes the 
action
\begin{equation}
 \label{eq:QFT_Action}
 S = \int \mathcal{L}(A_\mu)\, d^4x_\mu \,.
\end{equation}
In other words, it corresponds to the fields that satisfy $\delta S = 0$. However, as we saw in Section 
\ref{sec:path_integral}, this is only an approximation, which is very good in the classical limit.
For treating quantum systems, we have to allow for any possible configuration, associating with it a probability
$p = e^{-S}$. The procedure that follows is the same of Sections \ref{sec:path_integral} and \ref{sec:QM_Conn_w_SM}. We want 
to measure some observable $\mathcal{O}$. The expected value to be measured will be a weighted average over 
all field configurations, each configuration with weight\footnote{Notice we are already using the Euclidean action 
defined in Eq.\ \ref{eq:full_euclidean_action}.} $e^{-S}$
\begin{equation}
 \label{eq:Path_Integral_QFT}
 \langle \mathcal{O} \rangle =  \frac{1}{Z} \int \mathcal{D}[A_\mu] \mathcal{O}(A_\mu) \exp\left[-S \right]\,,
\end{equation}
where $Z$ stands for
\begin{equation}
 Z =  \int \mathcal{D}[A_\mu] \exp\left[-S\right]\,.
\end{equation}

The nonperturbative computation of expectation values as in Eq.\ \ref{eq:Path_Integral_QFT}
is the main challenge in the study of non-Abelian  gauge theories, such as QCD, discussed in the next section.
\section{Quantum Chromodynamics (QCD)}

So far we treated general aspects of gauge theories. However, we are interested in the case of QCD, which
has some particular features. Firstly we remark that the QCD symmetry group is SU(3),
whose generators are, e.g.\ the Gell-Mann matrices
\begin{align}
\lambda_1 \;&=\; \frac{1}{2}\begin{pmatrix} 0 & 1 & 0 \\ 1 & 0 & 0 \\ 0 & 0 & 0 \end{pmatrix} &\qquad 
\lambda_2 \;&=\; \frac{1}{2}\begin{pmatrix} 0 & -i & 0 \\ i & 0 & 0 \\ 0 & 0 & 0 \end{pmatrix}&\quad 
\lambda_3 \;&=\; \frac{1}{2}\begin{pmatrix} 1 & 0 & 0 \\ 0 & -1 & 0 \\ 0 & 0 & 0 \end{pmatrix}&\nonumber \\[2mm]
\lambda_4 \;&=\; \frac{1}{2}\begin{pmatrix} 0 & 0 & 1 \\ 0 & 0 & 0 \\ 1 & 0 & 0 \end{pmatrix} &\qquad
\lambda_5 \;&=\; \frac{1}{2}\begin{pmatrix} 0 & 0 & -i \\ 0 & 0 & 0 \\ i & 0 & 0 \end{pmatrix} &\quad
\lambda_6 \;&=\; \frac{1}{2}\begin{pmatrix} 0 & 0 & 0 \\ 0 & 0 & 1 \\ 0 & 1 & 0 \end{pmatrix} &\nonumber \\[2mm]
\lambda_7 \;&=\; \frac{1}{2}\begin{pmatrix} 0 & 0 & 0 \\ 0 & 0 & -i \\ 0 & i & 0 \end{pmatrix} &\qquad
\lambda_8 \;&=\; \frac{1}{2\sqrt{3}} \begin{pmatrix} 1 & 0 & 0 \\ 0 & 1 & 0 \\ 0 & 0 & -2 \end{pmatrix}\,.& \,
\end{align}

A trivial feature to be added is to expand the number of fermions of the theory.
In our discussion, we used just only one fermionic field, implying just one type of fermion. However, the fermions of
QCD are the quarks, which come in six different flavors. With this in mind, we may write the full Lagrangian of QCD as
\begin{equation}
\mathcal{L} \;=\; \sum_{f=1}^{N_f}{\overline{\psi}^f c^\dagger_f\left(i\slashed{D}-m_f\right)c_f\, \psi^f}
            \,-\,\frac{1}{2} \Tr \left(F_{\mu \nu}F^{\mu \nu}\right)\,.
\label{eq:Full_Lagrangian_QCD}
\end{equation}

There is one other peculiar characteristic that arises from QCD when it is quantized, which we discuss qualitatively. 
The bare coupling itself (see Section \ref{sec:Non-Abelian_Theory}) can be thought of as
being very weak. As happens in QED, the presence of an isolated charge polarizes the vacuum around it, i.e.\  
virtual particles ``fluctuating'' in the vacuum get attracted to the real particle. However, differently from QED, the 
vacuum polarization does not screen the charge in QCD. The non-Abelian structure of the theory induces an opposite phenomenon, 
forming a cloud of virtual particles around the charge, which strengthens it. This may result in a stronger charge when it is observed 
far away from the bare charge.

Therefore, when we probe color charges
at high energies, i.e.\ using particles with very small wavelengths, we do not perceive all the cloud and the particle will behave as
an almost free particle, interacting weakly with the fields around it. Conversely, if we use low-energy particles, with large 
wavelengths, we will measure the charge of the real particle plus the charge of the vacuum-polarization cloud, resulting 
in the perception of a stronger charge. This is the property of asymptotic freedom, which accounts for the observation of the behavior
of quarks as almost free particles in the deep-inelastic-scattering experiments performed at SLAC in the end of the 1960's. It is also 
in line with the property of confinement of quarks inside hadrons \cite{Wilczek_Nobel}.

To study the theory in the low-energy regime, perturbative methods will not be effective and we must resort to nonperturbative techniques,
such as discussed in the next chapter.

\cleardoublepage
\chapter{Lattice QCD and Monte Carlo Simulations}
\label{sec:LQCD_and_MCS}
\thispagestyle{capitulo}

\epigraph{``\textit{Unfortunately, I found myself lacking the detailed knowledge and skills required
to conduct research using renormalized non-Abelian gauge theories\ldots What
was I to do, especially as I was eager to jump into this research with as little delay as possible?\ldots
I decided I might find it easier to work with a lattice version of QCD\ldots}''}{Kenneth G. Wilson\\In ``The Origins of Lattice Gauge Theory'' [see Ref.\ \cite{Wilson:2004de}]}

In this chapter we show how to discretize non-Abelian gauge theories on a lattice, which will be an important
step in the nonperturbative study of QCD. Firstly, we compute the expectation value of the Wilson loop
in the strong-coupling limit. We use this to show that, in this limit, we expect a linear interaction 
potential between a quark and an antiquark. Then, we show how a simulation of a lattice gauge theory can be implemented 
and show some results of such a simulation.

\section{Lattice Gauge Theory}
\label{sec:LGT}

Although the path-integral formalism presented in Chap.\ \ref{sec:Overview_Gauge_Field} is very intuitive and presents
an easy way to quantize a gauge theory, the integral can be evaluated analytically only in
some simple cases, such as the harmonic oscillator. Most cases, such as QED and the high-energy limit of QCD, 
in which the coupling constant is small, may be treated by perturbation theory. However, this approach cannot be used in the
case of a strong coupling. The calculation of the 
path integral in Eq.\ \ref{eq:Path_Integral_QFT} can instead be performed by resorting to the lattice formulation, which involves
a discretization of the Lagrangian in Eq.\ \ref{eq:non-abelian_lagrangian} on a grid, or lattice, in Euclidean space-time. 
This formulation allows a characterization of
confinement in the so-called strong-coupling limit and a nonperturbative treatment in the physical limit through numerical simulations.
Also, it introduces a natural and mathematically well defined regularization scheme,
since we are excluding ultraviolet momenta. In fact, the maximum momentum allowed on the lattice is associated
with the inverse of the lattice spacing $a$.
This ultraviolet cutoff avoids infinities in the calculations and is removed when taking
the continuum limit, while also performing the renormalization of the theory. The careful extrapolation
to the continuum limit is therefore an essential step in numerical simulations of lattice QCD, in order to
avoid discretization errors. Note that when performing
computer simulations we will have to deal with finite lattices.
This corresponds to a infrared cutoff, which can be another source of systematic errors. The 
presence of such artifacts can be minimized
by using periodic boundary conditions for the lattice and by using large lattices.

\begin{figure}[h]
 \centering
\caption{Visual representation of the association of a group element $U_\mu(x)$ to a link of the lattice.}
\label{fig:link_def}
\begin{tikzpicture}
 \draw[help lines] (-2,-1) grid (2,2);
 \node[label=left:{$U_\mu(x)\!\!$}]     (Umu)   at (0,0.5)   { };
 \fill (0,0) circle (0.075) node[below]{$x$};
 \fill (0,1) circle (0.075) node[above]{$x+a\hat{\mu}$};
 \begin{scope}[very thick,decoration={
    markings,
    mark=at position 0.7 with {\arrow{latex}}}
    ] 
    \draw[postaction={decorate}] (0,0)--(0,1);
  \end{scope}
\end{tikzpicture}
\sourcebytheauthour
\end{figure}
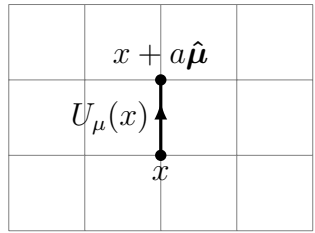

In Section \ref{sec:QGFT} we considered a particle traveling along a given path. Now we wish to 
describe this traveling particle on a space-time lattice of spacing
$a$ between neighboring sites. Also, we recall that in order to make a connection between the path-integral formalism
and statistical mechanics, as was done in Section \ref{sec:QM_Conn_w_SM}, it was necessary to perform the
Wick rotation $\tau = i t$, meaning that the lattice will be in Euclidean space.
The particle traveling through space-time can then be described as it hopping between neighboring sites 
on this lattice. We showed also in Section \ref{sec:QGFT} that a particle traveling along an infinitesimal
path gains a phase factor in its wave function, which is a group element (see Eq.\ \ref{eq:phase_equation}).
When discretized, the infinitesimal path becomes a single hop between neighboring sites. As the particle 
hops between these sites, it gains the corresponding phase factor. In this way, 
we can associate a $SU(N)$ group element $U_\mu(x)$ to each link between two adjacent sites (see Fig.\ \ref{fig:link_def}).
The notation used means that
this group element corresponds to the lattice link between the points $x$ and $x + a e_\mu$, where $ e_\mu$ stands
for a unit vector to indicate one of the four directions in (Euclidean) space.

Eq.\ \ref{eq:group_gauge_transformation} tells us directly how the group element $U_\mu(x)$ behaves under gauge
transformations. For a particular gauge transformation, we are allowed to associate a group
element $V(x)$ to each lattice point and $U_\mu(x)$ will transform as
\begin{equation}
  U_\mu(x) \;\to\; \, V(x) U_\mu(x)\, V^\dagger(x+a e_\mu)\,,
  \label{eq:lattice_gauge_transformation}
\end{equation}
where $V(x)$ is a $SU(N)$ group element as well.

These group elements are related to the field $A_\mu$, as seen in Eq.\ \ref{eq:phase_equation}.
Let us approximate the integral along the line to its value at the midpoint. We obtain
\begin{equation}
U_\mu(x) \;=\; \exp \Big[i g_0 a A_\mu(x+a e_\mu/2)\Big]\,.
\label{eq:relation_field_group_element}
\end{equation}
Thus, once we attribute a group element to every link of the lattice, we 
have fixed a field configuration. If a different configuration is desired, it will be enough to just change
the group elements associated with each link.
Also, Eq.\ \ref{eq:relation_field_group_element} defines a forward direction. With this, we can define oriented 
closed loops on the lattice, as was done in Eq.\ \ref{eq:def_continuum_Wilson_loop} in the continuum. This is 
achieved by multiplying the group elements of links in the closed loop and taking the trace
\begin{equation}
W(C) \;=\; \Tr\left[\prod_{x} U_\mu(x)\right]\,,
\end{equation}
where the variable $x$ in the product runs over all the lattice points in the loop.
Note that this will require some links to be oriented in the backward direction. When this happens, we must use the inverse 
element $U^{-1}$, or equivalently $U^\dagger$. It is important to notice that, as in the 
continuum case, these closed loops are gauge-invariant (see Eq.\ \ref{eq:lattice_gauge_transformation}). 
They are called Wilson loops and play a central role in lattice studies, both for analytical calculations 
and for computational simulations.

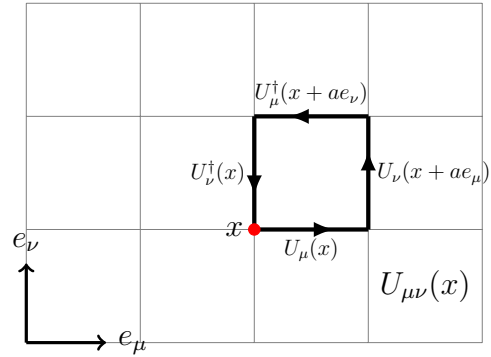
\begin{figure}[ht]
 \centering
\caption{A graphical representation of a plaquette.}
\label{fig:plaquette_def}%
\begin{tikzpicture}[scale=1.5]
 \draw[help lines] (-2,-1) grid (2,2);
 \draw[very thick,->] (-2,-1) -- (-1.3,-1) node[right]{$e_\mu$};
 \draw[very thick,->] (-2,-1) -- (-2,-0.3) node[above]{$e_\nu$};

\begin{scope}[ultra thick,decoration={
    markings,
    mark=at position 0.7 with {\arrow{latex}}}
    ] 
    \draw[postaction={decorate}] (0,0)--(1,0) node[below,scale=0.7] at (0.5,0) {$U_\mu(x)$};
    \draw[postaction={decorate}] (1,0)--(1,1) node[right,scale=0.7] at (1,0.5) {$U_\nu(x+ae_\mu)$};
    \draw[postaction={decorate}] (1,1)--(0,1) node[above,scale=0.7] at (0.5,1) {$U_\mu^\dagger(x+ae_\nu)$};
    \draw[postaction={decorate}] (0,1)--(0,0) node[left, scale=0.7] at (0,0.5) {$U_\nu^\dagger(x)$};
  \end{scope}
	\filldraw[red] (0,0) circle (0.05) node[left,black] {$x$};
 
  \node at (1.5,-0.5) {$U_{\mu \nu}(x)$};

\end{tikzpicture}
\sourcebytheauthour
\end{figure}

Finally, we need to build an action to use in the statistical weight. This action must correspond, in the 
limit $a \rightarrow 0$, to the integral
of Eq.\ \ref{eq:full_euclidean_action}. We note that $F_{\mu \nu}^a$ 
is a generalized form of curl. This information can be used to motivate the action in the following way. 
Firstly we define as a \emph{plaquette} the elementary square on a lattice, i.e.\ a  square of side $a$ 
(see Fig.\ \ref{fig:plaquette_def}).
Each one of these plaquettes corresponds to the smallest possible closed loop on the lattice and is 
described by Eq.\ \ref{eq:def_continuum_Wilson_loop}. We then relate the discretized curl to the
plaquettes. The resulting action, introduced by Wilson in 1974 \cite{Wils74}, is given by
\begin{equation}
 S \;=\; \sum_x{ \sum_{C(\mu,\nu)} S_{\mu \nu}(x)}\,,
 \label{eq:discrete_action}
\end{equation}
where $C(\mu,\nu)$ stands for the six possible combinations of $\mu$ and $\nu$, i.e.\ we have to 
consider the plaquettes at point $x$ on all the six planes \,[$(1,2)$, $(1,3)$, $(1,4)$, $(2,3)$, $(2,4)$, $(3,4)$].
We also sum over all lattice points $x$. Essentially, $S_{\mu \nu}(x)$ is a quantity that depends 
on a given plaquette and we are summing over all \emph{oriented} plaquettes on the lattice. We have
\begin{align}
 S_{\mu \nu}(x) & \;\equiv\; \beta\left\{1-\frac{1}{N}\Re \Tr\Big[U_\mu (x)\, U_\nu (x+a e_\mu)\, U_\mu^\dagger(x+a e_\nu)\, U_\nu^\dagger(x)\Big]\right\} \nonumber \\
                & \;\equiv\; \beta\left\{1-\frac{1}{N}\Re \Tr U_{\mu \nu} (x) \right\}\,,
 \label{eq:plaquette_action}
\end{align}
where $\Re$ indicates the real part of the trace and $N$ is the dimension of the group matrices, e.g.\ $N = 2$ for the $SU(2)$ group. Notice that $S_{\mu \nu}$
is clearly gauge-invariant and therefore the Wilson action preserves gauge symmetry exactly, for any value of $a$.
The factor $\beta$ is related to the bare coupling $g_0$ by
 \begin{equation}
  \beta \;=\; \frac{2N}{g_0^2}\,,
	\label{eq:beta_lattice_parameter}
 \end{equation}
in order to reproduce the gauge action in Eq.\ \ref{eq:full_euclidean_action} (with $\,j^{\mu,\,a} = 0$) as
we take the limit of zero lattice spacing. To verify this, let us consider a particular plane $(\mu,\nu)$. We use the 
Baker–Campbell–Hausdorff formula to reduce $U_{\mu \nu}$ (defined in Eq.\ \ref{eq:plaquette_action}) to a single exponential. 
We will keep in our 
expansion terms up to order $a^2$ and therefore we can ignore terms containing commutators of commutators. 
We Taylor expand $A_\mu(x)$ assuming small values of $a$. The first step consists in calculating $U_\mu(x) U_\nu(x+a e_\mu)$
\begin{align}
 U_\mu\left(x\right) U_\nu\left(x+a e_\mu\right)  \;&=\; e^{i g_0 a A_\mu\left(x+\frac{a}{2}e_\mu\right)}\,e^{i g_0 a A_\nu\left(x+a e_\mu +\frac{a}{2}e_\nu\right)} \nonumber \\[2mm]
							       \;&\approx\; e^{i g_0 a \left[A_\mu\left(x+\frac{a}{2}e_\mu\right) + A_\nu\left(x+a e_\mu + \frac{a}{2}e_\nu\right)\right]}
						       e^{\frac{g_0^2 a^2}{2} \left[A_\mu(x+\frac{a}{2}e_\mu),\,A_\nu(x+a e_\mu + \frac{a}{2}e_\nu)\right]}\,.
\label{eq:taylor_expansion_pot_vector}
\end{align}
For the calculation of $U_{\mu \nu}(x)$ we need also the expression for $U_\nu(x) U_\mu(x+a e_\nu)$. Notice that if we make the transformation in the indices
$\mu \to \nu$ and $\nu \to \mu$, we obtain that $U_\mu(x) U_\nu(x+a e_\mu) \to U_\nu(x) U_\mu(x+a e_\nu)$ and therefore these transformations allow us to get an expression 
for $U_\nu(x) U_\mu(x+a e_\nu)$ from Eq.\ \ref{eq:taylor_expansion_pot_vector}. We are now able to calculate $U_{\mu \nu}(x)$
\begin{align}
U_{\mu \nu}(x) \;=\;{}& \left[U_\mu\left(x\right) U_\nu\left(x+a e_\mu\right) \right]\left[U_\nu\left(x\right) U_\mu\left(x+a e_\nu\right) \right]^\dagger \nonumber \\[3mm]
         \approx\;{}& \exp\bigg\{i g_0 a A_\mu\left(x+a e_\mu /2\right) + i g_0 a A_\nu\left(x+a e_\mu + a e_\nu / 2\right) \nonumber \\[1mm]
		    &       -\frac{g_0^2 a^2}{2} \left[A_\mu\left(x+a e_\mu /2\right),\,A_\nu\left(x+a e_\mu + a e_\nu/2\right)\right] \nonumber \\[1mm]
		    &          - i g_0 a A_\nu\left(x+a e_\nu/2\right) - i g_0 a A_\mu\left(x + a e_\nu + a e_\mu /2\right) \nonumber \\[1mm]
		    &       -\frac{g_0^2 a^2}{2} \left[A_\nu\left(x+a e_\nu/2\right),\,A_\mu\left(x+a e_\nu + a e_\mu/2\right)\right]^\dagger\bigg\} \, \times \nonumber \\[1mm]
		    & \exp\bigg\{ g_0^2 a^2[A_\mu\left(x+a e_\mu /2\right),\,A_\nu\left(x+a e_\nu/2\right)] \nonumber \\[1mm]
		    & + g_0^2 a^2 [A_\mu\left(x+a e_\mu /2\right),\,A_\mu\left(x + a e_\nu + a e_\mu /2\right)] \nonumber \\[1mm]
		    & + g_0^2 a^2 [A_\nu\left(x+a e_\mu + a e_\nu / 2\right),\, A_\nu\left(x+a e_\nu/2\right)] \nonumber \\[1mm]
		    & + g_0^2 a^2 [A_\nu\left(x+a e_\mu + a e_\nu / 2\right),\, A_\mu\left(x + a e_\nu + a e_\mu /2\right)]\bigg\}\,.
\label{eq:plaquette_expansion_1}
\end{align}

We proceed to perform the Taylor expansion around small $a$. In the first exponential in Eq.\ \ref{eq:plaquette_expansion_1} we expand
up to first order in $a$, since each $A_\mu$ is accompanied by an $a$ and therefore this will result in terms of order up to $a^2$. For
the second exponential, each term on it is already of order $a^2$, so the expansion must be carried until zeroth order. It is easy to see
that with this expansion the argument of the second exponential will be zero and therefore we focus in the expansion on the first exponential. We obtain
\begin{align}
 U_{\mu \nu}(x) \;&=\; \exp\big\{-i g_0 a^2 \left[\partial_\mu A_\nu(x)-\partial_\nu A_\mu(x)\right]-g_0^2 a^2 \left[A_\mu(x),\,A_\nu(x)\right] +\mathcal{O}(a^4)\big\} \nonumber \\
		\;&=\; \exp\left[ -i g_0 a^2 F_{\mu\nu} +\mathcal{O}(a^3)\right]\,.           
\end{align}

We proceed to Taylor expand the exponential
\begin{equation}
 U_{\mu \nu}(x) \;=\; \dblone \,-\, i g_0 a^2 F_{\mu\nu} \,-\, \frac{g_0^2 a^4}{2} F_{\mu \nu}^2 \,+\, \mathcal{O}(a^6)\,.
 \label{eq:plaquette_expansion}
\end{equation}
We use Eq.\ \ref{eq:plaquette_expansion} in Eq.\ \ref{eq:plaquette_action} to obtain the 
contribution of a single plaquette to the action in the limit of small $a$
\begin{align}
 S_{\mu \nu}(x) \;&=\; \beta\left[1 - \frac{1}{N} \Re \left(\Tr \dblone\right) + \frac{g_0 a^2}{N} \Re(i \Tr F_{\mu \nu}) + \frac{g_0^2 a^4}{2 N} \Re\left(\Tr F_{\mu \nu}^2\right) \right] \nonumber \\[2mm]
                \;&=\; \beta\,\frac{g_0^2 a^4}{2 N} \Re\left(\Tr F_{\mu \nu}^2\right) 
                \;=\; \Tr\left[\frac{\beta g_0^2 a^4}{4 N}\left(F_{\mu \nu}F_{\mu \nu} + F_{\mu \nu}^* F_{\mu \nu}^*\right)\right] \nonumber \\[2mm]
                \;&=\; \frac{\beta g_0^2 a^4}{2 N} \Tr\left(F_{\mu \nu}F_{\mu \nu}\right)\,.
\label{eq:approx_single_plaq_action}
\end{align}
Notice that $S_{\mu \nu}$ is the contribution of a single plaquette to the action and therefore Einstein's sum rule does not apply for $F_{\mu \nu}F_{\mu \nu}$.
The following step is to sum over the plaquettes $S_{\mu \nu}(x)$ of the entire lattice and take the limit $a \to 0$
\begin{align}
\sum_x\frac{\beta g_0^2 a^4}{2N} \Tr\left[F_{\mu \nu}(x)F_{\mu \nu}(x)\right] \quad\to\quad \frac{\beta g_0^2}{2N} \int d^4 x\; \Tr\left[ F_{\mu \nu}(x)F_{\mu \nu}(x)\right]\,.
\end{align}

When we sum this result over all the six possible planes, we obtain
\begin{align}
  S \;=\; \frac{\beta g_0^2}{2N} \int d^4 x &\Tr\left[ F_{0 1}(x)F_{0 1}(x) + F_{0 2}(x)F_{0 2}(x) + F_{0 3}(x)F_{0 3}(x) \right. \nonumber \\[1mm]
 &+ F_{1 2}(x)F_{1 2}(x) + \left. F_{1 3}(x)F_{1 3}(x) + F_{2 3}(x)F_{2 3}(x) \right]\,,
\end{align}
which can now be written using Einstein's sum rule as
\begin{equation}
 S \;=\; \frac{\beta g_0^2}{2N} \int d^4 x \frac{1}{2}\Tr F_{\mu \nu}F_{\mu \nu}\,.
  \label{eq:gauge_euclidean_action}
\end{equation}
The factor one half comes from the fact that, when we sum over the indices, there will be terms 
$F_{\mu \nu}F_{\mu \nu}$ as well as $\,F_{\nu \mu}F_{\mu \nu} = F_{\mu \nu}F_{\mu \nu}$.

We compare the result in Eq.\ \ref{eq:gauge_euclidean_action} with Eq.\ \ref{eq:full_euclidean_action} (with the fermion fields $\psi$ set to zero)
\begin{equation}
  S \;=\; \frac{1}{2} \int d^4 x \frac{1}{2}\Tr F_{\mu \nu}F_{\mu \nu}\,.
\end{equation}
We see that the choice made in Eq.\ \ref{eq:beta_lattice_parameter} makes the lattice action equivalent to
the Euclidean action from Chapter \ref{sec:Overview_Gauge_Field}.

\section{Strong-Coupling Expansion}
\label{sec:strong_coupling}

We now take the limit of strong coupling, i.e.\ $g_0 \rightarrow \infty$ or, equivalently, $\beta \rightarrow 0$ \cite{gattringer2009quantum,Bali:2000gf}. In these
calculations we consider Euclidean time, as in Section \ref{sec:QM_Conn_w_SM}. When doing this, the weight in the 
path integral will have the same form as the Boltzmann weight and we can interpret $\beta$ as the inverse temperature. 
In fact, the method used here is the same as the high-temperature expansion in statistical mechanics \cite{binney1992theory}.
We start by considering a Wilson loop of sides $a I$ and $a J$, with $I$ and $J$ integer numbers. 
We will need to evaluate
\begin{equation}
\langle W(I,J) \rangle \;=\; Z^{-1} \int{\mathcal{D}U\, \exp{(-S)}\, \Tr\left[\prod_{x} U_\mu(x)\right]}\,,
\label{eq:expected_value_wilson_loop}
\end{equation}
where the product operator is a simplified notation to indicate that we must perform an ordered product of the matrices in
the Wilson loop.

The first term in Eq.\ \ref{eq:plaquette_action} can be neglected, since it will contribute with just a constant to 
the integral and to $Z$ and will therefore cancel out. Also, we can write the action as
\begin{equation}
S \;=\; -\frac{\beta}{2N}\sum_x \sum_{C(\mu,\nu)}\Big[ \Tr U_{\mu \nu} (x) + \Tr U_{\mu \nu}^\dagger (x)\Big]\,.
\label{eq:simplified_action}
\end{equation}

We proceed to expand $\exp({-S})$. Notice that, due to the term with the dagger of the plaquette, we will have 
in our computation terms containing the plaquette oriented in the reverse direction.
\begin{equation}
\exp{(-S)} \;=\; \sum_{i=0}^\infty{\sum_{j=0}^\infty{\,\frac{1}{i!}\frac{1}{j!}
\left[ \frac{\beta}{2N} \sum_x \sum_{C(\mu,\nu)} \Tr U_{\mu \nu}(x)\right]^i
\left[ \frac{\beta}{2N} \sum_x \sum_{C(\mu,\nu)} \Tr U_{\mu \nu}^\dagger(x)\right]^j
}}\,.
\label{eq:action_strong_coupling_expansion}
\end{equation}

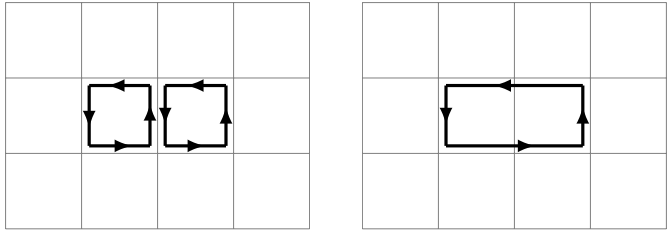
\begin{figure}[h]
 \centering
\caption{The process of integrating a paired link merge the two plaquettes forming a loop. 
In the left panel, two adjacent plaquettes with one paired link. In the right panel, 
plaquettes merged due to the integration.}
\label{fig:merging_plaquettes}
\subfigure
{

 \begin{tikzpicture}
  \draw[help lines] (1,2) grid (5,5);

  \begin{scope}[very thick,decoration={
    markings,
    mark=at position 0.65 with {\arrow{latex}}}
    ] 
    \draw[postaction={decorate}] (3.1,3.1)--(3.9,3.1);
    \draw[postaction={decorate}] (3.9,3.1)--(3.9,3.9);
    \draw[postaction={decorate}] (3.9,3.9)--(3.1,3.9);
    \draw[postaction={decorate}] (3.1,3.9)--(3.1,3.1);
  \end{scope}

	\begin{scope}[very thick,decoration={
    markings,
    mark=at position 0.7 with {\arrow{latex}}}
    ] 
    \draw[postaction={decorate}] (2.1,3.1)--(2.9,3.1);
    \draw[postaction={decorate}] (2.9,3.1)--(2.9,3.9);
    \draw[postaction={decorate}] (2.9,3.9)--(2.1,3.9);
    \draw[postaction={decorate}] (2.1,3.9)--(2.1,3.1);
  \end{scope}
\end{tikzpicture}}
\quad
\subfigure
{
\begin{tikzpicture}
  \draw[help lines] (1,2) grid (5,5);

  \begin{scope}[very thick,decoration={
    markings,
    mark=at position 0.65 with {\arrow{latex}}}
    ] 
    \draw[postaction={decorate}] (2.1,3.1)--(3.9,3.1);
    \draw[postaction={decorate}] (3.9,3.1)--(3.9,3.9);
    \draw[postaction={decorate}] (3.9,3.9)--(2.1,3.9);
    \draw[postaction={decorate}] (2.1,3.9)--(2.1,3.1);
  \end{scope}
\end{tikzpicture}}
\sourcebytheauthour
\end{figure}

Before we continue, let us demonstrate a useful property. We consider an integral of the type
\begin{equation}
 \int{dU\, \Tr U_{\mu \nu}(x) \, \Tr U_{\mu \nu}(x+a e_\mu)}\,,
\end{equation}
where the integration variable $U$ is the link shared between the plaquettes as shown in Fig.\ \ref{fig:merging_plaquettes}, left panel.
The shared link $U$ is $U_\nu(x+a e_\mu)$. Writing each trace as a sum, we obtain
\begin{equation}
\sum_{\alpha = 1}^N{\sum_{\beta = 1}^N{V_{\alpha \gamma} \left[\int{dU\, U_{\gamma \alpha}\, U^\dagger_{\beta \delta}}\right] W_{\delta \beta}}}
\;=\; \frac{1}{N}\sum_{\alpha = 1}^N{\sum_{\beta = 1}^N{V_{\alpha \gamma}\, \delta_{\alpha \beta}\, \delta_{\gamma \delta}\, W_{\delta \beta}}}\,,
\label{eq:integral_paired_links}
\end{equation}
where $V$ and $W$ are group elements as well\footnote{$V$ and $W$ are the result of the oriented 
multiplication of all other links not involved in the integral.} and we used Eq.\ \ref{eq:haar_integral_3} from
Appendix \ref{sec:group_theory_review} to obtain the expression on the RHS. We obtain therefore
\begin{equation}
 \int{dU\, \Tr U_{\mu \nu}(x)\, \Tr U_{\mu \nu}(x+a e_\mu)}\, \;=\; \frac{1}{N} \Tr\left( V W \right)\,.
 \label{eq:integral_paired_links_solved}
\end{equation}

The above result can be extended from a plaquette to a Wilson loop. The only requirement is that the loops 
run in opposite directions along the shared link between them. Let us define as \emph{paired links} the ones 
satisfying  this requirement. However, if this requirement is not fulfilled, the integral in 
Eq.\ \ref{eq:integral_paired_links} becomes similar to Eq.\ \ref{eq:haar_integral_2} and the result will be zero.

Notice that, after the integration, the two plaquettes behave as if they were a $1 \times 2$ loop. We say
that they \emph{merged} and we represent this schematically in the right panel of Fig.\ \ref{fig:merging_plaquettes}. This is 
not a property solely of plaquettes. If two loops have a paired link, i.e.\ a shared link such as the loops
run in opposite directions along it, integrating over this link will cause it to disappear and the loops 
will merge. We will only need to remember that this operation generates a multiplicative 
factor of $1/N$.

\begin{figure}[ht]
\centering
\quad
\caption{Two loops with more than one paired link and the result of integrating over one of them (in red).
Integrating over the other one then generates a trivial factor 1, as shown in Eq.\ \ref{eq:integral_paired_loops}.}
\label{fig:loops_two_paired_links}
\begin{tikzpicture}
  \draw[help lines] (0,0) grid (5,5);
  \draw[help lines] (7,0) grid (12,5); 

 \begin{scope}[very thick,decoration={
    markings,
    mark=at position 0.65 with {\arrow{latex}}}
    ] 
    \draw[postaction={decorate}] (1.9,3.9)--(1.1,3.9);
    \draw[postaction={decorate}] (1.1,3.9)--(1.1,2.1);
    \draw[postaction={decorate}] (1.1,2.1)--(1.9,2.1);
    \draw[postaction={decorate}] (1.9,2.1)--(1.9,3.1) node[pos=0.5,left,black] {$V$};
   \end{scope}
	
  \begin{scope}[very thick,decoration={
    markings,
    mark=at position 0.65 with {\arrow{latex}}}
    ] 
    \draw[postaction={decorate}] (2.1,3.1)--(2.1,1.1);
    \draw[postaction={decorate}] (2.1,1.1)--(3.9,1.1);
    \draw[postaction={decorate}] (3.9,1.1)--(3.9,3.9);
    \draw[postaction={decorate}] (3.9,3.9)--(2.1,3.9);
   \end{scope}
	
  \begin{scope}[very thick,red,decoration={
    markings,
    mark=at position 0.65 with {\arrow{latex}}}
    ] 
    \draw[postaction={decorate}] (2.1,3.9)--(2.1,3.1);
    \draw[postaction={decorate}] (1.9,3.1)--(1.9,3.9) node[pos=0.5,left,black] {$U$};
   \end{scope}
		  
  \begin{scope}[very thick,decoration={
    markings,
    mark=at position 1.0 with {\arrow{>}}}
    ] 
    \draw[postaction={decorate}] (5.5,2.5)--(6.5,2.5);
   \end{scope}

 \begin{scope}[very thick,decoration={
    markings,
    mark=at position 0.65 with {\arrow{latex}}}
    ] 
    \draw[postaction={decorate}] (8.1,3.9)--(8.1,2.1);
    \draw[postaction={decorate}] (8.1,2.1)--(9.0,2.1);
	
    \draw[postaction={decorate}] (9.0,2.1)--(9.0,1.1);
    \draw[postaction={decorate}] (9.0,1.1)--(10.9,1.1);
    \draw[postaction={decorate}] (10.9,1.1)--(10.9,3.9);
    \draw[postaction={decorate}] (10.9,3.9)--(8.1,3.9);
   \end{scope}
	
  \begin{scope}[very thick,red,decoration={
    markings,
    mark=at position 0.65 with {\arrow{latex}}}
    ] 
    \draw[postaction={decorate}] (2.1,3.9)--(2.1,3.1);
    \draw[postaction={decorate}] (1.9,3.1)--(1.9,3.9);
   \end{scope}
\end{tikzpicture}
\sourcebytheauthour
\end{figure}
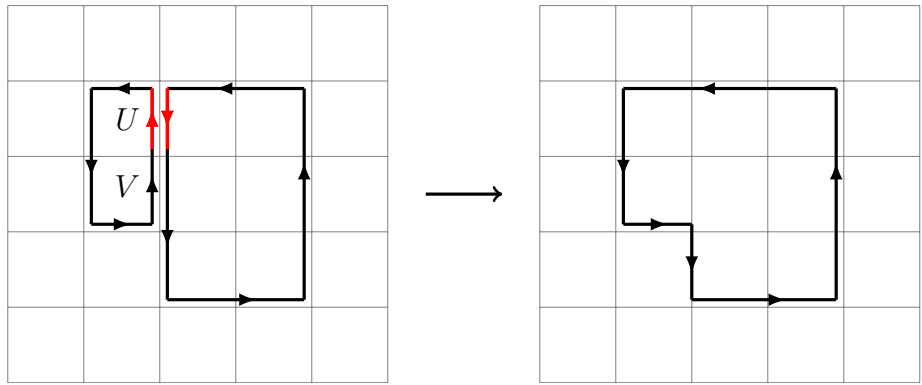

Finally, we need to know what happens when the loops share more than one paired link, as shown in Fig.\ \ref{fig:loops_two_paired_links}.
We call $V$ the additional (adjacent) link and $W$, $Y$ the results of the multiplication of the remaining links that form each loop.
Performing the integration we obtain
\begin{equation}
\int{dU\, dV\, \Tr\left(W V U\right) \Tr\left(U^\dagger V^\dagger Y\right)} \;=\; \frac{1}{N} \int{ dV\, \Tr\left(W V V^\dagger Y\right)} 
     \;=\; \frac{1}{N} \Tr\left(W Y\right) \,.
\label{eq:integral_paired_loops}
\end{equation}
Notice that, after the loops merge, the integration of additional (sequential) links in common does \emph{not}
generate a factor $1/N$ (see Eq.\ \ref{eq:Haar_measure_normalization}).

We conclude then that, through the integration of sequential common links of the loops, all such paired links 
are erased, i.e.\ the integrations merge the loops as shown in Fig.\ \ref{fig:loops_two_paired_links}, generating
a single multiplicative factor $1/N$. These properties will be fundamental for evaluating the leading term 
of our expansion.

We point out that, as mentioned above, any integral containing unpaired links will vanish (see Eq.\ \ref{eq:haar_integral_1}). This 
means that the zeroth-order term in Eq.\ \ref{eq:action_strong_coupling_expansion} will not contribute to the 
integral in Eq.\ \ref{eq:expected_value_wilson_loop}. The first-order contribution in $\beta$ selects one plaquette, 
which pairs at most two links with the Wilson loop and therefore all other unpaired links will make the integral vanish. 
We proceed to higher orders, selecting more plaquettes, until we have enough of them to pair all the links in 
the Wilson loop with links of the plaquettes\footnote{This happens at order $2I+2J-4$.}. With this many plaquettes 
we tile the perimeter of the Wilson loop, so that all is links are paired. However, this still leaves some 
plaquette links unpaired [see Fig.\ \ref{fig:tiling_wilson_loop}, left panel]. The only way to solve this is by proceeding 
to higher orders, selecting more plaquettes in the integration until we can tile the entire area of the Wilson loop 
[see Fig.\ \ref{fig:tiling_wilson_loop}, right panel].
\begin{figure}[ht]
\input{./img/tiled_wilson_loop}
\end{figure}

We know that only plaquettes oriented in the opposite direction of the Wilson loop will survive. For simplicity, we will
assume that the loop is oriented clockwise, so we do not have to carry the dagger symbol on the plaquettes. The term in 
the expansion corresponding to the right panel of Fig.\ \ref{fig:tiling_wilson_loop} is the first term that is able to combine as many 
plaquettes as necessary to fill the area of the Wilson loop, i.e.\ we will need $\,I J\,$ plaquettes. Therefore, the 
contributing term in Eq.\ \ref{eq:action_strong_coupling_expansion}, which corresponds to $i = IJ $ and $j = 0$, is\footnote{The
$SU(2)$ constitutes a special case, since $\Tr U = \Tr U^\dagger$ and therefore there will be a factor $2$ accompanying
$\beta$ in Eq.\ \ref{eq:simplified_action}, meaning that Eq.\ \ref{eq:first_order_contrib_WL} will be multiplied by a 
factor 2 if $N = 2$. This will have consequences on all the following results, where for the $SU(2)$ case the substitution
$\beta \to 2 \beta$ should be performed.}
\begin{equation}
\frac{1}{(I J)!}\left[ \frac{\beta}{2N} \sum_x \sum_{C(\mu,\nu)} \Tr U_{\mu \nu}(x) \right]^{I J}\,.
\label{eq:first_order_contrib_WL}
\end{equation}

Of course this expression will contain several combinations of plaquettes that do not correspond to the right panel of Fig.\ \ref{fig:tiling_wilson_loop},
such as terms in which the same plaquette appears multiple times and plaquettes that do not share links
with each other or with the loop. However, all these extra terms will vanish in the integration, leaving just the 
term containing the multiplication of the plaquettes inside the loop. If we perform a combinatorial analysis,
we see that there will be $\,(I J)!\,$ such terms and we deduce that the leading term in the expectation value of the
Wilson loop will be
\begin{equation}
\langle W(I,J) \rangle \;=\; Z^{-1} \left(\frac{\beta}{2N}\right)^{I J}\int{\mathcal{D}U\, \left(\prod_{x'} \Tr U_{\mu \nu}(x') \right) \Tr\left(\prod_{x} U_\mu(x)\right)}\,.
\end{equation}

We may carry out the following procedure to evaluate this integration:
\begin{itemize}
 \item Firstly, we merge all plaquettes in a  given column. This will generate a multiplicative factor $1/N^{J-1}$.
 \item We do this for all the $I$ columns. The total multiplicative factor is $1/N^{I(J-1)}$. There still remain 
 $I$ column shaped loops inside the Wilson loop.
 \item We proceed to merge these column loops. This will require $I-1$ integrations. The multiplicative factor by the
 end of the calculation is $1/N^{IJ-1}$. We will then have two Wilson loops, oriented in opposite
 directions.
 \item We integrate over just one of the common links between the loops, which generates a single factor $1/N$, 
 multiplying $\,\Tr \dblone = N$. Therefore this last integration is unity.
\end{itemize}

If we follow the above procedure, the leading order to the expected value of the Wilson loop will be
\begin{equation}
\langle W(I,J) \rangle \;=\; Z^{-1} \left(\frac{\beta}{2N}\right)^{I J}\left(\frac{1}{N}\right)^{I J-1}\,+\,\mathcal{O}(\beta^{I J + 1})\,.
\label{eq:Wilson_Loop_Strong_Expansion}
\end{equation}
The next-to-leading-order correction may be computed by substituting one of the tiling plaquettes by two plaquettes with the same orientation,
i.e.\ the opposite orientation of the Wilson loop \cite{creutz1983quarks}.

We still need to evaluate $Z$ to obtain the result. We use again the expansion in Eq.\ \ref{eq:action_strong_coupling_expansion}.
The integrand does not contain any loop or plaquettes and the zeroth-order contribution does not have any plaquette. We obtain that 
the zeroth-order contribution is a term equal to $1$. The first order selects a single plaquette and, therefore, the integration over
the links will vanish. This means that $Z$ is given by $1$ up to  order $\beta^2$. For the $SU(3)$ case we have
\begin{equation}
\langle W(I,J) \rangle \;=\; \frac{1}{2^{I J}\, 3^{2 I J-1}} \beta^{I J}\,+\,\mathcal{O}(\beta^{I J + 1})\,.
\label{eq:expected_value_wilson_loop_result}
\end{equation}

In the next section we will give a physical interpretation of this result.

\section{Linearly Rising Potential}
\label{sec:Linearly_Rising_Potential}

We consider the special case of a Wilson loop on a plane containing the time direction, i.e.\ $\mu = 4$, and a spatial direction
$i$, where $i = 1,2$ or $3$. In this case, a good choice of gauge will
simplify our interpretation of the Wilson loop. The chosen gauge is called the \emph{temporal gauge}\footnote{This choice is not possible in 
the case we use periodic boundary conditions, since it will break gauge invariance. See in Ref.\ \cite{gattringer2009quantum} details about maximal trees
and Ref.\ \cite[Chap.\ 10]{Roth92} details about the temporal gauge.} and it is characterized 
by setting
\begin{equation}
U_{4}(x) = \dblone\,.
\label{eq:temporal_gauge}
\end{equation}

We can verify that this is a valid choice since, starting at a point $x$, we have the freedom 
to choose $V(x+a e_4) = V(x) U_4 (x)$ in Eq.\ \ref{eq:lattice_gauge_transformation},
which will correspond to the desired gauge. In this gauge the Wilson loop becomes
\begin{equation}
W(|y-x|,t) \;=\; \Tr\left[L(x,y,t)\, L^\dagger(x,y,0)\right]\,,
\label{eq:wilson_loop_temporal_gauge}
\end{equation}
where $L(x,y,t)$ is a product of links taken along a spatial line at time $t$ connecting point $x$ to point $y$, i.e.
\begin{equation}
L(x,y,t) = U_{i}(x)U_{i}(x+a e_i)\dots U_{i}(y-a e_i)\,.
\end{equation}

The argument now is that the object $L(x,y,t)$ represents a pair of static, i.e.\ infinitely massive
fermionic sources (a quark and its antiquark), as shown in Ref.\ \cite[Chap.\ 5]{gattringer2009quantum}.
As a motivation, let us state that fermion fields (quarks) are associated with sites of the 
lattice and represented by $\psi^\alpha(x) = c \psi$ (see Section \ref{sec:Fermions}). Under gauge 
transformations, they behave as $\,c \psi(x) \rightarrow V(x)c\psi(x)$. The respective antiquark field is
represented by $\bar{\psi}(x)c^\dagger = \psi^\dagger(x)\gamma_0 c^\dagger$ and transforms as 
$\bar{\psi}(x)c^\dagger \rightarrow \bar{\psi}(x) c^\dagger V^\dagger(x)$. A static quark-antiquark 
pair would then be represented by $c\psi(x)\bar{\psi}(y)c^\dagger$, which has the same behavior as 
$L(x,y,t)$ under gauge transformations (see Eqs. \ref{eq:spinor_gauge_transform} and \ref{eq:spinor_dagger_gauge_transform}).
 It is thus natural to associate $L(x,y,t)$ to a quark-antiquark pair.

Eq.\ \ref{eq:wilson_loop_temporal_gauge} shows us that $\,\langle W(|y-x|,t)\rangle\,$ is a 
correlation function between $L(x,y,0)$ and $L(x,y,t)$, very similar to 
Eq.\ \ref{eq:2-point_connect_corr_func-def}. The main difference is that there is no subtraction 
of expectation values of the individual operators. We may thus perform the same procedure used to 
arrive at Eq.\ \ref{eq:gamma_n_c_large_time}, noting that the absence of the subtracted term will 
cause the sum to run over all possible states. Note that $n =0$ corresponds to the vacuum. We have then
\begin{align}
\langle W(|y-x|,t)\rangle \;&=\; \sum_{n}{\Big[ \left| \langle 0 \left|S(x,y)\right| n \rangle \right|^2\Big] e^{-(E_n-E_0) t}} \nonumber \\ 
			  \;&=\; \sum_{n}{\Big[ \left| \langle 0 \left|\psi(x)\bar{\psi}(y)\right| n \rangle\right|^2 \Big]e^{-(E_n-E_0)t}} \,.
\end{align}

The largest contribution will come from $n = 1$, which corresponds to the isolated 
pair in the vacuum. Higher-order contributions would come from introducing pairs of
virtual fermions. Since we are dealing with static fermions, we interpret that all 
the energy comes from the interaction potential and therefore $\,E_1 - E_0 = V(|y-x|)$.
Inverting the above equation, we obtain for large $t$
\begin{equation}
V(|y-x|) \;=\; -\frac{1}{t}\, \ln \left[\langle W(|y-x|,t)\rangle \right]\,.
\end{equation}

We substitute Eq.\ \ref{eq:expected_value_wilson_loop_result} into the above equation 
and take $t = a I$, $\,|y-x| = a J$ to obtain
\begin{equation}
V(|y-x|) \;=\; \frac{|y-x|}{a^2}\ln\left[\frac{18}{\beta}\right] + \mathcal{O}(\beta)\,.
\label{eq:Linear_Potential}
\end{equation}

Therefore, the potential rises linearly with the distance between the static quark sources.
We note that this result was obtained assuming small $\beta$.

We will now resort to Monte Carlo simulations in order to study the gauge theory for general values 
of the lattice parameter $\beta$. This kind of simulation is widely used in the study of statistical 
systems. We will therefore explain the procedure in this context and then describe how to adapt the 
method to simulate a gauge theory.

\section{Monte Carlo Methods}
\label{sec:MC_methods}

The nonperturbative tool that will enable us to compute the path integral for any value of $\beta$
is the numerical simulation, using Monte Carlo methods \cite{Creutz:1980gp,Newm99}. We will introduce 
these methods by discussing a statistical mechanical system.

Let us suppose that a classical system with many degrees of freedom has the Hamiltonian $H$. This 
Hamiltonian is a function of the configuration of the system $\mu$, i.e.\ when the system is in the 
configuration $\mu$, it possesses energy $E_\mu$. Due to the large number of degrees of freedom, we adopt
a statistical description. Typically, this system will fluctuate from 
configuration $\mu$ to a configuration $\nu$. We denote the probability of the system to be in a 
particular configuration $\mu$ at time $t$ by $p_\mu(t)$ and the transition probability to go from 
configuration $\mu$ to $\nu$ in a time interval $dt$ by $R(\mu \to \nu) dt$. We 
assume that the transition rate $R(\mu \to \nu)$ is time-independent. The time evolution of $p_\mu(t)$
is governed by the \emph{master equation} 
\begin{equation}
\frac{d p_\mu (t)}{dt} \;=\; \sum_\nu{\big[p_\nu(t)\,R(\nu \to \mu) \,-\, p_\mu(t)\,R(\mu \to \nu)\big]}\,.
\label{eq:master_equation}
\end{equation}

In particular, we wish to study the system in thermal equilibrium with a heat bath at temperature $T$. 
To be in equilibrium means that $p_\mu(t)$ will not be time-dependent anymore and from now on we 
will consider only the equilibrium distribution, which we denote by $p_\mu$. The fact that 
$p_\mu$ is time-independent means that $dp_\mu/dt = 0$. Furthermore, we will generally consider 
discrete time steps and define as $P(\mu \to \nu)$ the probability to go from configuration $\mu$ 
to $\nu$ in one time step. The master equation then becomes
\begin{equation}
   \sum_{\nu \neq \mu}\, p_\nu \,P(\nu \to \mu) \;=\; \sum_\nu p_\mu\, P(\mu \to \nu) \;=\; p_\mu \,.
\label{eq:equilibrium_equation}
\end{equation}
Additionally, in the stationary limit, the system will follow the Boltzmann distribution, i.e.\
\begin{equation}
p_\mu \;=\; e^{-\beta E_\mu}\,.
\label{eq:boltzmann_distribution}
\end{equation}
Note that $\,\beta = 1/K T$, where $K$ is the Boltzmann constant. This should not be confused with the
lattice parameter introduced in Eq.\ \ref{eq:plaquette_action}.

Let us assume that Hamiltonian describes accurately a physical system, and that someone studying it 
in a laboratory will try to measure an observable $A_\mu$. However, since the system will be 
fluctuating between configurations rapidly, what this person will record is an average 
$\langle A \rangle$ of this observable, given by the thermodynamic average
\begin{equation}
\langle A \rangle \;=\; \frac{\sum_\mu A_\mu\, e^{-\beta E_\mu}}{\sum_\mu e^{-\beta E_\mu}}\,.
\label{eq:average_observable}
\end{equation}

If we desire to theoretically compute $\langle A \rangle$, we could use several approaches. The naive 
approach would be to generate all possible configurations and then directly apply Eq.\ \ref{eq:average_observable}.
But typically the number of configurations possible in a system is huge and therefore this is not a viable 
approach. The alternative is to select a set of representative configurations and average over these configurations 
only. This will introduce an error in our results, which will be smaller the more configurations we consider.

The tool to generate the relevant configurations is the Monte Carlo technique. It gives a prescription 
of how to find $M$ relevant configurations to estimate the average $\langle A \rangle$. It says that 
we can take this set by sampling configurations following a certain probability distribution $\rho_\mu$. When 
doing this, one gives more importance to some configurations. This must be taken into account when 
estimating $\langle A \rangle$, by attributing a weight to these states that is the inverse of the probability
distribution. Therefore the estimate of the average is
\begin{equation}
\langle A \rangle_M \;=\; \frac{\sum_{\mu=1}^M A_\mu\, \rho_\mu^{-1}\, e^{-\beta E_\mu}}{\sum_{\mu=1}^M \rho_\mu^{-1}\, e^{-\beta E_\mu}}\,.
\label{eq:estimator}
\end{equation}

A clever choice therefore is to choose to sample the configurations following the own Boltzmann distribution itself,
i.e.\ $\rho_\mu = e^{-\beta E_\mu}$. This allows the weight to cancel out the Boltzmann factor and the estimator
assumes the simple form
\begin{equation}
\langle A \rangle_M \;=\; \frac{1}{M}\sum_{\mu=1}^M A_\mu\,.
\label{eq:estimator_MC}
\end{equation}

Another way to see that the estimator becomes a simple average when the configurations are sampled
according to the Boltzmann distribution is to realize that $e^{-\beta E_\mu}$ is a weight in the average.
Therefore, if we generate the configurations following the Boltzmann distribution directly we will not have to 
worry about the weight and we can estimate the average $\langle A \rangle_M$ by Eq.\ \ref{eq:estimator_MC}.

The idea of sampling just the relevant configurations by picking them according to the correct distribution 
is called \emph{importance sampling}.

Therefore the challenge is to generate configurations following the Boltzmann distribution. This is impossible to do for 
independent samples, but may be achieved if we implement our simulation as a \emph{Markov process}. A Markov process is 
a mechanism that randomly generates a new configuration $\nu$ based solely on the current configuration $\mu$, i.e.\ it 
has no memory of previous configurations. Mathematically, it means that $P(\mu \to \nu)$ depends solely on $\mu$ and $\nu$ 
as assumed above. Each change in the configuration is considered a step in simulation 
time, which is not related to the physical time. The chain of configurations that is formed is called a \emph{Markov chain}. 
Notice that
\begin{equation}
\sum_\nu P(\mu \to \nu) \;=\; 1\,.
\end{equation}

When building the Markov process, we will need to be careful to build it in such a way that, given an initial configuration,
it will converge to the stationary distribution in Eq.\ 
\ref{eq:boltzmann_distribution} \cite{Newm99}. Therefore, even though it would be impossible to sample $p_\mu$ for a realistic 
system directly, we may implement this Markov chain with a convenient transition probability $P(\mu \to \nu)$ and, after a 
transient period, approach the equilibrium distribution $p_\mu$ (taken as $e^{- \beta E_\mu}$). One often imposes a 
stronger condition in order to determine the transition probabilities, the \emph{detailed balance} condition
\begin{equation}
 \label{eq:detailed_balance}
 p_{\mu} \,P(\mu \rightarrow \nu) \;=\; p_{\nu}\, P(\nu \rightarrow \mu)\,,
\end{equation}
which clearly implies Eq.\ \ref{eq:equilibrium_equation}. Rearranging it we get
\begin{equation}
 \label{eq:transition_probability_constraint}
 \frac{p_{\mu}}{ p_{\nu}} \;=\; \frac{P(\mu \rightarrow \nu)}{P(\nu \rightarrow \mu)} \;=\;  e^{-\beta (E_{\mu}-E_{\nu})}\,.
\end{equation}
This constraint will ensure that at some point in our Markov chain we will start generating configurations that will follow 
the Boltzmann distribution.

This leaves us with some freedom of choice for the transition probabilities. The Metropolis algorithm 
consists in making a choice for them. This choice is
\begin{equation}
  \label{eq:trans_prob_metropolis}
  P(\mu \rightarrow \nu) \;=\; \min[1,e^{-\beta (E_{\nu}-E_{\mu})}]\,. 
\end{equation}

The interpretation is that when we propose a new configuration, we accept it if there is a decrease in the energy. Otherwise, 
we accept it with probability $e^{-\beta (E_{\nu}-E_{\mu})}$. We use this rule to generate an ensemble of
configurations.

So far we assumed that we already have a starting configuration to apply our method. But this is not the case for the first step. 
We need to choose a suitable starting configuration and then apply our algorithm. Ideally we should choose the starting 
configuration to already be relevant to the Boltzmann distribution. However, this is rarely possible. What is done is to 
choose an arbitrary initial configuration. As this initial configuration will usually be far away from equilibrium, 
when proposing the second configuration in the Markov chain, if it brings the system nearer to the equilibrium, we have 
a large probability of accepting it. Otherwise the probability of accepting the change will be quite small. The result is that, 
as expected, the chain will reach equilibrium after a finite amount of steps. The process of leaving a non-equilibrium 
configuration to reach an equilibrium one is called \emph{thermalization}. This will happen for any initial configuration proposed. 
The reason for this is that we built our Markov chain to be \emph{ergodic}, i.e.\ from a configuration $\mu$, the system may 
go to any configuration $\nu$, given enough simulation time. Of course, one should check that the chain has reached equilibrium
before selecting the $M$ representative states. Also, if we select consecutive configurations, they may be highly correlated. 
Ideally one should skip configurations over a 
determined time interval to get independent samples.

If we consider that we generated $M$ independent configurations from the ensemble of all possible configurations and that the 
sequence is thermalized after generating the configuration $n_{therm}$, the expected value of an observable quantity $A$ will be
\begin{equation}
 \label{eq:measuring_observable}
 \langle A \rangle \;=\; \frac{1}{M} \sum_{\mu = 1}^{M} A_\mu\,,
\end{equation}
where the samples $\mu$ are thermalized and separated by $k$ steps, assuming that it takes $k$ steps in the chain to get independent 
samples. Thus, $M$ is given by $M = (N_{time}-n_{therm})/k$.

\section{Simulating a Gauge Theory with the Metropolis Algorithm}
\label{sec:Metropolis_for_SU(N)}

The above method can be applied to simulate a gauge field theory as well. For a non-Abelian theory, the simplest case is $SU(2)$. 
It can also be used as basis for calculations in $SU(3)$ theory. Therefore this is the case we will study here. For now, the 
measurable quantity will be the average action per plaquette $\langle S_{\mu \nu} \rangle$ (see Eq.\ \ref{eq:plaquette_action}).

We will start by setting up a four-dimensional lattice. A configuration will be characterized by group elements of the $SU(2)$
group situated in the links between the lattice sites. Also, we identify the exponential of the system's action with the Boltzmann 
factor.

We proceed in the following way to implement a program to make these simulations.

\begin{itemize}
 \item \textbf{Group parametrization}:  For the $SU(2)$ group we parametrize a group element as
 \begin{equation}
  U \;=\; \left\{a_0\,+\,i \vect{a}\cdot \vect{\sigma}\; |\; a_0^2\,+\,\vect{a}\cdot\vect{a}\;=\;1 \right\}\,,
	\label{eq:SU2_parametrization}
 \end{equation}
 where the components of $\vect{\sigma}$ are the Pauli matrices.
 
 \item \textbf{Lattice initialization}: To each link of the lattice we associate a group element. Our initial choice is
 $\,a_0 = 1$ and $\vect{a} = 0$. This initial condition is called a cold lattice initialization, since it resembles the 
 configuration of a spin model at zero temperature. It is also possible to make a hot initialization, in which random 
 numbers are drawn for the parameters at each site uniformly on the sphere $\,a_0^2+\vect{a}\cdot\vect{a}=1$.
 
 \item \textbf{Storing group elements}: For simplicity, we store all components of the vector $a$ at each site.
 Also, we associate to each point just $D$ links, where $D$ is the dimension of the lattice. This avoids double counting 
 the links. The first link stored for a given site will be the link between it and the site in front of it along 
 direction 1. The second link will be the one in front of it in direction 2 and so on.
 
 \item \textbf{Plaquette definition}: As discussed in Section \ref{sec:LGT}, a plaquette is the smallest square we can build on 
 the lattice. For a $D=2$ lattice it is easy to see that there are $N^2$ plaquettes ($N$ is the number of points on the lattice). 
 We can build higher-dimension lattices by stacking these lattices one on the top of the other. For a $D=3$ lattice we obtain then 
 immediately $N N^2 = N^3$ plaquettes. However, this considers just the plane $1,2$. If we look at the lattices on the planes 
 $2,3$ and $1,2$ we get the same amount of plaquettes on each plane, giving a total of $3 N^3$ plaquettes. By stacking spaces 
 we can build a $D=4$ lattice. We repeat the process looking at all the six possible $D=2$ lattices inside our $D=4$ lattice to 
 conclude that the number of plaquettes is $6 N^4$. We can continue this process for higher dimensions and by induction prove 
 that for a $D$-dimensional lattice the number of plaquettes is $\,C(D,2)N^D = \dfrac{D(D-1)}{2} N^D$.

 \item \textbf{Plaquette Action}: To each plaquette we associate an elementary action given by Eq.\ \ref{eq:plaquette_action}
 \begin{equation}
   S_{\mu \nu} \;=\; \beta\left\{1-\frac{1}{N}\,\Re \Tr\Big[U_\mu (x)\, U_\nu (x+a e_\mu)\, U_\mu^\dagger(x+a e_\nu)\, U_\nu^\dagger(x)\Big]\right\}\,.
 \end{equation}
 
 \item The system's action will be given by Eq.\ \ref{eq:discrete_action}
 \begin{equation}
     S \;=\; \sum_x{ \sum_{C(\mu,\nu)} S_{\mu \nu}(x)}\,.
 \end{equation}

 \item \textbf{Perform a Metropolis hit on each lattice link}. We pick a link to change the value of its parameters. 
 Since we will be proposing a change to this link, the value of the action $S$ must be updated. To this end, we subtract
 from it all terms due to plaquettes that contain this link (in $D=4$ this will be six plaquettes). Then we draw four random numbers 
 to be our new parameters of the group element along this link. We must draw the parameters $\,(a_0,\vect{a})\,$ uniformly
 distributed on the surface of the 4-sphere  defined by the parametrization of the group elements in Eq.\ \ref{eq:SU2_parametrization}.
 This is accomplished by drawing each component of $\,(a_0,\vect{a})\,$ following a normal distribution in the range $[-1,1]$ and then 
 normalizing the components [see Refs.\ \cite{Muller:1959:NMG:377939.377946} and \cite{MathWorld_HPP}]. We use the Ziggurat algorithm \cite{Marsaglia:Tsang:2000:JSSOBK:v05i08}
 to generate the normal distribution. One implementation of it in FORTRAN may be found in Ref.\ \cite{Ziggurat}. After this process, 
 we add the contribution of the plaquettes again to the action. Once this is done, we accept the change or not, using conditions for 
 the Metropolis algorithm in Eq.\ \ref{eq:trans_prob_metropolis}. This process is repeated over the entire lattice.

 \item \textbf{Run the Metropolis Algorithm several times}. After the previous step is finished, we will have a new state, 
 to which we can associate a new value of the desired observable. We repeat the previous step several times to obtain the ensemble
 of desired states. We do not write out the entire configuration. The only output is the value of the observable 
 at each of these steps.
 
 \item \textbf{Determine the thermalization time and make the measurement}. Once the simulation is completed, we can plot
 the value of the observable as a function of the simulation time. It is possible, using this graph, to visually determine after
 how many steps of the Markov chain the system has thermalized. The initial transient is discarded and then we apply 
 Eq.\ \ref{eq:measuring_observable}. This gives us a Monte Carlo estimate for the expectation value $\langle A \rangle$. This 
 comprises the average and its error, which we compute as a standard deviation. We note that due to the correlations errors 
 mentioned above, we will not use all samples generated by this method in the computation of $\langle A \rangle$, but just 
the samples distant enough to be considered decorrelated.
\end{itemize} 
 
\section{Some Results}
\label{sec:some_results}

Using the method described in Sections \ref{sec:MC_methods} and \ref{sec:Metropolis_for_SU(N)}, we calculated the average of 
$P = S_{\mu \nu}/\beta$ as a function of the parameter $\beta$ for the $SU(2)$ case and $D=4$. For this calculation we used a lattice of side $N_s = 10$. The 
first step consists in finding the thermalization time.
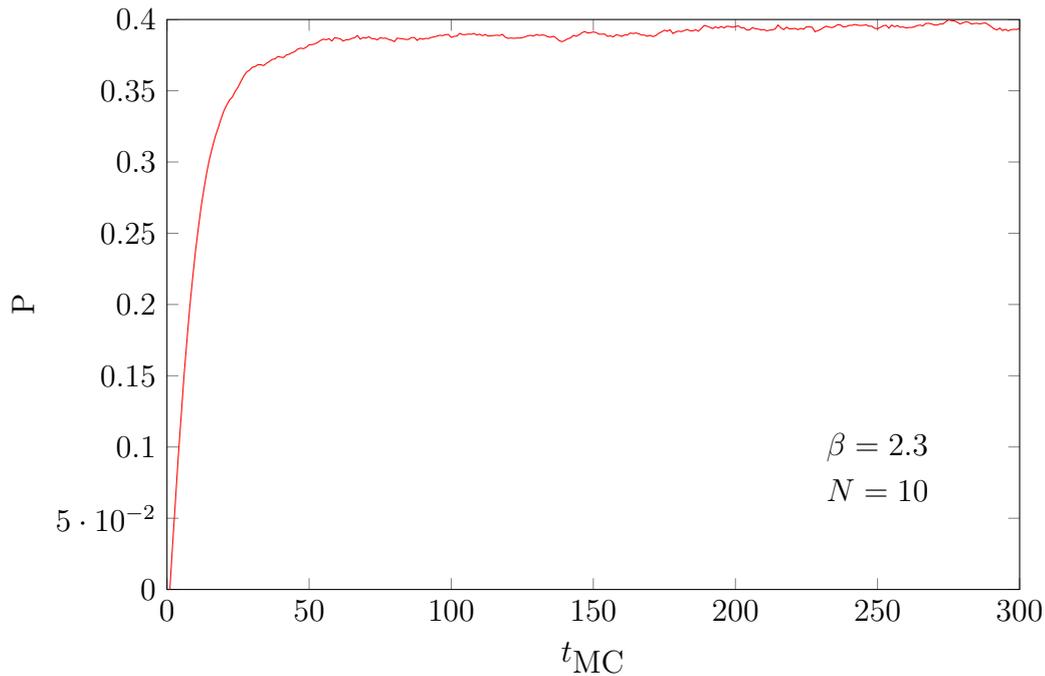
\begin{figure}[ht]
\centering
\caption{ $P$ as function of Monte Carlo time steps ($t_{\mbox{MC}}$). Notice that we can consider the system  to be thermalized 
after 100 steps.}
\label{fg:thermalization_time} 
\begin{tikzpicture}
    \begin{axis}[width=.8\textwidth,height=.37\textheight,
		xmin=0,xmax=300,ymin=0,ymax=.4,title={},
		xtick = {0,50,...,300},
    xlabel={$t_{\mbox{MC}}$}, ylabel={P},
		]
		    \pgfplotstableread{./img/Data_N=10/S_vs_MCT.out}{\SvsMCT}
        \addplot[red] table[x index=0, y index=1]{\SvsMCT};
				\node at (axis cs:250,0.1) {$\beta = 2.3$};
				\node at (axis cs:250,0.07) {$N = 10$};
    \end{axis}
\end{tikzpicture}
\sourcebytheauthour
\end{figure}

As can be seen in Fig.\ \ref{fg:thermalization_time}, we can consider the system to be thermalized after 100 steps. We did these
calculations for other values of $\beta$ and the longest thermalization time obtained was around 150 steps. We used this last 
value as thermalization time for all our simulations. As mentioned above, there is one more thing that needs attention. When we 
built our algorithm, we used a previous configuration to determine the following configuration. This introduces correlations 
between the configurations and it is not true anymore that they are independently drawn. Therefore, we need to compute the correlation
functions
\begin{equation}
f(\tau) \;=\; \frac{\langle S_{\mu \nu}(t) S_{\mu \nu}(t+\tau)\rangle - \langle S_{\mu \nu}(t)\rangle^2}{\langle S_{\mu \nu}(t)^2\rangle- \langle S_{\mu \nu}(t)\rangle^2}\,,
\label{eq:corr_func_MC_time}
\end{equation}
where $t$ and $\tau$ are Monte Carlo times (integers).

We expect that $f(\tau) = e^{-\tau/\tau_0}$ \cite{Newm99}, where $\tau_0$ is a constant. We use the criterion that
two samples are decorrelated when the simulation time between them is larger than $\tau_0$. For averaging $S_{\mu \nu}$ 
we will use only decorrelated samples. For estimating the error associated with this estimate, we then use the
standard deviation
\begin{equation}
\Delta S \;=\; \sqrt{\frac{\langle \left( S_{\mu \nu} - \langle S_{\mu \nu} \rangle \right)^2 \rangle}{M}}\,.
\end{equation}
       
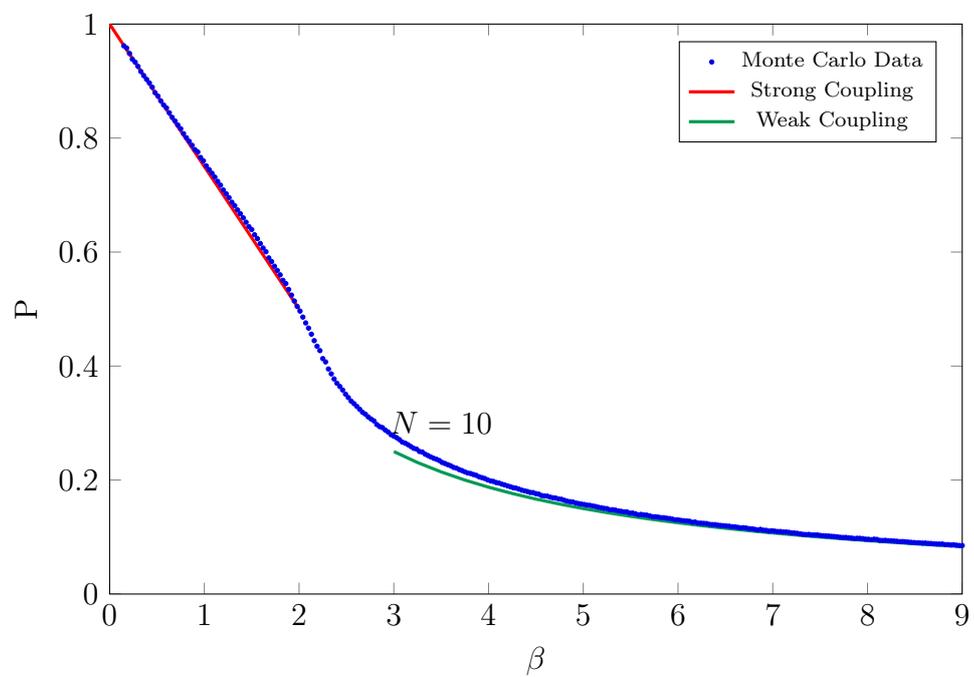
\begin{figure}[ht]
\centering
\caption{ Results of $\langle P \rangle$ as a function of $\beta$ in a simulation of a pure $SU(2)$ gauge theory on a
lattice of size $10^4$. Error bars are omitted due to their negligible size.}
\label{fig:avrg_action_vs_beta} 
\begin{tikzpicture}
    \begin{axis}[width=.8\textwidth,height=.37\textheight,
		xmin=0,xmax=9,ymin=0,ymax=1,title={},
    xlabel={$\beta$}, ylabel={P},
    legend pos = {north east},    
		legend style = {align=justify,font=\scriptsize},
		]
		    \pgfplotstableread{./img/Data_N=10/S_vs_beta.dat}{\SvsBeta}
        \addplot+[only marks,mark = *,blue,mark size = .8pt,fill=blue]
				table[x index=0, y index=1]{\SvsBeta};
				\addplot[red,very thick,domain=0:2]{1-x/4};
  			\addplot[ForestGreen,very thick,domain=3:9]{3/(4*x)};
				\addlegendimage{only marks, mark = *, blue,fill=blue};
				\addlegendimage{red,very thick};
				\addlegendimage{ForestGreen,very thick};
        \addlegendentry{Monte Carlo Data};
				\addlegendentry{Strong Coupling};
				\addlegendentry{Weak Coupling};
				\node at (axis cs:3.5,0.3) {$N = 10$};
    \end{axis}
\end{tikzpicture}
\sourcebytheauthour
\end{figure}

The next step is to measure $\langle S_{\mu \nu} \rangle$ as function of 
$\beta$. The result of this calculation is shown in Fig.\ \ref{fig:avrg_action_vs_beta},
together with analytic forms for the strong-coupling (calculated at Section \ref{sec:strong_coupling}) and weak-coupling limit
(extracted from Ref.\ \cite{creutz1983quarks}), i.e.\ for small and
large $\beta$ respectively.

The strong-coupling form comes from noticing that the average of the plaquette is the same as the average of a
$1 \times 1$ Wilson loop. We have only to notice that, for the $SU(2)$ case, $\Tr U_{\mu \nu} = \Tr U_{\mu \nu}^\dagger$ and
therefore the RHS of Eq.\ \ref{eq:action_strong_coupling_expansion} will carry a factor 2. This factor
will be accompanying $\beta$ until the final result and Eq.\ \ref{eq:Wilson_Loop_Strong_Expansion} becomes, in the $SU(2)$
case
\begin{equation}
 \langle W(I,J) \rangle \;=\; \left(\frac{\beta}{N}\right)^{I J} \left(\frac{1}{N}\right)^{I J - 1} + \mathcal{O}(\beta^{IJ+1}),\quad N \;=\; 2\,.
\end{equation}
We then obtain $\langle P \rangle$ in the strong-coupling limit
\begin{equation}
 \langle P \rangle \;=\; 1 - \frac{1}{2}\langle W(1,1) \rangle \;=\; 1 - \frac{\beta}{4}\,,
\end{equation}
which corresponds to the red line in the region of small $\beta$ in Fig.\ \ref{fig:avrg_action_vs_beta}.
The other region can be described by the weak-coupling expansion (for large $\beta$) and is given in the $SU(2)$ case by [see 
Ref.\ \cite[Chap.\ 18]{creutz1983quarks}]
\begin{equation}
 \langle P \rangle \;=\; \frac{3}{4\beta}\,.
\end{equation}
Details about the weak-coupling expansion can be found in Ref.\ \cite[Chap.\ 11]{creutz1983quarks}.

\cleardoublepage
\chapter{Potential Models for Heavy Quarkonia}
\label{sec:Pot_Model_Heavy_Quark}
\thispagestyle{capitulo}

\epigraph{``\textit{It is often stated that of all the theories proposed in this century, the silliest is quantum theory. In fact, some say that the only thing that quantum theory has going for it is that it is unquestionably correct.}''}{Michio Kaku\\Hyperspace (1995), p. 263} 

In this chapter we will briefly review the usual treatment for a nonrelativistic
system using a potential-model approach \cite{Lucha:1991vn,Bali:2005fu}. We motivate this study by the fact that 
it is possible to extract the Coulomb potential from perturbative QED, in the 
nonrelativistic limit. We then show that this method can be generalized to QCD and 
will yield a similar result, assuming that the gluon propagator has the same form 
as the photon propagator. With this assumption, the property of confinement is not 
reproduced in the resulting bound states. We therefore make use of the motivation in 
Chapter \ref{sec:LQCD_and_MCS} to justify the addition of a linear term to the 
potential in order to model confinement. Lastly, we discuss how to use the gluon propagator obtained 
from lattice simulations of pure gauge theory with $SU(2)$ symmetry to obtain a 
different potential. Also in this case, even though the propagator is nonperturbative,
confinement cannot be reproduced without the addition, by hand, of a linear term in
the potential.

\section{Brief Review of Potential Models}
\label{sec:Pot_Model_Review}

The use of potential models in the study of heavy quarkonia is based on the 
assumption that the interaction between a heavy quark (namely the charm or the 
bottom quark) and its antiquark may be described by a potential.
This is inspired by the fact that the Coulomb potential, which may be 
obtained as a limiting case from QED, explains with great accuracy 
bound states of nonrelativistic systems such as atoms or the positronium. 

\begin{figure}
 \centering
 \hspace{-6.0mm}
\caption{Feynman diagrams leading to the Coulomb potential in QED.}
\label{fig:Feynman_Graph_Electrons}
 \subfigure{
  \centering
  \raisebox{0.25\height}{\includegraphics[scale=0.45]{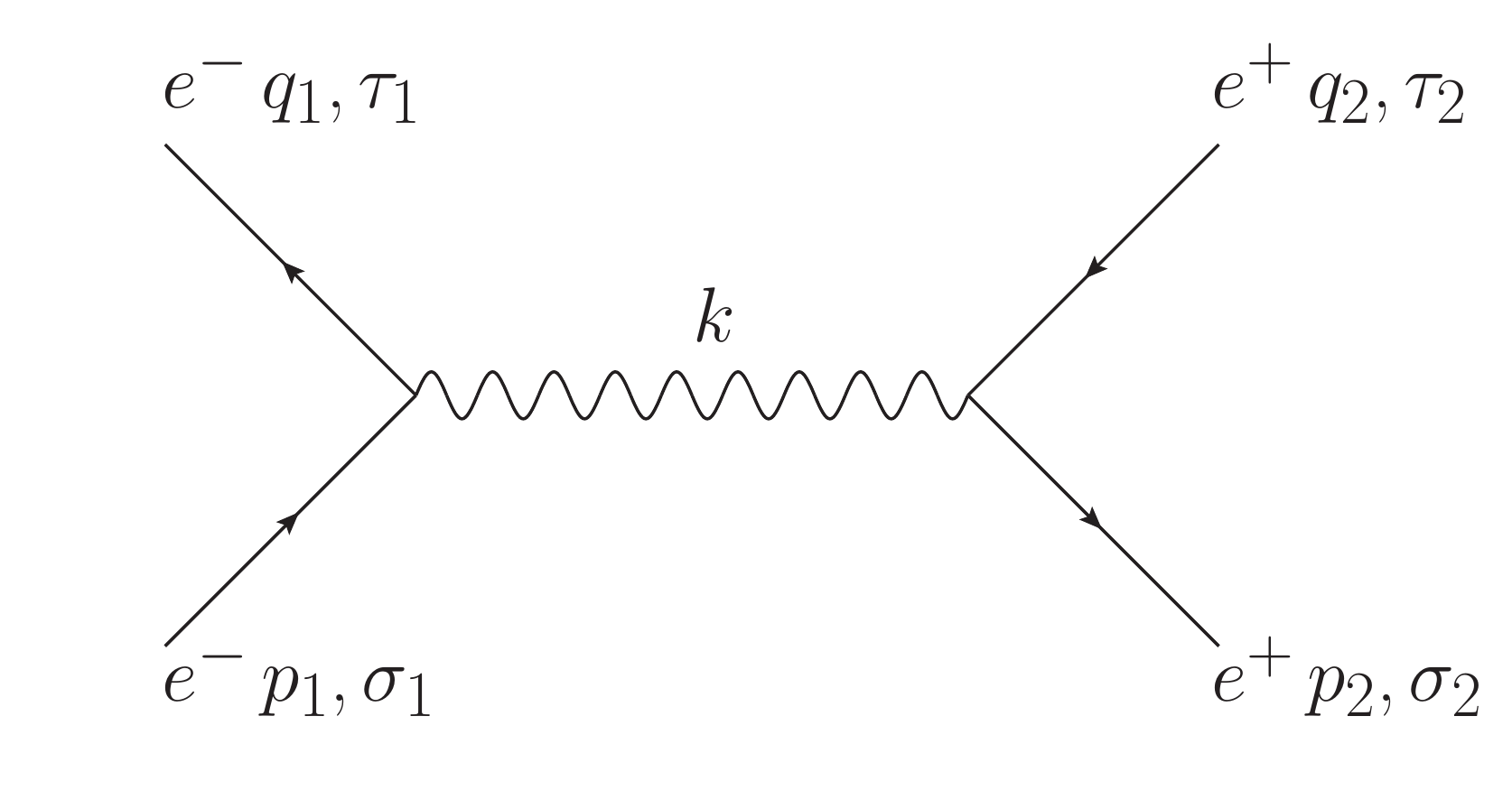}}
  \label{fig:Feynman_Graph_1_Electrons}
 }\qquad
 \subfigure{
  \centering
  \includegraphics[scale=0.45]{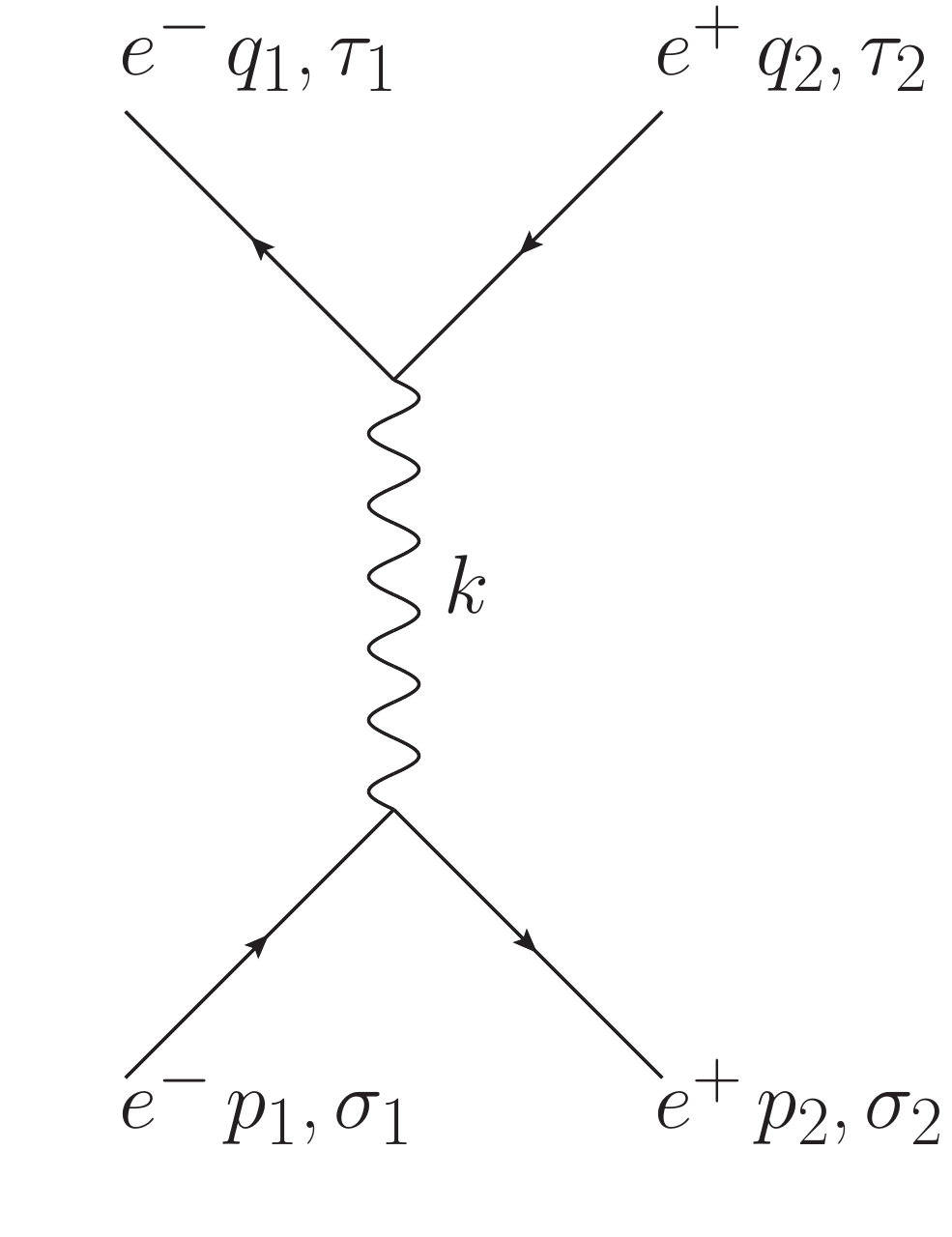}
  \label{fig:Feynman_Graph_2_Electrons}
 }
\sourcebytheauthour
\end{figure}

To obtain the Coulomb potential, one starts by considering an {\em elastic} 
$e^-e^+$ scattering process. Applying perturbation theory in the first-order 
Born approximation, we obtain the scattering-matrix element $S_{fi}$
\begin{equation}
 S_{fi} \;\equiv\; \langle f | i \rangle \;=\; \delta_{fi} \,+\, 
i (2 \pi)^4 \, \delta^{(4)}(Q-P)\, T_{fi} \,,
\end{equation}
where $Q$ and $P$ correspond respectively to the final and initial total
momentum and $T_{fi}$ is the scattering amplitude, which can be computed 
through Feynman rules. 
There are two Feynman diagrams contributing to it, which 
are shown in Fig.\ \ref{fig:Feynman_Graph_Electrons}.

These diagrams result in the following scattering amplitude
\begin{align}
 \label{eq:QED_Scattering_Amplitude}
 T_{fi} \;={}&\; \frac{1}{(2 \pi)^6} \frac{m^2}{\sqrt{E_{p_1}E_{p_2}E_{q_1}E_{q_2}}} \times \nonumber\\[2mm]
 & \left[
  \,+\,e^2 \,\overline{u}(q_1,\tau_1)\,\gamma^\mu 
  \,u(p_1,\sigma_1)\;P_{\mu \nu}(k)\;\overline{v} 
  (p_2,\sigma_2)\,\gamma^\nu\,v(q_2,\tau_2) \right. \nonumber \\[1mm]
  & \left. \;\,-\,e^2\,\overline{v}(p_2,\sigma_2)\,\gamma^\mu
  \,u(p_1,\sigma_1)\;P_{\mu \nu}(k)\;\overline{u}
  (q_1,\tau_1)\,\gamma^\nu v(q_2,\tau_2)\,\right] \;,
\end{align}
where $E_p$ is the energy of a particle with four-momentum $p$.
We follow the notation in \cite{bjorken1965relativistic, Lucha:1991vn}, i.e., $u$ represents 
the spinor of a free fermion propagating forward in time and $v$
a fermion propagating backwards in time. In other words, if a fermion
is described by the Dirac equation in Eq.\ \ref{eq:Dirac_equation}, there
will be four linearly independent solutions for $\psi(x)$, given by
\begin{equation}
 e^{-i p \cdot x} u_\sigma\,, \quad e^{i p \cdot x} v_\sigma\,,
\end{equation}
where $\sigma = 1,2$.
Therefore $u$ and $v$ are column vectors of dimension four, given by
\begin{equation}
u_\sigma \;=\; \sqrt{\frac{E_p+m}{2m}}\,
  \begin{pmatrix}
	\chi_\sigma \\[3mm]
	\dfrac{ \vect{\sigma} \cdot \bf{p} }{E_p+m}\chi_\sigma
  \end{pmatrix},\;
v_\sigma \;=\; \sqrt{\frac{E_p+m}{2m}}\,
  \begin{pmatrix}
	\dfrac{ \vect{\sigma} \cdot \bf{p} }{E_p+m}\chi_\sigma^c \\[3mm] \chi_\sigma^c
  \end{pmatrix}\,,
\label{eq:spinors}
\end{equation}
where $\vect{\sigma}$ is the vector of Pauli matrices.
We use $\chi_\sigma$ as a shorthand to write the four components
of the spinors. It represents a column vector of dimension 2 that can 
assume the values
\begin{equation}
 \chi_{1} \;=\; \begin{pmatrix} 1 \\ 0 \end{pmatrix}, \quad \chi_{2} \;=\; \begin{pmatrix} 0 \\ 1 \end{pmatrix}\,.
\end{equation}
We relate $\,\chi_\sigma^c$ to $\,\chi_\sigma$ by $\,\chi_\sigma^c =
-i\sigma_2 \chi_\sigma$, where $\sigma_2$ is the second Pauli matrix
\begin{equation}
\sigma_2 \;=\; \begin{pmatrix}
	0 & -i \\
	i & 0
\end{pmatrix}\,.
\label{eq:pauli_matrix_2}
\end{equation}

We use $p_i$ to denote the relativistic momentum of the i-th incoming fermion 
and $q_i$ is reserved for outgoing fermions. Each of these fermions is in an 
initial spin state $\sigma_i$ and leaves the scattering in a spin state $\tau_i$.
Also, $\gamma_\mu$ are the Dirac matrices and $P_{\mu \nu}(k)$ represents the 
gauge boson, i.e.\ the photon in QED, whose propagator is given in Feynman gauge by $P_{\mu \nu}(k) 
= - g_{\mu \nu}/k^2$. The transferred momentum $k$ is defined in such a way that 
total momentum is conserved at each vertex.

We then make the nonrelativistic approximation, i.e.\ we impose the kinetic 
energy of the system to be much smaller than its rest energy 
($ p \ll m \cong E$). When imposing this approximation on the spinors we have
\begin{align}
u \;={} &\; \sqrt{\frac{E_p+m}{2m}}\,
  \begin{pmatrix}
	1 \\[2mm]
	\dfrac{ \vect{\sigma} \cdot \bf{p} }{E_p+m}
  \end{pmatrix}\,
	\chi_\sigma	\;\cong \;
	\begin{pmatrix}
		\chi_\sigma \\ 0
	\end{pmatrix} \nonumber \\[2mm]
v \;={} &\; \sqrt{\frac{E_p+m}{2m}}\,
  \begin{pmatrix}
	   \dfrac{ \vect{\sigma} \cdot \bf{p} }{E_p+m} \\[4mm] 1
  \end{pmatrix}\,
	\chi_\sigma^c	\;\cong \;
	\begin{pmatrix}
		0 \\ \chi_\sigma^c
	\end{pmatrix}\,.
\label{eq:spinors_non_relativistic}
\end{align}

We proceed by calculating the first term in Eq.\ \ref{eq:QED_Scattering_Amplitude}, 
which we will name $t_\text{exch}$. We adopt the center-of-momentum frame for the calculation.
Since the Dirac matrices can be written in block-diagonal form, it is simple to 
deduce that
\begin{equation}
t_{\text{exch}} \,=\, e^2\,\delta^{\mu 0} \delta_{\sigma_1 \tau_1} \,P_{\mu \nu}(k) \,\delta^{\nu 0} \delta_{\sigma_2 \tau_2} \,=\, - e^2\, P_{0 0}(k)\, \delta_{\sigma_1 \tau_1} \delta_{\sigma_2 \tau_2}\,,
\label{eq:texch}
\end{equation}
with $k$ given by
\begin{equation}
k \,=\, p_1\, - \,q_1\, = \begin{pmatrix} 0 ,& \bf{k}\end{pmatrix}\,.
\label{eq:transferred_momentum}
\end{equation}

The absence of the temporal component is justified from our approximation.
The temporal components of $p_1$ and $q_1$ are simply $m$ and therefore they
cancel each other in the subtraction.
 
We now turn our attention to the annihilation term ($t_{\text{annih}}$). When 
considering conservation of momentum at the vertices, we conclude that the 
boson will need to carry the entire system's momentum, and therefore must assume
the form
\begin{equation}
k \,=\, \begin{pmatrix} 2 m ,& 0\end{pmatrix}\,.
\label{eq:transferred_momentum_annih}
\end{equation}

For the specific case of QED, the propagator is inversely proportional to $\vect{k}^2$.
Since the spinors do not introduce a dependence on the momentum when using the 
nonrelativistic approximation, the exchange term will be proportional to
$1/\vect{k}^2$, while the annihilation term will be proportional to $1/4 m^2$.
This implies that the first term is much larger than the second and allows us to
infer that the effects of annihilation between particles must be small. We therefore
keep only the first term in Eq.\ \ref{eq:QED_Scattering_Amplitude}.

Considering the usual photon propagator $P_{\mu \nu}(k) = - g_{\mu \nu}/\vect{k}^2$ we obtain
the following scattering amplitude
\begin{equation}
T_{fi} \;=\; \frac{1}{(2 \pi)^6}\frac{e^2}{\vect{k}^2}\,.
\label{eq:scatter_amp}
\end{equation}
 
The last step consists in performing an inverse spatial Fourier transform, which will 
lead to the familiar Coulomb potential\footnote{Here we should not confuse the norm of the vector $\abs{\vect{k}} = k$ with the
four-vector $k$.}
\begin{align}
 V(\vect{r}) 	\;&=\; - (2 \pi)^3 \int \exp(-i \vect{k}\cdot\vect{r})\;T_{fi}(k)\,d^3 k \nonumber \\
		\;&=\; - \frac{e^2}{(2 \pi)^3}\int_0^\infty \int_0^\pi \int_0^{2 \pi}  e^{-i k r \cos \theta} \frac{1}{k^2} k^2 \sin \theta\, d \varphi\, d \theta\, d k \nonumber \\[2mm]
		\;&=\; - \frac{e^2}{i r (2 \pi)^2}\int_0^\infty \frac{e^{i k r} - e^{-i k r}}{k}\, d k \;=\; - \frac{e^2}{r (2 \pi)^2}\int_{-\infty}^\infty \frac{\sin(k r)}{k}\, d k\;=\;- \frac{e^2}{4 \pi 	r}\,.
\label{eq:Couloumb_Pot}
\end{align}

This potential may be used to model the interaction in the positronium. 
Since we know that the binding energies will be of the order of \si{\electronvolt}, while the 
electron and positron masses are approximately 0.5 \si{\mega \electronvolt}, we can expect that 
the nonrelativistic approximations for the potential will hold. We may use 
this potential in the Schrödinger equation to obtain the 
energy spectrum of the system. Notice that the nonrelativistic 
approximation completely removes spin dependencies from the potential.

In our work we follow the same procedure, replacing the electron-positron
pair by a quark-antiquark pair and the photon by a gluon. The Dirac spinors
are kept intact, however they will be accompanied by a three-component column vector
$c_{1,2}$ encoding the color state of the quark, as in 
Section \ref{sec:Fermions}. Due to the SU(3) symmetry of QCD, the 
Dirac matrices will be accompanied by factors $\lambda^a$, where $\lambda^a$
is one of the Gell-Mann matrices. The scattering amplitude can be expressed as
\begin{align}
 \label{eq:QCD_Scattering_Amplitude}
 T_{fi} \;&=\; \frac{1}{(2 \pi)^6} \frac{m^2}{
    \sqrt{E_{p_1}E_{p_2}E_{q_1}E_{q_2}}} \times \nonumber\\[2mm] & \left[
              \,+\,g_0^2\,\overline{u}(q_1,\tau_1)\,c_{1,\,f}^\dagger\,\lambda^a\gamma^\mu \,c_{1,\,i}\,u(p_1,\sigma_1)\;P_{\mu \nu}^{a b}(k)\;
	\overline{v} (p_2,\sigma_2)\,c_{2,\,i}^\dagger\,\lambda^b\gamma^\nu\,c_{2,\,f}\,v(q_2,\tau_2) \right. \nonumber \\[1mm]
  & \left. \;\,	-\,g_0^2\,\overline{v}(p_2,\sigma_2)\,c_{2,\,f}^\dagger\,\lambda^a\gamma^\mu  \,c_{1,\,i}\,u(p_1,\sigma_1)\;P_{\mu \nu}^{a b}(k)\;
	\overline{u}(q_1,\tau_1)\,c_{1,\,f}^\dagger\,\lambda^b\gamma^\nu c_{2,\,f}\,v(q_2,\tau_2)\,\right] \,.
\end{align}

The effect of these modifications will be only a 
numerical factor in each of the contributions of the scattering amplitude
(exchange and annihilation). We use that the gluon propagator will be 
color-diagonal and that the two quarks are in a color-singlet state
(the meson must be colorless), and proceeds to calculate the numerical factor for the
exchange and annihilation scattering amplitude. We have the constraint that incoming and outgoing
quarks must be in a singlet state, i.e.\ $c_{1,i} = c_{2,i}$ and  $c_{1,f} = c_{2,f}\,$, and
we will have to sum over all the three possible states for $c_i$, $c_f$ [we are already assuming the 
group of the gauge symmetry is $SU(3)$]
\begin{equation}
 c_r \;=\;
 \begin{pmatrix}
  1 \\ 0 \\ 0
 \end{pmatrix}\,,\quad
 c_g \;=\;
 \begin{pmatrix}
  0 \\ 1 \\ 0
 \end{pmatrix}\,,\quad
 c_b \;=\;
 \begin{pmatrix}
  0 \\ 0 \\ 1
 \end{pmatrix}\,.
\end{equation}
Also, we will have to normalize these states by $1/\sqrt{3}$, in order to obtain the normalization
$(c_{1,i})_k (c_{1,i}^\dagger)_k (c_{2,i}^\dagger)_l (c_{2,i})_l = 1$. We obtain
\begin{align}
\label{eq:color_factor_exch}
 c^\dagger_{1,f}\,\lambda^a\,c_{1,i}\;c^\dagger_{2,i}\,\lambda^a\,c_{2,f}
 \;&=\; \frac{1}{\sqrt{3}}\sum_{\alpha=1}^3{c^\dagger_{1,f}\,\lambda^a\,c_{\alpha}\;c^\dagger_{\alpha}\,\lambda^a\,c_{2,f}} \nonumber \\[2mm]
 \;&=\; \frac{1}{3} \sum_{\alpha=1}^3 \sum_{\beta=1}^3{c^\dagger_{\beta}\,\lambda^a\,c_{\alpha}\;c^\dagger_{\alpha}\,\lambda^a\,c_{\beta}}
 \;=\; \frac{1}{3} \Tr \lambda^a \lambda^a \;=\; \frac{\delta^{a a}}{6} \;=\; \frac{4}{3}
\end{align}

For the annihilation diagram, we proceed the same way
\begin{align}
\label{eq:color_factor_annih}
c^\dagger_{2,i}\,\lambda^a\,c_{1,i}\;c^\dagger_{1,f}\,\lambda^a\,c_{2,f}
 \;&=\; \frac{1}{\sqrt{3}}\sum_{\alpha=1}^3{c^\dagger_{\alpha}\,\lambda^a\,c_{\alpha}\;c^\dagger_{1,f}\,\lambda^a\,c_{2,f}} \nonumber \\[2mm]
 \;&=\; \frac{1}{3} \sum_{\alpha=1}^3 \sum_{\beta=1}^3{c^\dagger_{\alpha}\,\lambda^a\,c_{\alpha}\;c^\dagger_{\beta}\,\lambda^a\,c_{\beta}}
 \;=\; \frac{1}{3} (\Tr \lambda^a) (\Tr \lambda^a) \;=\; 0
\end{align}

This implies that, regardless of the propagator used, the term coming from the 
annihilation does not need to be considered. One may now assume that the gluon 
propagator will behave as the photon propagator\cite{griffiths2008introduction}, i.e.
\begin{equation}
P_{\mu \nu}^{a b} \;=\; -\frac{g_{\mu \nu}\, \delta^{a b}}{k^2}\,. 
\end{equation}

This means that the perturbative inter-quark potential will be a Coulomb potential
with a factor $4/3$ multiplying it
\begin{equation}
V(r) \;=\; -\frac{4}{3}\frac{\alpha_s}{r}\,.
\label{eq:QCD_pert_pot}
\end{equation}

Notice that the potential in Eq.\ \ref{eq:QCD_pert_pot} is a non-confining one.
This is expected, since it is result of a perturbative calculation, while confinement 
is a nonperturbative phenomenon. To account for confinement, we use the result of Section
\ref{sec:Linearly_Rising_Potential} as a motivation to add a linear term to the potential. 
\begin{equation}
V(r) \;=\; -\frac{4}{3}\frac{\alpha_s}{r} \,+\, F_0 r\,.
\label{eq:cornell_pot}
\end{equation}

This potential is called Coulomb-plus-linear potential --- also known as Cornell
potential\cite{PhysRevD.17.3090}. It describes surprisingly well the states of the bottomonium and charmonium
as we will show in Chapter \ref{sec:results}.
\section{Potential from Lattice Propagator}

The Cornell potential is one of the most successful potentials
to describe bound states of heavy quarks. However, it is 
constructed in an ``ad hoc'' manner and it is not completely precise. We seek to improve it by changing the gluon 
propagator used. The calculation of the scattering amplitude will be performed 
using the gluon propagator obtained from fits of lattice data for a
pure $SU(2)$ gauge theory in Landau gauge\footnote{We note that the propagator in the $SU(2)$ and 
$SU(3)$ case have essentially the same behavior apart from a global constant \cite{Cucchieri:2007zm}.
}, given in Ref.\ \cite{Cucchieri:2011ig}. 
\begin{gather}
 P_{\mu \nu}^{a b}(k) \;=\; \frac{C\,(s+k^2)}{t^2+u^2 k^2+k^4} \left(\delta_{\mu \nu} - \frac{k_\mu k_\nu}{k^2}\right) \delta^{a b}\,, \\[2mm]
 \label{eq:lattice_gluon_propagator}
  \begin{aligned}
   C & \;=\; \num{0.784},\; & s &  \;=\; \SI{2.508}{\giga \electronvolt \squared}\,, \nonumber \\
   u & \;=\; \SI{0.768}{\giga \electronvolt},\; & t & \;=\; \SI{0.720}{\giga \electronvolt \squared}\,.
  \end{aligned}
\end{gather}

From Eq.\ (\ref{eq:texch}) we see that, with our approximation, just 
the component $P_{0 0}(k)$ survives. Also, we must use that the transferred
momentum has the form in Eq.\ (\ref{eq:transferred_momentum}). This makes
the tensorial structure $k_\mu k_\nu/k^2$ vanish. Lastly, 
we remember that this propagator was obtained in Euclidean space, while
we are working in Minkowski space. We make then the transformation
$\delta_{\mu \nu} \rightarrow -g_{\mu \nu}$ and obtain 
\begin{equation}
 P_{0 0}^{a b}\big(\vect{k}\,\big) \;=\; \frac{C\,\big(s+\vect{k}^2\big)}{t^2+u^2 \vect{k}^2+\vect{k}^4}\, \delta^{a b}\,,
\label{eq:Nonrelativistic_Lattice_Propagator}
\end{equation}

From the calculations done in the previous section we get the scattering amplitude
\begin{equation}
T_{fi} \;=\; \frac{4}{3}\frac{g_0^2}{(2 \pi)^6}\frac{C\,\big(s+\vect{k}^2\big)}{t^2+u^2 \vect{k}^2+\vect{k}^4}\,.
\label{eq:scatter_amp_QCD}
\end{equation}

As was done in the positronium case, to obtain the interaction potential we
perform a Fourier transform of the scattering amplitude. The dependence of the
propagator solely on the magnitude of $\vect{k}$ hints us to use spherical coordinates
and the angular integration is trivial if we use $\vect{r} = r\, \hat{z}$.
\begin{align}
V(\vect{r}\,) \;&=\; -\frac{4}{3}\,\frac{g_0^2 C}{(2 \pi)^3} \int_0^\infty{ \int_0^\pi \int_0^{2 \pi} \frac{s+k^2}{t^2+u^2 k^2+k^4} e^{-i k r \cos \theta}\;k^2\, \sin \theta\, d\varphi\, d\theta\,dk} \nonumber \\[2mm]
           \;&=\; -\frac{4}{3}\,\frac{g_0^2 C}{2i(2 \pi)^2 r} \int_{-\infty}^\infty{\left(\frac{s+k^2}{t^2+u^2 k^2+k^4}\right)\left(e^{i k r}-e^{-i k r}\right)k\,dk}\,.
\label{eq:angular_integration}
\end{align}

The radial integral, although more elaborate then in the positronium case, can be easily
calculated through residue integration. The poles entering the integration are 
the same of Eq.\ (\ref{eq:scatter_amp_QCD}) and are indexed in the following way
\begin{align}
k_{m,n} \;=\; & (-1)^m\, i \sqrt{t}\, \exp{\left[(-1)^n\,	 i \frac{\theta}{2}\right]}\,,\quad 
 \theta \;\equiv\; \arctan \left(\frac{\sqrt{4 t^2-u^4}}{u^2}\right)\,,\quad m, n = 0,1\,.
\end{align}

Let us make an analytic continuation of the integrand to the complex plane. Also we modify
the integral to follow a path $C_i$ in the complex plane. This path will be a semi-circle
of radius $R$ and must be defined in such a way that when we take the limit $R \to \infty$
the contribution to the integration due the arc part of the integration must vanish, therefore
reducing to the integration on the real line.

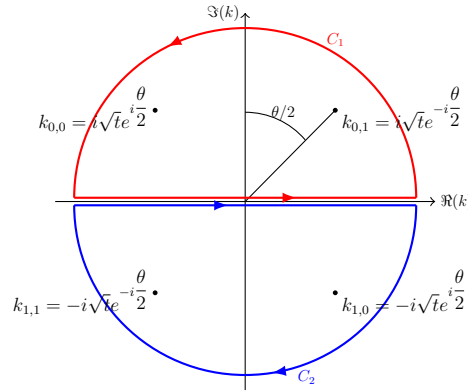
\begin{figure}[h]
\caption{Poles of the lattice gluon propagator along the contours used for the evaluation of the integrals in Eq.\ \ref{eq:angular_integration}.}
 \label{fig:poles_gluon_prop}
 \centering
\begin{tikzpicture}[scale = .5]

  \draw (0,0) -- (2.37,2.42);
  \draw (2.37/1.5,2/1.5+.3) arc [radius = sqrt(2.37*2.37+4)/1.5, start angle = atan(2/2.37), end angle = 90] node[above,scale=.5] at (1,2) {$\theta / 2$};

  \draw[->] (-5,0) -- (5,0) node[right,scale = .5] at (5,0) {$\Re(k)$};
  \draw[->] (0,-5) -- (0,5) node[left,scale = .5] at (0,5) {$\Im(k)$};
       
  \begin{scope}[thick,red,decoration={
    markings,
    mark=at position 0.65 with {\arrow{latex}}}
    ]]
      \draw[postaction={decorate}] (4.5,0.1) arc [radius = 4.5, start angle = 0, end angle = 180] node[right,scale=.5] at (2,4.3) {$C_1$};
      \draw[postaction={decorate}] (-4.5,0.1) -- (4.5,0.1);
  \end{scope}
       
  \begin{scope}[thick,blue,decoration={
    markings,
    mark=at position 0.45 with {\arrow{latex}}}
    ]]
      \draw[postaction={decorate}] (4.5,-.1) arc [radius = -4.5, start angle = 180, end angle = 0] node[left,scale=.5] at (2,-4.65) {$C_2$};
      \draw[postaction={decorate}] (-4.5,-.1) -- (4.5,-.1);
  \end{scope}
  \draw[fill] (2.37,2.42)   circle [radius=0.05] node[right,scale=.6] {$k_{0,1} = i \sqrt{t} e^{-i \dfrac{\theta}{2}}$};
  \draw[fill] (-2.37,2.42)  circle [radius=0.05] node[left,scale=.6] {$k_{0,0} = i \sqrt{t} e^{i \dfrac{\theta}{2}}$};
  \draw[fill] (2.37,-2.42)  circle [radius=0.05] node[right,scale=.6] {$k_{1,0} = -i \sqrt{t} e^{i \dfrac{\theta}{2}}$};
  \draw[fill] (-2.37,-2.42) circle [radius=0.05] node[left,scale=.6] {$k_{1,1} = -i \sqrt{t} e^{-i \dfrac{\theta}{2}}$};
\end{tikzpicture}
 \sourcebytheauthour
\end{figure}

We notice that the radial integration in Eq.\ \ref{eq:angular_integration} is actually two 
integrations. Each of these integrals will be associated with a different path. The 
integral having $e^{i k r}$ in its integrand will be over the path $C_1$, illustrated in
Fig.\ \ref{fig:poles_gluon_prop}. The other integral, which has $e^{-i k r}$ in its 
integrand, will be over the path $C_2$. These choices satisfy the requirements
presented above.

The residue theorem says that, since these paths are closed, the result of the integration will
be independent of the radius of these paths and equal to the residues inside the contour \cite{butkov1968mathematical}.
Therefore, it will be enough to compute the residue in each pole
\begin{align}
 \Res \left[\left(\frac{s+k^2}{t^2+u^2 k^2+k^4}\right)k\,e^{\pm i k r},\, k_{m,n}\right] 
 \;&=\; \lim_{k \to k_{m,n}}\frac{\left(k-k_{m,n}\right)\left(s+k^2\right)k\,e^{\pm i k r}}{t^2+u^2 k^2+k^4} \nonumber \\[2mm]
 \;&=\; \frac{1}{2}\,\frac{\left(s+k_{m,n}^2\right)\,e^{\pm i k_{m,n} r}}{ u^2+2 k_{m,n}^2}\,.
\end{align}
The result of the integration will be
\begin{align}
 V(r) \;=\; -\frac{4}{3}\frac{g_0^2\, C}{2i\,(2 \pi)^2} \frac{2 \pi i}{r}\,\frac{1}{2} &\left[
  \frac{\left(s+k_{0,0}^2\right)\,e^{ i k_{0,0} r}}{ u^2+2 k_{0,0}^2} + 
  \frac{\left(s+k_{0,1}^2\right)\,e^{ i k_{0,1} r}}{ u^2+2 k_{0,1}^2} \right. \nonumber \\[1mm]  & +  \left.
  \frac{\left(s+k_{1,1}^2\right)\,e^{ - i k_{1,1} r}}{ u^2+2 k_{1,1}^2} +
  \frac{\left(s+k_{1,0}^2\right)\,e^{ - i k_{1,0} r}}{ u^2+2 k_{1,0}^2}
 \right]\,.
\end{align}

This can be simplified if we notice that $\,k_{1,0} = -k_{0,0}$, $\,k_{0,1}=-k_{1,1}$ and $\,k_{1,1}=k_{0,0}^*$. The expression
for the potential becomes then quite concise
\begin{equation}
V(r) \;=\; -\frac{4}{3} \frac{2\alpha_s}{r}
 \Re \left[ \frac{C(s+k_{0,0}^2)\,e^{i k_{0,0} r}}{u^2+2 k_{0,0}^2}\right]\,,
\label{eq:Potential_QCD}
\end{equation}
where $\alpha_s = g_0^2/4 \pi$. We can expand the term inside the real part to get
\begin{align}
  V(r) \;&=\; -\frac{4}{3} \frac{\alpha_s}{r} \frac{C e^{-r \sqrt{t} \cos \frac{\theta}{2}}}{\sqrt{\Delta}}
         \left[\sqrt{\Delta}\cos\left(r \sqrt{t} \sin \frac{\theta}{2}\right)+\left(2s-u^2\right) \cos\left(r \sqrt{t} \sin \frac{\theta}{2}\right)\right]\,, \\
\Delta \;&\equiv\; 4 t^2 - u^4\,.
\end{align}

As before, this method has the limitation that, since we apply perturbation theory 
to obtain the potential $V(r)$, it is not expected to be a confining potential, 
even though the propagator used is obtained nonperturbatively. We use the already mentioned
linearly rising term $F_0 r$ to model confinement. The resulting potential will be
$V_{\mbox{LGP}}=V(r)+F_0 r$. Also, the nonrelativistic approach
removes spin-dependent interactions. This means that, in this model, states that are
spin-dependent will be degenerate.

\begin{figure}[H]
\centering
\caption{In red, the first term of the potential obtained from the lattice propagator ($V_{\mbox{LGP}}-F_0 r$). In blue, is the Coulomb-like potential (color factor included).}
\label{fig:Comparison_Potentials}
\begin{tikzpicture}[scale=0.9]
	\begin{axis}[ 
		xlabel = $r$,
		ylabel = $V(r)$,
		axis lines = left,
	  	axis x line=center,
		x unit=\si{\femto \metre},
		y unit=\si{\giga \electronvolt},
	  	grid=both,
	 	minor y tick num = {2},
		minor x tick num = {2},
		ymin=-1,ymax=.1,
		legend pos = {south east},    
		legend style = {align=left},
		legend cell align = left,
		legend entries = {\small First term of $V_{\mbox{LGP}}$, \small Coulomb-like potential},
	]

	\def\C{0.784};
	\def\s{2.508};
	\def\u{0.768};
	\def\t{0.720};
	\def\Delt{4*\t*\t-\u*\u*\u*\u};
	\def\hc{0.197326971}
	\def\thet{atan(sqrt(\Delt)/\u/\u)};
	\addplot[red,domain=.05:5.0,samples=300]{-(4/3) * (0.1185/(x/\hc)) * (\C/sqrt(\Delt)) * exp(-(x/\hc)*sqrt(\t)*cos(deg(\thet/2)))
		*( sqrt(\Delt)*cos(deg((x/\hc)*sqrt(\t)*sin(\thet/2))) - (\u*\u-2*\s)*sin(deg((x/\hc)*sqrt(\t)*sin(\thet/2))) )};
	\addplot[blue,domain=.05:5.0,samples=300]{-(4/3) * (0.1185/(x/\hc))};
	\end{axis}
\end{tikzpicture}
\sourcebytheauthour
\end{figure}
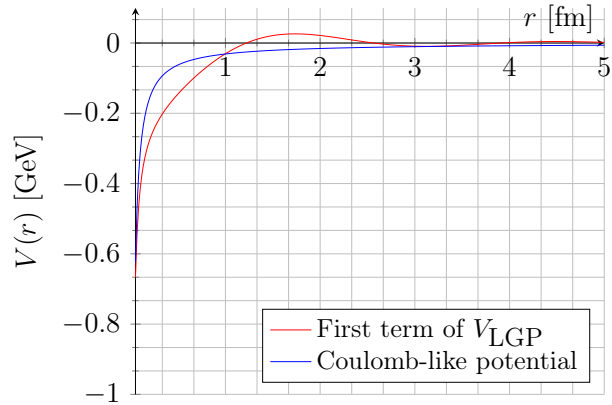

In Fig.\ \ref{fig:Comparison_Potentials} we compare this potential with the Coulomb-like potential.

Let us mention that a study using a similar method was carried out in
Refs.\ \cite{Gonzalez:2011zc, Gonzalez:2012hx} to propose a potential 
for quarkonium states. In that case, the gluon propagator was taken from
a study of Schwinger-Dyson equations \cite{PhysRevD.78.025010}.
This propagator is in qualitative agreement with the lattice results we
use. The main difference with respect to our study is that these authors
do not include the linear term in the potential, but consider an additive
contribution to the one-gluon-exchange potencial, in such a way that the 
zero of the proposed potential coincides with the Cornell one. 
Clearly, this procedure is not able to generate a linearly rising potential, 
associated with confinement in the static case \cite{Vento:2012wp}. Their 
obtained spectrum is in general agreement with the expected values.

We point out that if the propagator used is already renormalized there is
no need to implement a renormalization procedure. In our case, we should
ensure that the expression in Eq.\ \ref{eq:lattice_gluon_propagator} approaches $1/k^2$ in the ultraviolet limit,
i.e.\ we should rescale the corresponding term in the potential by a
suitable constant. At present, this has not been implemented.

\newpage
\section{Method for Obtaining Quarkonium Masses}
\label{sec:Methods}

As shown previously, our choice of propagator resulted in a central potential. 
Since our system contains only two particles, the Hamiltonian will be 
essentially the same as the one of the hydrogen atom, written in terms of 
relative coordinates. We use separation of variables to isolate the angular 
dependence of the wave function (which will be given by the
spherical harmonics) form the radial wave function $R(r)$. We then use the variable substitution
$R(r) = f(r)/r$ to obtain the Ordinary Differential Equation (ODE) for $f(r)$
\begin{equation}
\label{eq:Radial_Equation}
\frac{d^2 f}{dr^2}\,+\,2\mu\left[E - V(r) -2m - \frac {l\left(l+1\right)}{2 \mu r^2} \right] f\left(r\right) \;=\; 0 \,,
\end{equation}
where $\mu$ is the reduced mass
\begin{equation}
\mu \;=\; \frac{m}{2}\,,
\end{equation}
and $m$ is the mass of the heavy quark, e.g.\ the bottom quark.
We use units for which $c = \hbar = 1$ (see Appendix \ref{sec:notation}).
Notice the addition of the rest mass of the particles, which will allow us to 
compare the eigenenergy directly with masses in Ref.\ \cite{Beringer:1900zz}.

The present problem is a boundary-condition problem. One of these conditions is that
$f(0) = 0$. This comes from the requirement of $R(0)$ to be non-singular. The other
requirement is that $f(r \to \infty) = 0$ and comes from the fact that $R(r)$ is
normalized, i.e.
\begin{equation}
\int_0^\infty \left| R(r) \right|^2 r^2 dr \;=\; \int_0^\infty \left| f(r) \right|^2 dr = 1\,.
\end{equation}

Since the potential is arbitrary, it will not usually be possible to find an 
analytic expression for the eigenenergies. We therefore use a numerical approach. 
The algorithm consists in the following steps:
\begin{enumerate}
  \item Finding a likely range for the eigenenergies and discretizing this 
interval in $N$ steps separated by $dE$. (We fix the range using the 
experimental values for the lowest and the highest energy states.)
  \item Numerically solving the ODE in Eq.\ (\ref{eq:Radial_Equation}) 
to obtain the function $f(r)$ for each energy. We use the Numerov method,
explained below, which is given by [see Ref.\ \cite[Chapter 3]{koonin1998computational}]
\begin{equation}
f_{n+1} \;=\; \frac{2 \left(1 + \dfrac{5 dr^2}{12}g_{n}\right) f_n \,-\, \left(1 - \dfrac{dr^2}{12} g_{n-1}\right) f_{n-1}}{1 \,-\, \dfrac{dr^2}{12}g_{n+1}} \,+\, \mathcal{O}(dr^6)\,,
\label{eq:Numerov_method}
\end{equation}
where $f_n \equiv f(r_n)$ and $g_n$ is given by
\begin{equation}
g_n \; \equiv \; -2\mu\left[E - V(r_n) -2m_b - \frac {l\left(l+1\right)}{2 \mu r_n^2} \right]\,.
\label{eq:def_g}
\end{equation}
The variable $r$ is discrete, varying in steps $dr$. Therefore, 
we use the index $n$ to distinguish between different values of $r$. 
We use $r_1 = 0$ and $r_m = r_{\rm MAX}$, where $r_{\rm MAX}$ is a large
value where the wave function is expected to be approximately zero. 
Notice as well that the Numerov method requires knowledge of the first two points of
the wave function. The first point is given by the contour condition. The second
point can be found using a second order Runge-Kutta method. However, this method
will require knowledge of the derivative of the wave function at the origin. We do not
have this information, but it is easy to verify that if the derivative is zero, we
will obtain a wave function that is zero at all points. However, the exact value
will only influence the normalization. Therefore we use $f'(0) = 1$ and
keep in mind that the wave function will need to be normalized if we want to compute
probabilities from it. The motivation for this method comes from considering the 
formula for the three-point difference
\begin{equation}
\frac{f_{n+1}-2f_n+f_{n-1}}{dr^2} \;=\; \frac{d^2 f}{dr^2} \,+\, \frac{dr^2}{12}\, \frac{d^4 f}{dr^4} \,+\, \mathcal{O}(dr^4)\,.
\end{equation}
and using the ODE and the second derivative of the ODE to obtain discrete 
expressions for $f^{(2)}(x)$ and $f^{(4)}(x)$.

  \item Estimating the eigenenergy using the boundary conditions. The 
  functions $f(r)$ will generally diverge to $\pm \infty$, since our guess 
  for $E$ in Eq.\,(\ref{eq:Radial_Equation}) is not an eigenenergy. If we find 
  that the sign of this divergence is reversed when changing from $E_{n,i}$
  to $E_{n+1,i}$, the $i$-th eigenenergy will be estimated by 
  $(E_{n,i}+E_{n+1,i})/2$. The error is taken as $dE/2$.
\end{enumerate}

Note that the only free parameter in the potential is the string force $F_0$,
but we also leave the mass of the bottom quark free since, at present,
it is not well determined. In fact, different approaches give different 
results for $m_b$ [for instance, Ref.\ \cite{Beringer:1900zz} has two values for it]. 
To find the best values for these parameters, we adopt a similar strategy 
used in the calculation of the eigenenergies described above: we set a 
range where it is believed the values of the parameters may be and discretize 
it. We then compute the eigenenergies for each proposed set of parameters and select 
the one that best describes the observed spectrum. The criterion for choosing 
a set of parameters with this property is to look for the set that minimizes 
the residual
\begin{equation}
 R(\text{Parameters}) \;=\; \sum_{i} (E_{i}-E_{i,\text{Experimental}})^2\,.
 \label{eq:residual_criteria}
\end{equation}

Notice that this method of determining the best parameters is general for any potential
used. One can, for example. use a few known energy levels in the fit to obtain optimal 
values for the ODE parameters and then use these values to predict additional energy levels.
The method enables us to test, using the same algorithm, the Coulomb plus linear potential
and the one extracted from the lattice gluon propagator, or other potentials, if desired.
The result obtained with this algorithm is the subject of the next chapter.
\newpage

\cleardoublepage
\chapter{Results}
\label{sec:results}
\thispagestyle{capitulo}

\epigraph{``\textit{All science is either physics or stamp collecting.}''}{Ernest Rutherford\\Quoted in Rutherford at Manchester (1962) by J. B. Birks}

We devote this chapter to presenting the results obtained using the method explained in Section \ref{sec:Methods}. As an initial stage, we test our method by 
applying it to the well known problem of a particle interacting with a Coulomb-like potential. Once it is tested, we proceed to the study of the spectra of bottomonium 
and charmonium. In both cases we make some considerations about the validity of our methods to study these systems and then present our results. At the end of 
the chapter, we present a brief consideration about these results as well as about other methods available to the study of bottomonium and 
charmonium spectra.

\section{Code Testing}

Our first step is to test the method discussed in Section \ref{sec:Methods} for the well known case of 
two particles with opposite electric charge, interacting through a Coulomb potential
\begin{equation}
V(r) \;=\; -\frac{\alpha}{r}\,,
\end{equation}
where $\alpha$ is a constant proportional to the charge of the particles\footnote{In this case we will consider the ODE in Eq.\ \ref{eq:Radial_Equation}
without the term $2m$.}.
This allows us to test the validity of the method and better understand its limitations, since we know the analytic solution for the
eigenenergies and the corresponding wave functions in this case. The eigenenergies are given by
\begin{equation}
 E_n \;=\; -\frac{\mu}{2}\frac{\alpha^2}{n^2}\,,
\end{equation}
where $n \;=\; l+1,l+2,\dots$ and $\mu$ is the reduced mass.

Instead of choosing the parameters so as to reproduce the hydrogen atom, i.e.\ to pick $\mu$ as the electron mass and $\alpha$ 
as the fine-structure constant, for simplicity,
we choose the reduced mass $\mu$ to be $\num{1.0}$ in some energy units and take
$\alpha = \num{1.0}$. With these parameters, the eigenenergies are given by
\begin{equation}
 E_n \;=\; -\frac{0.5}{n^2}\,.
\label{eq:Coulomb_Pot_Energies}
\end{equation}

The ODE in Eq.\ \ref{eq:Radial_Equation} has as parameters the reduced mass $\mu$ and the parameters in the potential. 
In the present case, the potential parameter is $\alpha$. To test our methods, we will consider a hypothetical 
situation in which we do not know the value of $\mu$ and $\alpha$, but know some energy levels (that we calculate
from Eq.\ \ref{eq:Coulomb_Pot_Energies}). We will apply the method described at the end of Section \ref{sec:Methods} 
to fit the values for the parameters that best describe the spectrum given in Eq.\ \ref{eq:Coulomb_Pot_Energies}, 
i.e.\ $E_{i,\text{Experimental}}$ in Eq.\ \ref{eq:residual_criteria} is given by Eq.\ \ref{eq:Coulomb_Pot_Energies}.
The values found for $\mu$ and $\alpha$ can be seen as estimates of the ``exact'' values quoted above.

We search for eigenenergies between the values of \num{-1.5} and \num{0}, sampling uniformly \num{10000} values in this interval, 
i.e.\ two consecutive energy values differ by \num{1.5E-4}, corresponding to a precision of \num{7.5E-5}. We run the code for several values of the discretization step $dr$ of
the wave functions to study discretization effects on the eigenenergies found. The determination of the range over which 
the free parameters $\mu$ and $\alpha$ will run is by trial and error. We start by guessing a range and, 
if the result that minimizes $R$ (see Eq.\ \ref{eq:residual_criteria}) is at one end of the range, we increase it in this 
direction. In our case, we vary $\alpha$ from \num{0.9} to \num{1.1} and $2\mu$ from \num{1.9} to \num{2.1}. In both cases we 
choose a step size between two parameters of $\num{0.01}$, meaning that we try a total of \num{400} parameter values. We use twice the
 step size as an estimation of our error in the parameter determination.

For the fitting, we choose $l = 0,\,1$ and $2$. For each $l$, we will choose the first five energy levels. This means we have a total
of fifteen predicted points to fit. We follow the usual labeling for hydrogen-atom states, i.e.\ they are labeled in the format 
$n l$, where $l = \,$S, P or D. We relate $n$ to $l$ by $n = l + k$ and $k$ is an integer such as $k >0$. In our application
for heavy quarks, we will select some known states and try to describe the others. But for this test case, we will not try 
make any predictions, since we are aiming at testing the fit method. The obtained results are shown in Table \ref{table:Results_Hidrogen}.
\renewcommand{\arraystretch}{1.3}
\begin{table}[h]
\centering
\caption{Results for a Coulomb-like potential using two different step sizes to calculate the wave functions. 
The header of the first column contains the parameters used to generate the points to which we fit the results of our computation.}
\label{table:Results_Hidrogen}
\resizebox{16cm}{!}{
\begin{tabu}{||c|c||c|c||c|c||}
\hline
\multicolumn{ 2}{||c||}{ $\alpha = \num{1.0}$}	& \multicolumn{ 1}{c}{                  }	& $\alpha = 1.04(2)$    	& \multicolumn{ 1}{c}{                 }    	& $\alpha = \num{0.98(2)E-3}$\\
\multicolumn{ 2}{||c||}{ $\mu = \num{1.0}$}    	& \multicolumn{ 1}{c}{ $dr = \num{1E-1}$}	& $\mu = 0.955(10)$     	& \multicolumn{ 1}{c}{$dr = \num{1E-2}$}    	& $\mu = \num{1.045(10)}$ \\
\multicolumn{ 2}{||c||}{                }     	& \multicolumn{ 1}{c}{                  }	& $R = \num{4.38E-5}$   	& \multicolumn{ 1}{c}{                 }    	& $R  = \num{6.25E-7}$ \\
\hline
{\bf System}	& \multicolumn{ 1}{c||}{{\bf Predicted}}	& {\bf Calculated Energy} & \multicolumn{ 1}{c||}{{\bf Deviation from}}	& {\bf Calculated Energy}	& \multicolumn{ 1}{c||}{{\bf Deviation from}} \\
{\bf State}		& \multicolumn{ 1}{c||}{{\bf Energy}}   	& {\bf ($\num{+-7.5E-5}$)}& \multicolumn{ 1}{c||}{{\bf Analytic Result\footnotemark}}& {\bf ($\num{+-7.5E-5}$)}& \multicolumn{ 1}{c||}{{\bf Analytic Result\footnotemark[\value{footnote}]}}\\
\hline\rowfont{\large}
1S 		      	& -0.500000																& -0.496425								&     0.003575 	(0.7150\%)										&   -0.499725						&     0.000275 (0.0550\%)\\
\hline\rowfont{\large}
2S 		      	& -0.125000																& -0.126525								&     0.001525	(1.2200\%)										&   -0.125175						&     0.000175 (0.1400\%)\\
\hline\rowfont{\large}
3S 		      	& -0.055556																& -0.056625								&     0.001069  (1.9242\%)										&   -0.055725						&     0.000169 (0.3042\%)\\
\hline\rowfont{\large}
4S 		      	& -0.031250																& -0.032025								&     0.000775 	(2.4800\%)										&   -0.031275						&     0.000025 (0.0800\%)\\
\hline\rowfont{\large}
5S 		      	& -0.020000																& -0.020475								&     0.000475 	(2.3750\%)										&   -0.020025						&     0.000025 (0.1250\%)\\
\hline\rowfont{\large}
2P 		      	& -0.125000																& -0.129075								&     0.004075 	(3.2600\%)										&   -0.125475						&     0.000475 (0.3800\%) \\
\hline\rowfont{\large}
3P 		      	& -0.055556																& -0.057375								&     0.001819 	(3.2742\%)										&   -0.055725						&     0.000169 (0.3042\%)\\
\hline\rowfont{\large}
4P 		      	& -0.031250																& -0.032325								&     0.001075	(3.4400\%)										&   -0.031425						&     0.000175 (0.5600\%) \\
\hline\rowfont{\large}
5P 		      	& -0.020000																& -0.020625								&     0.000625 	(3.1250\%)										&   -0.020025						&     0.000025 (0.1250\%)\\
\hline\rowfont{\large}
6P 		      	& -0.013889																& -0.014325								&     0.000436 	(3.1392\%)										&   -0.013875						&     0.000014 (0.1008\%) \\
\hline\rowfont{\large}
3D 		      	& -0.055556																& -0.057375								&     0.001819	(3.2742\%)										&   -0.055725						&     0.000169 (0.3042\%)\\
\hline\rowfont{\large}
4D 		      	& -0.031250																& -0.032325								&     0.001075	(3.4400\%)										&   -0.031425						&     0.000175 (0.5600\%) \\
\hline\rowfont{\large}
5D 		      	& -0.020000																& -0.020625								&     0.000625 	(3.1250\%)										&   -0.020025						&     0.000025 (0.1250\%) \\
\hline\rowfont{\large}
6D 		      	& -0.013889																& -0.014325								&     0.000436	(3.1392\%)										&   -0.013875						&     0.000014 (0.1008\%) \\
\hline\rowfont{\large}
7D 		      	& -0.010204																& -0.009975								&     0.000229	(2.4422\%)										&   -0.009825						&     0.000379 (3.7142\% )\\
\hline
\end{tabu}
}
\sourcebytheauthour
\end{table}
\footnotetext{In parentheses, we give the relative error.}

We notice that, as we improve the precision of our computed wave functions, we obtain an improvement 
in the description of the spectrum as well. We can clearly see this by the value obtained for $R$, which for
$dr = \num{1E-2}$ is two orders of magnitude bellow $R$ when using $dr = \num{1E-1}$.
Surprisingly, this does not imply that the parameters used in the ODE are nearer to
expected values. We obtain that we reduced the error for $\alpha$ from 4 \% to 2 \%, but for
$\mu$ we have an increase in the error from 0.5 \% to 4.5 \%.  Also, there is a limit
to how much we can increase the precision of the calculation of the wave function.
Not only does the code become increasingly slow, but also the divergence of the potential
at $r = 0$ induces divergent values in this region, introducing numerical 
imprecision to the wave function. The result is that for $dr = \num{1e-3}$ our 
algorithm is unable to find several states inside the specified range. This probably happens
because, for small $dr$, we start the integration of $f(r)$ too close to the singularity of
the potential $V(r)$.

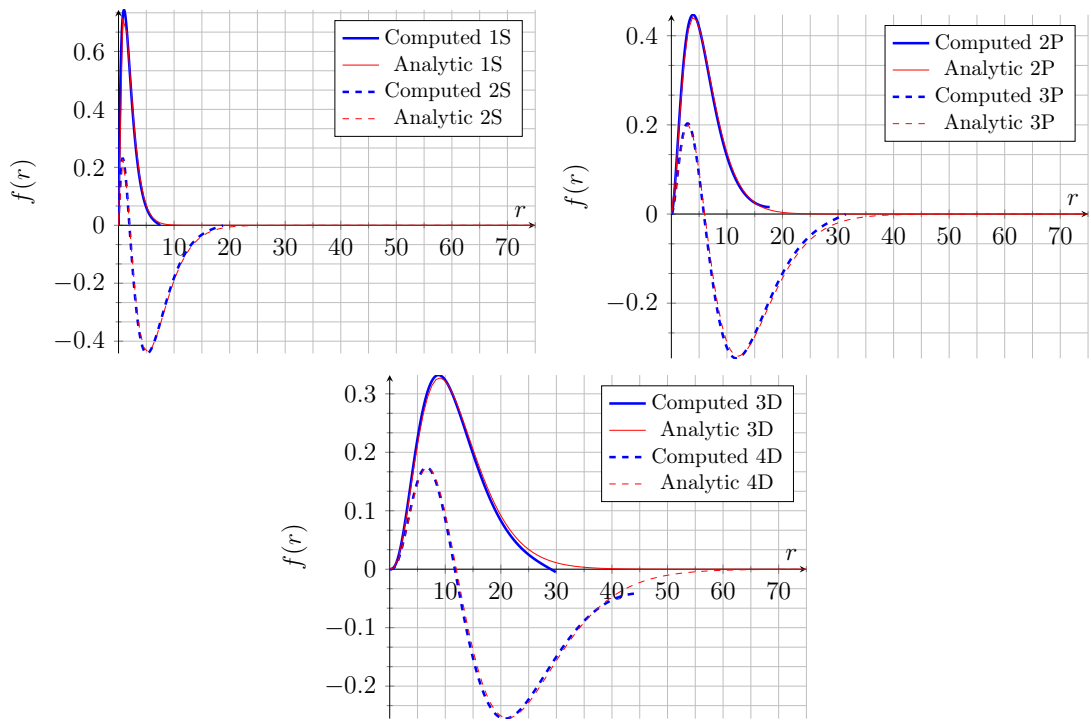
\begin{figure}[ht]
\caption{Comparison of wave functions for a Coulomb-like potential obtained through our computations with the analytic solutions.
}
\label{fig:wf_plots_H_atom}
  \pgfplotstableread{./img/Hidrogen_Data/Normalized_0_wave_function_0.dat}{\zwfz}
  \pgfplotstableread{./img/Hidrogen_Data/Normalized_0_wave_function_1.dat}{\zwfo}

  \pgfplotstableread{./img/Hidrogen_Data/Normalized_1_wave_function_0.dat}{\owfz}
  \pgfplotstableread{./img/Hidrogen_Data/Normalized_1_wave_function_1.dat}{\owfo}

  \pgfplotstableread{./img/Hidrogen_Data/Normalized_2_wave_function_0.dat}{\twfz}
  \pgfplotstableread{./img/Hidrogen_Data/Normalized_2_wave_function_1.dat}{\twfo}

  \centering
\subfigure
{
  \centering
  \label{fig:wf_1s_2s}
\begin{tikzpicture}[scale=0.8]
  \begin{axis}
     [xlabel=$r$,
      ylabel=$f(r)$,
      axis lines = left,
      axis x line=center,
      grid=both,
      minor y tick num = {2},
      minor x tick num = {1},
			legend pos = {north east},    
			legend style = {align=justify,font=\footnotesize},
			legend entries = {Computed 1S,Analytic 1S,Computed 2S,Analytic 2S},]
    \addplot [blue,very thick] table {\zwfz};
    \addplot [red,domain=0:75,samples=100]{2.0*x*exp(-x)};
		\addplot[blue,very thick,style=dashed] table {\zwfo};
		\addplot[red,domain=0:75,samples=100,style=dashed]{(0.7071067810)*x*(1-x/2)*exp(-x/2)};
		
  \end{axis}
\end{tikzpicture}}\;%
\subfigure

	{
  \centering
  \label{fig:wf_2p_3p}
\begin{tikzpicture}[scale=0.8]
  \begin{axis}
     [xlabel=$r$,
      ylabel=$f(r)$,
      axis lines = left,
      axis x line=center,
      grid=both,
      minor y tick num = {2},
      minor x tick num = {1},
			legend pos = {north east},    
			legend style = {align=justify,font=\footnotesize},
			legend entries = {Computed 2P,Analytic 2P,Computed 3P,Analytic 3P},]
    \addplot [blue,very thick] table {\owfz};
    \addplot [red,domain=0:75,samples=100]{0.2041241452*exp(-x/2)*x^2};
		\addplot [blue,very thick,style=dashed] table {\owfo};
		\addplot [red,domain=0:75,samples=100,style=dashed]{0.1209624564*exp(-x/3)*(x^2)*(1.-x/6)};
  \end{axis}
\end{tikzpicture}}

\subfigure

{
  \centering
  \label{fig:wf_3d_4d}
\begin{tikzpicture}[scale=0.8]
  \begin{axis}
     [xlabel=$r$,
      ylabel=$f(r)$,
      axis lines = left,
      axis x line=center,
      grid=both,
      minor y tick num = {2},
      minor x tick num = {1},
			legend pos = {north east},    
			legend style = {align=justify,font=\footnotesize},
			legend entries = {Computed 3D,Analytic 3D,Computed 4D,Analytic 4D},]
    \addplot [blue,very thick] table {\twfz};
    \addplot [red,domain=0:75,samples=100]{0.009016009177*exp(-x/3)*x^3};

		\addplot [blue,very thick,style=dashed] table {\twfo};
		\addplot [red,domain=0:75,samples=100,style=dashed]{0.006987712428*exp(-x/4)*(x^3)*(1.-x/12)};

  \end{axis}
\end{tikzpicture}}
\sourcebytheauthour
\end{figure}

In Fig.\ \ref{fig:wf_plots_H_atom}, we plot the obtained wave functions (blue lines) for the 
two lowest energy states  for $l = 0,\, 1, \, 2$ (upper left panel, upper right panel and center panel respectively).
We compare them with the analytic solution (red lines). The lowest energy state for each $l$ is presented with a 
solid line and the first excited state with a dashed line. Both are normalized and we omit in these plots 
the divergence of our computed wave functions for large $r$. Notice that we get good
agreement between them, the errors being largest for $l = 2$\,.

For comparison, we changed the integration algorithm to a second-order Runge-Kutta method \cite{press2007numerical}, i.e.
the point $f_{n+1}$ is given by\footnote{We use the notation from Eqs.\ \ref{eq:Numerov_method} and \ref{eq:def_g}, defining
$g_{n+1/2}\equiv g(r_n+dr/2)$.}
\begin{align}
c_1 \;&=\; dr\, f^\prime_n\,, \nonumber \\
c_2 \;&=\; dr\, g_n f_n\,, \nonumber \\
c_3 \;&=\; dr (f^\prime_n+c_2/2)\,, \nonumber \\
c_4 \;&=\; dr\, g_{n+1/2} (f_n+c_1/2)\,, \nonumber \\
f_{n+1} \;&=\; f_n + c_3\,, \nonumber \\ 
f^\prime_{n+1} \;&=\; f^\prime_n + c4\,.
\label{eq:RK2_method}
\end{align}

We use the same setup as in the Numerov case. We ran on a CPU Intel i5-3230M under Windows Seven Professional and this
search took \SI{22.4}{\minute} using the Numerov algorithm with the integration step $dr = \num{1E-2}$. The same search using 
the second-order Runge-Kutta (RK2) method given in Eq.\ \ref{eq:RK2_method} took \SI{10.6}{\minute}.
This is not surprising, since the RK2 method requires just two calls to the function $g(r)$, while the 
Numerov requires three calls. However, the precision of the Numerov algorithm is of $\mathcal{O}(dr^6)$, while
the RK2 one is of $\mathcal{O}(dr^3)$. One would expect therefore to get much better results for the Numerov method.
Surprisingly, this is not observed. The RK2 yields the same result for the set of parameters obtained when using the
Numerov method. We obtain a spectrum with the same order of accuracy, with $R = \num{5.51E-7}$ (compare with the results in 
Table \ref{table:Results_Hidrogen}).

This led us
to conclude that the RK2 is more efficient and has enough precision for our application.

\section{Bottomonium}

Our aim is to compute the masses for the various states of the bottomonium. Ref.\ \cite{Beringer:1900zz} gives us two 
different values for the bottom quark mass. We use these values to estimate that this mass lies between 
\SI{4.18}{\giga \electronvolt} and \SI{4.66}{\giga \electronvolt}. Also, from data available in Ref.\ \cite{Beringer:1900zz}, 
we obtain the mass of the bottomonium in the lowest energy state and in the highest observed excited state. We can use these values 
to estimate the validity of our non-relativistic approximation. The criterion for the system to be non-relativistic will be 
that $\,\gamma = 1/\sqrt{1-v^2}$ be near 1. We estimate $\gamma$ by remembering that $E = \gamma m\,$ and therefore 
$E/m$ is a good estimate for $\gamma$.

The best-case scenario occurs for the ground state, which has a mass of approximately \SI{9.43}{\giga \electronvolt} and, 
considering the bottom mass to be \SI{4.66}{\giga \electronvolt}, we have
\begin{equation}
\gamma \;=\; \frac{E}{m} \;=\; \frac{E}{2 m_b} \;\approx\; \frac{9.43}{9.32} \;\approx\; 1.01\,.
\end{equation}
This shows that, in the ground state, we will be dealing with a non-relativistic system to a very good approximation.

For the worst-case scenario, we consider the state with highest energy and the bottom mass as \SI{4.18}{\giga \electronvolt},
to obtain
\begin{equation}
\gamma \;=\; \frac{E}{2 m_b} \;\approx\; \frac{11.02}{8.36} \;\approx\; 1.32\,.
\end{equation}
If the particle were a classical particle, this would correspond to it traveling at about 43\% of the speed of light, and 
therefore relativistic effects should be more visible in this case. Therefore, we expect that our approximation will not 
hold too well in this limit.

Before proceeding to computing the masses using our algorithm, we will briefly review some properties of quarkonia, such
as the nomenclature scheme used and the relation between their quantum numbers.

We are interested in knowing the orbital angular momentum $l$ of the system, since this is one of the parameters needed to solve
Schrödinger’s equation. In some cases $l$ is already known and the notation adopted by Ref.\ \cite{Beringer:1900zz} is to 
include it in the particle name using the usual spectroscopic notation, e.g.\ $\Upsilon(1S)$, $\Upsilon(2S)$ etc.
For the cases in which $l$ is unknown, an approximate value of the particle mass is used, e.g.\ $\Upsilon(10860)$.
However, these particles have other measured quantum numbers, which gives us hints about $l$. Usually, these are the total 
angular momentum $j$, the particle parity quantum number $P$ and the charge-conjugation quantum number $C$. These quantities 
are related to $l$ and the total spin of the two quarks $s$ by
\begin{align}
P \;&=\; (-1)^{(l+1)}\,, \\
C \;&=\; (-1)^{(l+s)}\,.
\end{align}
Since we know that quarks have spin one half, we know that $s$ is constrained to $s = 0,1$\,. This allows us to identify if 
the particle has $l$ even or odd. In fact, this actually helps define the name of the particle. The adopted naming scheme 
for bottomonium is presented in Table \ref{tb:naming_scheme_bottomonium}.
\begin{table}[h]%
\centering
\caption{Naming scheme for bottomonium states.}
\label{tb:naming_scheme_bottomonium}
\begin{tabu}{||c|c|c||}
\hline
Name				& s & l 		\\
\hline
$\eta_b$		& 0 & Even	\\
$h_b$				& 0 & Odd		\\
$\Upsilon$	& 1 & Even	\\
$\chi_b$		& 1 & Odd		\\
\hline
\end{tabu}\vspace*{10pt}\\Source: BERINGER et al.\ \cite{Beringer:1900zz}.
\end{table}

For the cases in which the main quantum numbers are not known and it is not possible to assign one of these
names, the state is denoted by an $X$.

As inputs in our fit, we will use only particles for which $l$ is known. For the bottomonium this means the states 
$1S$ to $4S$, $1P$ to $3P$ and $1D$. This will leave for our model to make predictions just two states (see Table \ref{tb:bottomonium_experimental}).
Therefore, in addition to this fit using eight states, we perform another fit using
four states, namely the $1S$, $2S$, $1P$ and $2P$.

Since our potential is not sensitive to $j$ and $s$, we need a prescription 
for how to average over the masses of particles that have the same $l$, but different $j$ and $s$. The way to do this 
is to remember that the particles are detected as resonances in scattering processes, with their masses given by the peak of 
the scattering amplitude $\sigma_{M_i,\,\Gamma_i}(m)$. We model these resonances by normal distributions centered at 
the mass of the state and with standard deviation equal to the decay width\footnote{The distribution usually used is the relativistic Breit–Wigner distribution [see Ref.\ \cite{Beringer:1900zz}].
However, it is not possible to extract the particle mass using a simple operation (such as averaging over the distribution) and therefore we choose to model
them with normal distributions.}
\begin{equation}
\sigma_{M_i,\,\Gamma_i}(m) \;=\; \frac{1}{\Gamma_i \sqrt{2 \pi}} \exp\left[ - \frac{(m-M_i)^2}{2 \Gamma_i^2} \right]\,.
\end{equation}
If we want to describe $n$ states as a single one, we average these distributions
\begin{equation}
\overline{\sigma}(m) \;=\; \sum_{i=1}^n \frac{\sigma_{M_i,\,\Gamma_i}(m)}{n}\,.
\label{eq:normal_dist_average}
\end{equation}
We define $M_i$ as
\begin{equation}
M_i \;\equiv\; \int_{-\infty}^\infty m\, \sigma_{M_i,\,\Gamma_i}(m) \, dm\,.
\end{equation}
It is immediate from this definition and Eq.\ \ref{eq:normal_dist_average} that the mass $M$ will be
\begin{equation}
M \;=\; \sum_{i=1}^n \frac{M_i}{n}\,.
\end{equation}

We use as an estimate of the mass error the decay width of the particle. For the case where we have to consider several states
as a single state, we proceed in a similar way as for the mass. The width will be given by
\begin{equation}
\overline{\Gamma} \;=\; \sqrt{\int_{-\infty}^\infty m^2 \overline{\sigma}(m)\, dm - \left(\int_{-\infty}^\infty m\, \overline{\sigma}(m)\, dm \right)^2}
       \;=\ \sqrt{\sum_{i=1}^n \frac{\Gamma_i^2}{n}}\,.
\end{equation}
Unfortunately, we do not have the full widths for all states. If we have to average over states with missing information for 
the decay width, the error of the averaged state is taken as one half of the difference between the mass of the most massive 
state and the least massive one. If we have missing information about the decay width, but there is no need to perform an average over the states, we use the error associated
with the uncertainty of the location of the peak of the resonance, given in Ref.\ \cite{Beringer:1900zz}.

In Table \ref{tb:bottomonium_experimental} we summarize the data extracted from Ref.\ \cite{Beringer:1900zz} and
how they are related to the input used in our program.

We implement the algorithm described in Section \ref{sec:Methods}, but using the RK2 method, since our tests showed no gain in 
the use of the Numerov method. We vary the quark mass from \SI{4.1}{\giga \electronvolt}
through \SI{4.7}{\giga \electronvolt} [which includes both masses listed in Ref.\ \cite{Beringer:1900zz}] in steps of $\SI{0.01}{\giga \electronvolt}$. For the string-tension 
parameter $F_0$, we search from \SI{0.0}{\giga \electronvolt \squared} through \SI{1.0}{\giga \electronvolt \squared} in steps of $\SI{0.01}{\giga \electronvolt \squared}$. We apply 
the same algorithm using the Cornell potential for comparison. The step used for the integration of the wave function
is $dr = \SI{5e-3}{\per \giga \electronvolt}$ and the energy range in which we seek our eigenvalues goes from \SI{9}{\giga \electronvolt}
to \SI{11}{\giga \electronvolt} in steps of \SI{2e-4}{\giga \electronvolt}. The fitting method is the one described in Section 
\ref{sec:Methods} and its results, using eight states ($1S$ to $4S$, $1P$ to $3P$ and $1D$) can be found in Table \ref{table:Results_bottom_8_states}.

\begin{table}[h!]%
\centering
\caption{Experimental spectrum for the bottomonium and its preparation (see text) as input for our calculations.}
\label{tb:bottomonium_experimental}
\resizebox{16cm}{!}{
\begin{tabu}{||c|c|c|c|c|c||}
\hline
Particle Name 		& Mass (\si{\giga \electronvolt}) & Decay width (\si{\giga \electronvolt})	& $l$ 							& Mass Input (\si{\giga \electronvolt}) & Mass Error\footnotemark(\si{\giga \electronvolt})\\ \hline
$\eta_b(1S)$			& 9.398														& \num{1.08E-2}														& \multirow{2}{*}{0}& \multirow{2}{*}{9.429}							& \multirow{2}{*}{0.008} \\
$\Upsilon(1S)$		& 9.4603													& \num{5.402E-5}													& & & \\ \hline
$\eta_b(2S)$			& 9.999														& $ < \num{2.4E-2}$												& \multirow{2}{*}{0}& \multirow{2}{*}{10.01113}							& \multirow{2}{*}{$0.017$} \\
$\Upsilon(2S)$		& 10.02326												& \num{3.198E-5}													& & & \\ \hline
$\Upsilon(3S)$		& 10.3552													& \num{2.032E-5}													& 0									& 10.3552																& \num{2.03E-5} \\ \hline
$\Upsilon(4S)$		& 10.5794													& \num{2.05E-2}														&	0									&	10.5794																& 0.021 \\ \hline
$\chi_{b0}(1P)$ 	& 9.85944													& -																				&	\multirow{4}{*}{1}& \multirow{4}{*}{9.89093}							& \multirow{4}{*}{0.02638} \\
$\chi_{b1}(1P)$		& 9.89278													& -																				& & & \\
$h_b(1P)$					& 9.8993													& -																				& & & \\
$\chi_{b2}(1P)$		& 9.91221													& -																				& & & \\ \hline
$\chi_{b0}(2P)$ 	& 10.2325													& -																				& \multirow{4}{*}{1}& \multirow{4}{*}{10.2541}							& \multirow{4}{*}{0.0181} \\
$\chi_{b1}(2P)$		& 10.25546												& -																				& & & \\
$h_b(2P)$					& 10.2598													& -																				& & & \\
$\chi_{b2}(2P)$		& 10.26865												& -																				& & & \\ \hline
$\chi_b(3P)$			& 10.534													& -																				& 1									& 10.534																& 0.009 \\ \hline
$\Upsilon(1D)$		& 10.1637													& -																				& 2									& 10.1637																& 0.0014 \\ \hline
$\Upsilon(10860)$	& 10.876													& \num{5.5E-2}														& Even							& 10.876 																& 0.055 \\ \hline
$\Upsilon(11020)$	&	11.019													& \num{1.15E-2}														& Even							& 11.019 																& 0.011 \\ \hline
\end{tabu}
}
\vspace*{10pt}\\Source: Adapted from BERINGER et al.\ \cite{Beringer:1900zz}.
\end{table}
\footnotetext{We considered $\Gamma_{\eta_b(2S)} = \SI{2.4E-2}{\giga \electronvolt}$ for the input mass error on the second row.}

\begin{figure}[h!]
\centering
\caption{The mass spectrum of the bottomonium, along with the computed spectrum using the two different potentials and 8 states to fit the parameters (data from Table \ref{table:Results_bottom_8_states}; see text for details).}
\label{fig:Bottomonium_Mass_Spectrum_8_states}
 \def\Ei{9.42915}
 \def\Eii{10.01113}
 \def\Eiii{10.3552}
 \def\Eiv{10.5794}
 \def\Ev{9.89093}
 \def\Evi{10.2541}
 \def\Evii{10.530}
 \def\Eviii{10.1637}
 
 \def\Eix{10.6072}
 \def\Ex{10.6522}
 
 \def\Exi{10.876}
 \def\Exii{11.019}

 \def\LEi{9.5383}
 \def\LEii{10.0049}
 \def\LEiii{10.3579}
 \def\LEiv{10.6621}
 \def\LEv{9.8445}
 \def\LEvi{10.2191}
 \def\LEvii{10.5359}
 \def\LEviii{10.0773}
 
 \def\LEix{10.9373}
 \def\LEx{11.1923}
 
 \def\LExi{10.8195}
 \def\LExii{11.0809}
 \def\LExiii{11.3259}
 
 \def\LExiv{10.4071}
 \def\LExv{10.6995}
 \def\LExvi{10.9677}
 \def\LExvii{11.2179}
 \def\LExviii{11.4543}

 \def\CEi{9.5367}
 \def\CEii{9.9931}
 \def\CEiii{10.3461}
 \def\CEiv{10.6521}
 \def\CEv{9.8361}
 \def\CEvi{10.2085}
 \def\CEvii{10.5263}
 \def\CEviii{10.0743}
  
 \def\CEix{10.9291}
 \def\CEx{11.1855}
 
 \def\CExi{10.8117}
 \def\CExii{11.0747}
 \def\CExiii{11.3209}
 
 \def\CExiv{10.3955}
 \def\CExv{10.6903}
 \def\CExvi{10.9603}
 \def\CExvii{11.2121}
 \def\CExviii{11.4499}
 \begin{tikzpicture}[scale=2.5]

  \draw (2.4,9.6) -- ++(2.6,0) -- ++(0,-.65) -- ++(-2.6,0) -- ++(0,.65);
	\draw[very thick] (2.5,9.5) -- +(0.5,0) node[right] {\scriptsize Fit input};
	\draw[very thick,lightgray] (2.5,9.35) -- +(0.5,0) node[right] {\scriptsize \textcolor{black}{Experimental data}};
	\draw[thick,dashed,red] (2.5,9.2) -- +(0.5,0) node[right] {\scriptsize \textcolor{black}{Lattice propagator potential results}};
	\draw[thick,dotted,blue] (2.5,9.05) -- +(0.5,0) node[right] {\scriptsize \textcolor{black}{Cornell potential results}};

    \draw[very thick,->,>=latex] (0,8.5)--(0,12.2) node[right] (0,11.5) {Energy};

  \foreach \y in {9,...,12}
  	\draw (1pt,\y) -- (-3pt,\y) node[anchor=east] {\y};
  	
  \draw[very thick] (0.5,\Ei) -- (1.5,\Ei);
  \draw[very thick] (0.5,\Eii) -- (1.5,\Eii);
  \draw[very thick] (0.5,\Eiii) -- (1.5,\Eiii);
  \draw[very thick] (0.5,\Eiv) -- (1.5,\Eiv);
  
  \draw[very thick] (2.0,\Ev) -- (3.0,\Ev);
  \draw[very thick] (2.0,\Evi) -- (3.0,\Evi);
  \draw[very thick] (2.0,\Evii) -- (3.0,\Evii);
  
  \draw[very thick] (3.5,\Eviii) -- (4.5,\Eviii);

  \draw[lightgray,very thick] (0.5,\Exi) -- (4.5,\Exi);
	\draw (4.525,\Exi) -- (5,\Exi-0.08) node[right] {\scriptsize $\Upsilon(10860)$};
  \draw[lightgray,very thick] (0.5,\Exii) -- (4.5,\Exii);
	\draw (4.525,\Exii) -- (5,\Exii+0.08) node[right]  {\scriptsize $\Upsilon(11020)$};

  \draw[thick,dashed,red] (0.5,\LEi) -- (1.5,\LEi);
  \draw[thick,dashed,red] (0.5,\LEii) -- (1.5,\LEii);
  \draw[thick,dashed,red] (0.5,\LEiii) -- (1.5,\LEiii);
  \draw[thick,dashed,red] (0.5,\LEiv) -- (1.5,\LEiv);
  
  \draw[thick,dashed,red] (2.0,\LEv) -- (3.0,\LEv);
  \draw[thick,dashed,red] (2.0,\LEvi) -- (3.0,\LEvi);
  \draw[thick,dashed,red] (2.0,\LEvii) -- (3.0,\LEvii);

  \draw[thick,dashed,red] (3.5,\LEviii) -- (4.5,\LEviii);
  
  \draw[thick,dashed,red] (0.5,\LEix) -- (1.5,\LEix);
  \draw[thick,dashed,red] (0.5,\LEx) -- (1.5,\LEx);
  
  \draw[thick,dashed,red] (2.0,\LExi) -- (3.0,\LExi);
  \draw[thick,dashed,red] (2.0,\LExii) -- (3.0,\LExii);
  \draw[thick,dashed,red] (2.0,\LExiii) -- (3.0,\LExiii);
  
  \draw[thick,dashed,red] (3.5,\LExiv) -- (4.5,\LExiv);
  \draw[thick,dashed,red] (3.5,\LExv) -- (4.5,\LExv);
  \draw[thick,dashed,red] (3.5,\LExvi) -- (4.5,\LExvi);
  \draw[thick,dashed,red] (3.5,\LExvii) -- (4.5,\LExvii);
  \draw[thick,dashed,red] (3.5,\LExviii) -- (4.5,\LExviii);
  
  \draw[thick,dotted,blue] (0.5,\CEi) -- (1.5,\CEi);
  \draw[thick,dotted,blue] (0.5,\CEii) -- (1.5,\CEii);
  \draw[thick,dotted,blue] (0.5,\CEiii) -- (1.5,\CEiii);
  \draw[thick,dotted,blue] (0.5,\CEiv) -- (1.5,\CEiv);
  
  \draw[thick,dotted,blue] (2.0,\CEv) -- (3.0,\CEv);
  \draw[thick,dotted,blue] (2.0,\CEvi) -- (3.0,\CEvi);
  \draw[thick,dotted,blue] (2.0,\CEvii) -- (3.0,\CEvii);

  \draw[thick,dotted,blue] (3.5,\CEviii) -- (4.5,\CEviii);
  
  \draw[thick,dotted,blue] (0.5,\CEix) -- (1.5,\CEix);
  \draw[thick,dotted,blue] (0.5,\CEx) -- (1.5,\CEx);
  
  \draw[thick,dotted,blue] (2.0,\CExi) -- (3.0,\CExi);
  \draw[thick,dotted,blue] (2.0,\CExii) -- (3.0,\CExii);
  \draw[thick,dotted,blue] (2.0,\CExiii) -- (3.0,\CExiii);
  
  \draw[thick,dotted,blue] (3.5,\CExiv) -- (4.5,\CExiv);
  \draw[thick,dotted,blue] (3.5,\CExv) -- (4.5,\CExv);
  \draw[thick,dotted,blue] (3.5,\CExvi) -- (4.5,\CExvi);
  \draw[thick,dotted,blue] (3.5,\CExvii) -- (4.5,\CExvii);
  \draw[thick,dotted,blue] (3.5,\CExviii) -- (4.5,\CExviii);

  \node at (1.0,8.75) {S states};
  \node at (2.5,8.75) {P states};
  \node at (4.0,8.75) {D states};
 \end{tikzpicture}
\sourcebytheauthour
\end{figure}
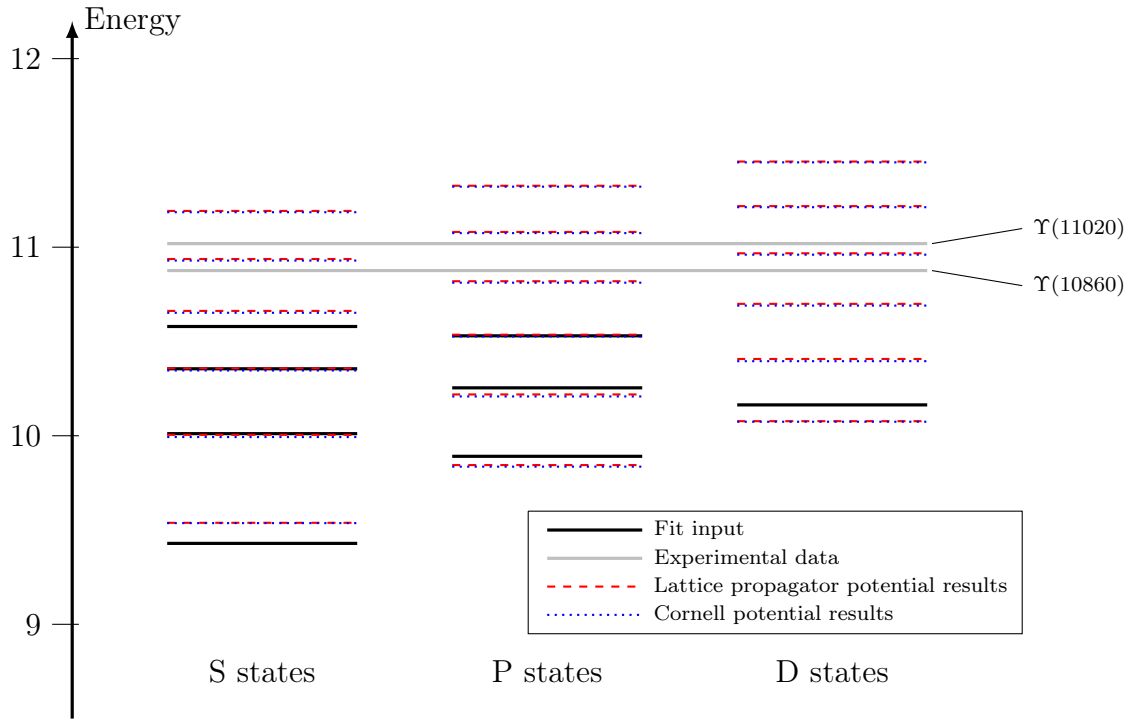

The data of Table \ref{table:Results_bottom_8_states} may be visualized in figure \ref{fig:Bottomonium_Mass_Spectrum_8_states}. Black lines
represents states of the observed spectrum used as input in the fit while gray lines are not used in the fit. Red lines 
are the results from our simulation using the potential from the lattice propagator and blue lines from the Coulomb plus
linear propagator.

\begin{figure}[h!]
\caption{In the left panel, we compare the potential obtained using the lattice propagator ($V_{\mbox{LGP}}$, in blue)
with the Cornell potential ($V_{\mbox{Cornell}}$, in red). The difference between these two potentials is shown in the right panel.}
\label{fig:Potential}
\centering
\def\C{0.784}
\def\s{2.508}
\def\u{0.768}
\def\t{0.720}
\def\Delt{4*\t*\t-\u*\u*\u*\u}
\def\thet{atan(sqrt(\Delt)/\u/\u)}
\def\hc{0.197326971}
\begin{tikzpicture}[scale=.8]
	\begin{axis}
	[ 
		xlabel = r,
		ylabel = V,
		axis lines = left,
	  	axis x line=center,
		x unit=\si{\femto \meter},
		y unit=\si{\giga \electronvolt},
	  	grid=both,
	 	minor y tick num = {2},
		minor x tick num = {2},
		every axis x label/.style={
    at={(ticklabel* cs:1.05)},
    anchor=south,},
		legend pos = {north west,font=\small},    
		legend style = {align=justify},
		legend cell align = left,
		legend entries = {$V_{\mbox{LGP}}$,$V_{\mbox{Cornell}}$},
	]
	\addplot[blue,domain=.01:5.0,samples=300]{-(4/3) * (0.1185/(x/\hc)) * (\C/sqrt(\Delt)) * exp(-(x/\hc)*sqrt(\t)*cos(deg(\thet/2)))*( sqrt(\Delt)*cos(deg((x/\hc)*sqrt(\t)*sin(\thet/2))) - (\u*\u-2*\s)*sin(deg((x/\hc)*sqrt(\t)*sin(\thet/2))) )+0.24*(x/\hc)};
	\addplot[red,domain=.01:5.0,samples=300]{-(4/3) * (0.1185/(x/\hc))+0.24*(x/\hc)};
	\end{axis}
\end{tikzpicture}
\begin{tikzpicture}[scale=.8]
	\begin{axis}
	[ 
		xlabel=$r$,
		ylabel=$V_{\mbox{LGP}} - V_{\mbox{Cornell}}$,
		axis lines = left,
	  	axis x line=center,
		x unit=\si{\femto \meter},
		y unit=\si{\giga \electronvolt},
	  	grid=both,
	 	minor y tick num = {2},
		minor x tick num = {2},
		every axis x label/.style={
    at={(ticklabel* cs:1.05)},
    anchor=south,},
	]
	\addplot[red,domain=.01:5.0,samples=300]{-(4/3) * (0.1185/(x/\hc)) * (\C/sqrt(\Delt)) * exp(-(x/\hc)*sqrt(\t)*cos(deg(\thet/2)))*( sqrt(\Delt)*cos(deg((x/\hc)*sqrt(\t)*sin(\thet/2))) - (\u*\u-2*\s)*sin(deg((x/\hc)*sqrt(\t)*sin(\thet/2))) )+0.24*(x/\hc) -(-(4/3) * (0.1185/(x/\hc))+0.24*(x/\hc))};
	\end{axis}
\end{tikzpicture}
\sourcebytheauthour
\end{figure}
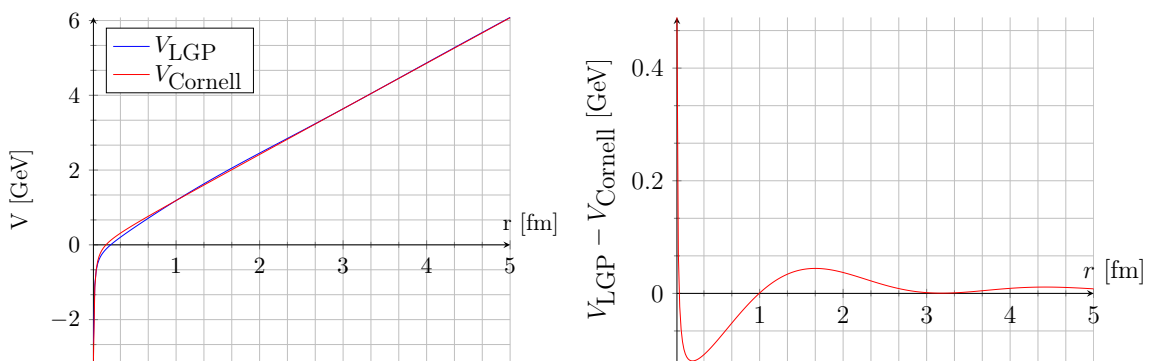

It is possible to see from Fig.\ \ref{fig:Bottomonium_Mass_Spectrum_8_states} and from the data in Table \ref{table:Results_bottom_8_states}
that the two potentials behave very similarly. However, the smaller value of $R$ obtained for the
potential from the lattice propagator indicates that it yields a better result. We also 
remark that, in both cases, our obtained value for the quark mass agrees much better 
with the $m_b$(1S) than with $m_b$($\overline{MS}$)  (see header in Table \ref{table:Results_bottom}).
The small difference between 
the results for our potential and for the Cornell one can be traced to 
the fact that the two potentials are nearly identical,
as show in Fig.\ \ref{fig:Potential}.

\begin{table}[h!]
\centering
\caption{Comparison between the results obtained for the potential extracted 
using the lattice propagator and the usual Cornell potential, using 8 states to fit the parameters.}
\label{table:Results_bottom_8_states}
\resizebox{16cm}{!}{
\begin{tabu}{||c|c||c|c||c|c||}
\hline
\multicolumn{ 2}{||c||}{$m_b(\overline{\text{MS}}) = \SI{4.18(3)}{\giga\electronvolt}$}	& \multicolumn{ 1}{c}{Potential from}	 & $F_0 = \SI{0.24(2)}{\giga \electronvolt \squared}$	& \multicolumn{ 1}{c}{Cornell}	& $F_0 = \SI{0.24(2)}{\giga \electronvolt \squared}$\\

\multicolumn{ 2}{||c||}{$m_b(\text{1S}) = \SI{4.66(3)}{\giga\electronvolt}$}		& \multicolumn{ 1}{c}{Lattice Propagator}& $m_b = \SI{4.57(2)}{\giga \electronvolt}$		& \multicolumn{ 1}{c}{Potential}	& $m_b = \SI{4.56(2)}{\giga \electronvolt}$ \\

\multicolumn{ 2}{||c||}{See \cite{Beringer:1900zz}}					& \multicolumn{ 1}{c}{}			 & $R = 0.0297$						& \multicolumn{ 1}{c}{}			& $R  = 0.0322$ \\
\hline
{\bf Particle}	& \multicolumn{ 1}{c||}{{\bf Experimental}}	& {\bf Calculated Mass} 			& \multicolumn{ 1}{c||}{{\bf Deviation from}}	& {\bf Calculated Mass}					& \multicolumn{ 1}{c||}{{\bf Deviation from}} \\
{\bf State}	& \multicolumn{ 1}{c||}{{\bf Mass(\si{\giga \electronvolt})}}	& {\bf ($\SI{+-1.e-4}{\giga \electronvolt}$)}	& \multicolumn{ 1}{c||}{{\bf Experiment (\si{\giga \electronvolt})}\footnotemark}	& {\bf ($\SI{+-1.5e-4}{\giga \electronvolt}$)}		& \multicolumn{ 1}{c||}{{\bf Experiment (\si{\giga \electronvolt})}\footnotemark[\value{footnote}]} \\
\hline\rowfont{\large}
1S		& 9.429(8) 							&     9.5383					&     0.1109  (1.18\%)	&    9.5367 						&     0.1076 (1.14\%)\\
\hline\rowfont{\large} 
2S 		& 10.011(17)						&    10.0049 					&     0.0062  (0.06\%)	&    9.9931 						&     0.0180 (0.18\%)\\
\hline\rowfont{\large} 
3S\footnotemark& 10.3552(5)				&    10.3579 					&     0.0027	(0.03\%)	&    10.3461 						&     0.0091 (0.09\%)\\
\hline\rowfont{\large} 
4S 		& 10.5794(21) 					&    10.6621 					&     0.0827	(0.78\%)	&    10.6521 						&     0.0727 (0.69\%)\\
\hline\rowfont{\large} 
1P		& 9.89093(2638)					&     9.8445 					&     0.0464	(0.47\%)	&     9.8361 						&     0.0548 (0.55\%)\\
\hline\rowfont{\large} 
2P		& 10.2541(181) 					&    10.2191 					&     0.0350	(0.34\%)	&    10.2085 						&     0.0456 (0.44\%)\\
\hline\rowfont{\large} 
3P 		& 10.530(9)							&    10.5359 					&     0.0059	(0.06\%)	&    10.5263 						&     0.0037 (0.04\%)\\
\hline\rowfont{\large} 
1D 		& 10.1637(14)			 			&    10.0773 					&     0.0864	(0.85\%)	&    10.0643 						&     0.0994 (0.98\%)\\
\hline\rowfont{\large}
5S\footnotemark	& ---							&    10.9373					&     ---								&    10.9291						&     --- \\
\hline\rowfont{\large}
6S\footnotemark[\value{footnote}]	& ---							&    11.1923					&     ---								&    11.1855						&     --- \\
\hline\rowfont{\large}
4P\footnotemark[\value{footnote}]	& ---							&    10.8195					&     ---								&    10.8117						&     --- \\
\hline\rowfont{\large}
5P\footnotemark[\value{footnote}]	& ---							&    11.0809					&     ---								&    11.0747						&     --- \\
\hline\rowfont{\large}
6P\footnotemark[\value{footnote}]	& ---							&    11.3259					&     ---								&    11.3209						&     --- \\
\hline\rowfont{\large}
2D\footnotemark[\value{footnote}]	& ---							&    10.4071					&     ---								&    10.3955						&     --- \\
\hline\rowfont{\large}
3D\footnotemark[\value{footnote}]	& ---							&    10.6995					&     ---								&    10.6903						&     --- \\
\hline\rowfont{\large}
4D\footnotemark[\value{footnote}]	& ---							&    10.9677					&     ---								&    10.9603						&     --- \\
\hline\rowfont{\large}
5D\footnotemark[\value{footnote}]	& ---							&    11.2179					&     ---								&    11.2121						&     --- \\
\hline\rowfont{\large}
6D\footnotemark[\value{footnote}]	& ---							&    11.4543					&     ---								&    11.4499						&     --- \\
\hline
\end{tabu}
}
\sourcebytheauthour
\end{table}
\vspace{-.5cm}

\addtocounter{footnote}{-2}
\addtocounter{Hfootnote}{-2}
\footnotetext[\value{footnote}]{Relative errors are in parentheses.}
\addtocounter{footnote}{1}
\addtocounter{Hfootnote}{1}
\footnotetext[\value{footnote}]{In this case, the uncertainty in the position of the center of the resonance is greater than the resonance width, so we use the former as error estimation.}
\addtocounter{footnote}{1}
\addtocounter{Hfootnote}{1}
\footnotetext[\value{footnote}]{These states are not used in the fit.}

\begin{figure}[h!]
\centering
\caption{The computed spectrum using the two different potentials and 4 states to fit the parameters (data from Table \ref{table:Results_bottom}; see text for details).}
\label{fig:Bottomonium_Spectrum}
 \def\Ei{9.42915}
 \def\Eii{10.01113}
 \def\Eiii{10.3552}
 \def\Eiv{10.5794}
 \def\Ev{9.89093}
 \def\Evi{10.2541}
 \def\Evii{10.530}
 \def\Eviii{10.1637}
 
 \def\Eix{10.6072}
 \def\Ex{10.6522}
 
 \def\Exi{10.876}
 \def\Exii{11.019}

 \def\LEi{9.4719}
 \def\LEii{10.0207}
 \def\LEiii{10.4405}
 \def\LEiv{10.8029}
 \def\LEv{9.8275}
 \def\LEvi{10.2733}
 \def\LEvii{10.6509}
 \def\LEviii{10.1023}
 
 \def\LEix{11.1307}
 \def\LEx{11.4347}
 
 \def\LExi{10.9891}
 \def\LExii{11.3009}
 \def\LExiii{11.5931}
 
 \def\LExiv{10.4959}
 \def\LExv{10.8451}
 \def\LExvi{11.1651}
 \def\LExvii{11.4637}
 \def\LExviii{11.7457}

 \def\CEi{9.4639}
 \def\CEii{10.0133}
 \def\CEiii{10.4411}
 \def\CEiv{10.8127}
 \def\CEv{9.8201}
 \def\CEvi{10.2717}
 \def\CEvii{10.6581}
 \def\CEviii{10.0955}
  
 \def\CEix{11.1493}
 \def\CEx{11.4615}
 
 \def\CExi{11.0053}
 \def\CExii{11.3253}
 \def\CExiii{11.6251}
 
 \def\CExiv{10.4981}
 \def\CExv{10.8567}
 \def\CExvi{11.1853}
 \def\CExvii{11.4921}
 \def\CExviii{11.7815}
 \begin{tikzpicture}[scale=2.5]
 
  \draw (2.4,9.6) -- ++(2.6,0) -- ++(0,-.65) -- ++(-2.6,0) -- ++(0,.65);
	\draw[very thick] (2.5,9.5) -- +(0.5,0) node[right] {\scriptsize Fit input};
	\draw[very thick,lightgray] (2.5,9.35) -- +(0.5,0) node[right] {\scriptsize \textcolor{black}{Experimental data}};
	\draw[thick,dashed,red] (2.5,9.2) -- +(0.5,0) node[right] {\scriptsize \textcolor{black}{Lattice propagator potential results}};
	\draw[thick,dotted,blue] (2.5,9.05) -- +(0.5,0) node[right] {\scriptsize \textcolor{black}{Cornell potential results}};

    \draw[very thick,->,>=latex] (0,8.5)--(0,12.2) node[right] (0,11.5) {Energy};

  \foreach \y in {9,...,12}
  	\draw (1pt,\y) -- (-3pt,\y) node[anchor=east] {\y};
  	
  \draw[very thick] (0.5,\Ei) -- (1.5,\Ei);
  \draw[very thick] (0.5,\Eii) -- (1.5,\Eii);
  \draw[lightgray,very thick] (0.5,\Eiii) -- (1.5,\Eiii);
  \draw[lightgray,very thick] (0.5,\Eiv) -- (1.5,\Eiv);
  
  \draw[very thick] (2.0,\Ev) -- (3.0,\Ev);
  \draw[very thick] (2.0,\Evi) -- (3.0,\Evi);
  \draw[lightgray,very thick] (2.0,\Evii) -- (3.0,\Evii);
  
  \draw[lightgray,very thick] (3.5,\Eviii) -- (4.5,\Eviii);

  \draw[lightgray,very thick] (0.5,\Exi) -- (4.5,\Exi);
	\draw (4.525,\Exi) -- (5,\Exi-0.08) node[right] {\scriptsize $\Upsilon(10860)$};
  \draw[lightgray,very thick] (0.5,\Exii) -- (4.5,\Exii);
	\draw (4.525,\Exii) -- (5,\Exii+0.08) node[right]  {\scriptsize $\Upsilon(11020)$};

  \draw[thick,dashed,red] (0.5,\LEi) -- (1.5,\LEi);
  \draw[thick,dashed,red] (0.5,\LEii) -- (1.5,\LEii);
  \draw[thick,dashed,red] (0.5,\LEiii) -- (1.5,\LEiii);
  \draw[thick,dashed,red] (0.5,\LEiv) -- (1.5,\LEiv);
  
  \draw[thick,dashed,red] (2.0,\LEv) -- (3.0,\LEv);
  \draw[thick,dashed,red] (2.0,\LEvi) -- (3.0,\LEvi);
  \draw[thick,dashed,red] (2.0,\LEvii) -- (3.0,\LEvii);

  \draw[thick,dashed,red] (3.5,\LEviii) -- (4.5,\LEviii);
  
  \draw[thick,dashed,red] (0.5,\LEix) -- (1.5,\LEix);
  \draw[thick,dashed,red] (0.5,\LEx) -- (1.5,\LEx);
  
  \draw[thick,dashed,red] (2.0,\LExi) -- (3.0,\LExi);
  \draw[thick,dashed,red] (2.0,\LExii) -- (3.0,\LExii);
  \draw[thick,dashed,red] (2.0,\LExiii) -- (3.0,\LExiii);
  
  \draw[thick,dashed,red] (3.5,\LExiv) -- (4.5,\LExiv);
  \draw[thick,dashed,red] (3.5,\LExv) -- (4.5,\LExv);
  \draw[thick,dashed,red] (3.5,\LExvi) -- (4.5,\LExvi);
  \draw[thick,dashed,red] (3.5,\LExvii) -- (4.5,\LExvii);
  \draw[thick,dashed,red] (3.5,\LExviii) -- (4.5,\LExviii);
  
  \draw[thick,dotted,blue] (0.5,\CEi) -- (1.5,\CEi);
  \draw[thick,dotted,blue] (0.5,\CEii) -- (1.5,\CEii);
  \draw[thick,dotted,blue] (0.5,\CEiii) -- (1.5,\CEiii);
  \draw[thick,dotted,blue] (0.5,\CEiv) -- (1.5,\CEiv);
  
  \draw[thick,dotted,blue] (2.0,\CEv) -- (3.0,\CEv);
  \draw[thick,dotted,blue] (2.0,\CEvi) -- (3.0,\CEvi);
  \draw[thick,dotted,blue] (2.0,\CEvii) -- (3.0,\CEvii);

  \draw[thick,dotted,blue] (3.5,\CEviii) -- (4.5,\CEviii);
  
  \draw[thick,dotted,blue] (0.5,\CEix) -- (1.5,\CEix);
  \draw[thick,dotted,blue] (0.5,\CEx) -- (1.5,\CEx);
  
  \draw[thick,dotted,blue] (2.0,\CExi) -- (3.0,\CExi);
  \draw[thick,dotted,blue] (2.0,\CExii) -- (3.0,\CExii);
  \draw[thick,dotted,blue] (2.0,\CExiii) -- (3.0,\CExiii);
  
  \draw[thick,dotted,blue] (3.5,\CExiv) -- (4.5,\CExiv);
  \draw[thick,dotted,blue] (3.5,\CExv) -- (4.5,\CExv);
  \draw[thick,dotted,blue] (3.5,\CExvi) -- (4.5,\CExvi);
  \draw[thick,dotted,blue] (3.5,\CExvii) -- (4.5,\CExvii);
  \draw[thick,dotted,blue] (3.5,\CExviii) -- (4.5,\CExviii);

  \node at (1.0,8.75) {S states};
  \node at (2.5,8.75) {P states};
  \node at (4.0,8.75) {D states};
 \end{tikzpicture}
\sourcebytheauthour
\end{figure}
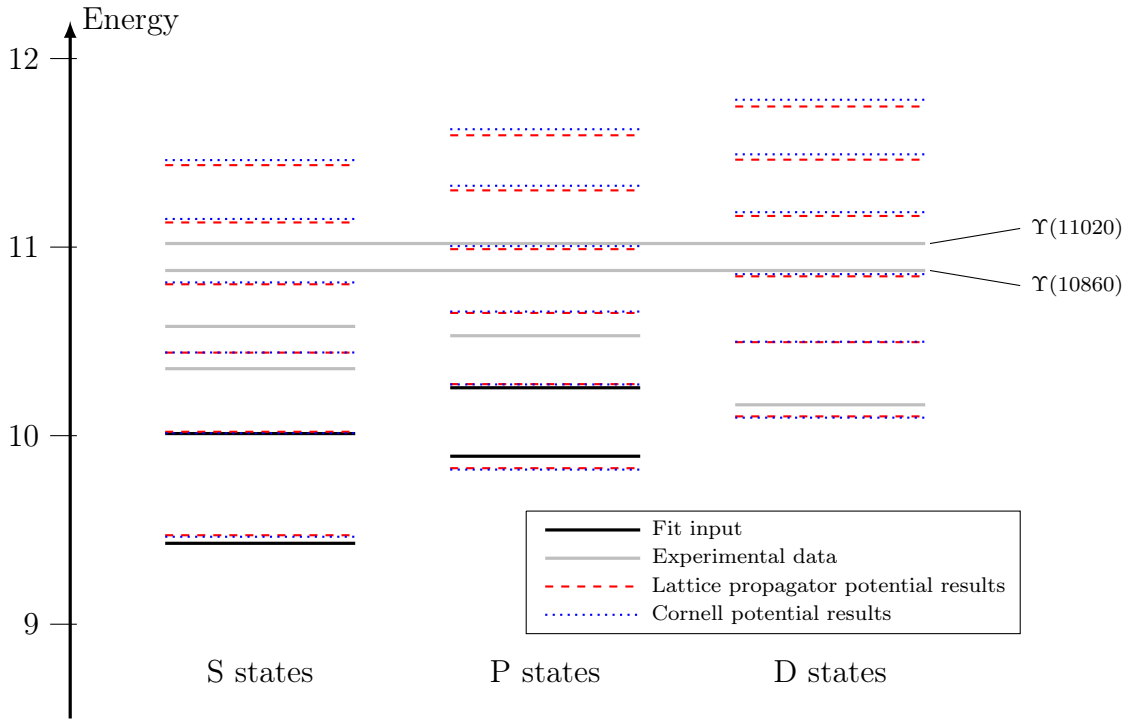

We proceed to compute other excited states using the found parameters values and comparing them with the experimental values of $\Upsilon(10860)$ and $\Upsilon(11020)$ present
in Table \ref{tb:bottomonium_experimental} (gray lines in Fig.\ \ref{fig:Bottomonium_Spectrum}).
We search for $l = 0$, $1$, $2$, $3$ and for six states with each $l$. For $l = 3$ we do not find any eigenstate.
The other states found can be seen in red (dashed lines for the potential from the lattice propagator and 
dotted lines for the Cornell potential) in Fig.\ \ref{fig:Bottomonium_Spectrum}.

We note that the state $4D$ is near the average of $\Upsilon(10860)$ and $\Upsilon(11020)$. However, since there are
no observed particles that could be identified with the lower energy levels, it seems unlikely
that these states would correspond to a $D$-wave state. The $P$ states are also excluded, since they are incompatible with the
quantum numbers of $\Upsilon$. The state $5S$ seems to be in an average position between both particles. 
However, with this assumption, they would have the same angular and spin quantum numbers, leading us to believe that they 
have different radial quantum numbers and therefore we cannot average over them. We believe therefore that our model is 
able to describe the $\Upsilon(10860)$ as the $5S$ state but is not able to describe the $\Upsilon(11020)$.

So far we used eight states to perform the fit, leaving almost no space to predictions. To allow the model to
have more predictions, we reduce the number of states used to perfom the fit to four. We will be using the states
$1S$, $2S$, $1P$ and $2P$. These states were chosen because they are less prone to relativistic corrections, as
was explained above. Also, when we perform the same study for the charmonium in the next section, we will be using
the charmonium states with the same quantum numbers in the fit, making error comparisons between bottomonium and 
charmonium easier.

The method and parameters used to obtain the wavefunctions and eigenenergies are the same, and the results can be found
in Table \ref{table:Results_bottom}. A graphical representation of this data may be found in Fig.\
\ref{fig:Bottomonium_Spectrum}. We use the same color legend as in Fig.\ \ref{fig:Bottomonium_Mass_Spectrum_8_states}.

\begin{table}[h]
\centering
\caption{Comparison between the results obtained for the potential extracted 
using the lattice propagator and the usual Cornell potential, using 4 states to fit the parameters.}
\label{table:Results_bottom}
\resizebox{16cm}{!}{
\begin{tabu}{||c|c||c|c||c|c||}
\hline
\multicolumn{ 2}{||c||}{$m_b(\overline{\text{MS}}) = \SI{4.18(3)}{\giga\electronvolt}$}	& \multicolumn{ 1}{c}{Potential from}	 & $F_0 = \SI{0.31(1)}{\giga \electronvolt \squared}$	& \multicolumn{ 1}{c}{Cornell}	& $F_0 = \SI{0.32(1)}{\giga \electronvolt \squared}$\\

\multicolumn{ 2}{||c||}{$m_b(\text{1S}) = \SI{4.66(3)}{\giga\electronvolt}$}		& \multicolumn{ 1}{c}{Lattice Propagator}& $m_b = \SI{4.48(1)}{\giga \electronvolt}$		& \multicolumn{ 1}{c}{Potential}	& $m_b = \SI{4.48(1)}{\giga \electronvolt}$ \\

\multicolumn{ 2}{||c||}{See \cite{Beringer:1900zz}}					& \multicolumn{ 1}{c}{}			 & $R = \SI{0.00631}{\giga \electronvolt \squared}$						& \multicolumn{ 1}{c}{}			& $R  = \SI{0.00653}{\giga \electronvolt \squared}$ \\
\hline
{\bf Particle}	& \multicolumn{ 1}{c||}{{\bf Experimental}}	& {\bf Calculated Mass} 			& \multicolumn{ 1}{c||}{{\bf Deviation from}}	& {\bf Calculated Mass}					& \multicolumn{ 1}{c||}{{\bf Deviation from}} \\
{\bf State}	& \multicolumn{ 1}{c||}{{\bf Mass(\si{\giga \electronvolt})}}	& {\bf ($\SI{+-1.e-4}{\giga \electronvolt}$)}	& \multicolumn{ 1}{c||}{{\bf Experiment (\si{\giga \electronvolt})}\footnotemark}	& {\bf ($\SI{+-1.5e-4}{\giga \electronvolt}$)}		& \multicolumn{ 1}{c||}{{\bf Experiment (\si{\giga \electronvolt})}\footnotemark[\value{footnote}]} \\
\hline\rowfont{\large}
1S\footnotemark		& 9.429(8) 							&     9.4719					&     0.0429  (0.45\%)	&    9.4639 						&     0.0349 (0.37\%)\\
\hline\rowfont{\large} 
2S\footnotemark[\value{footnote}] 		& 10.011(17)						&    10.0207 					&     0.0097  (0.10\%)	&    10.0133 						&     0.0023 (0.02\%)\\
\hline\rowfont{\large} 
1P\footnotemark[\value{footnote}]		& 9.89093(2638)					&     9.8275 					&     0.0634	(0.64\%)	&     9.8201 						&     0.0708 (0.72\%)\\
\hline\rowfont{\large} 
2P\footnotemark[\value{footnote}]		& 10.2541(181) 					&    10.2733 					&     0.0192	(0.19\%)	&    10.2717 						&     0.0176 (0.17\%)\\
\hline\rowfont{\large} 
3S\footnotemark		& 10.3552(5)						&    10.4405 					&     0.0853	(0.82\%)	&    10.4411 						&     0.0859 (0.83\%)\\
\hline\rowfont{\large} 
4S 		& 10.5794(21) 					&    10.8029 					&     0.2235	(2.11\%)	&    10.8127 						&     0.2333 (2.21\%)\\
\hline\rowfont{\large}
5S		& ---										&    11.1307					&     ---								&    11.1493						&     --- \\
\hline\rowfont{\large}
6S		& ---										&    11.4347					&     ---								&    11.4615						&     --- \\
\hline\rowfont{\large} 
3P 		& 10.530(9)							&    10.6509 					&     0.1209	(1.15\%)	&    10.6581 						&     0.1281 (1.22\%)\\
\hline\rowfont{\large}
4P		& ---										&    10.9891					&     ---								&    11.0053						&     --- \\
\hline\rowfont{\large}
5P		& ---										&    11.3009					&     ---								&    11.3253						&     --- \\
\hline\rowfont{\large}
6P		& ---										&    11.5931					&     ---								&    11.6251						&     --- \\
\hline\rowfont{\large} 
1D 		& 10.1637(14)			 			&    10.1023 					&     0.0614	(0.60\%)	&    10.0955 						&     0.0682 (0.67\%)\\
\hline\rowfont{\large}
2D		& ---										&    10.4959					&     ---								&    10.4981						&     --- \\
\hline\rowfont{\large}
3D		& ---										&    10.8451					&     ---								&    10.8567						&     --- \\
\hline\rowfont{\large}
4D		& ---										&    11.1651					&     ---								&    11.1853						&     --- \\
\hline\rowfont{\large}
5D		& ---										&    11.4637					&     ---								&    11.4921						&     --- \\
\hline\rowfont{\large}
6D		& ---										&    11.7457					&     ---								&    11.7815						&     --- \\
\hline
\end{tabu}
}
\sourcebytheauthour
\end{table}
\vspace{-.5cm}

\addtocounter{footnote}{-2}
\addtocounter{Hfootnote}{-2}
\footnotetext[\value{footnote}]{Relative errors are in parentheses.}
\addtocounter{footnote}{1}
\addtocounter{Hfootnote}{1}
\footnotetext[\value{footnote}]{These states are used in the fit.}
\addtocounter{footnote}{1}
\addtocounter{Hfootnote}{1}
\footnotetext[\value{footnote}]{In this case, the uncertainty in the position of the center of the resonance is greater than the resonance width, so we use the former as error estimation.}

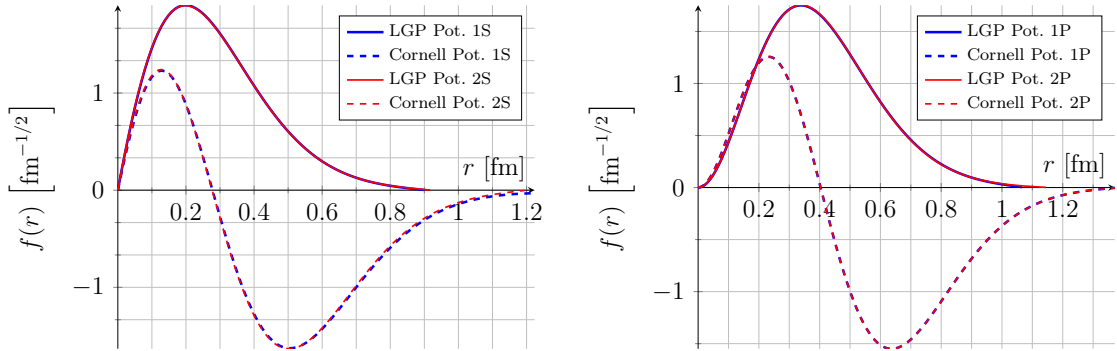
\begin{figure}[h!]
\centering
\caption{Plot of wave functions for the bottomonium obtained through our computations, using the potential from the lattice propagator (in blue) and the Cornell potential (in red).}
\label{fig:wf_plots_Bottomonium}
  \pgfplotstableread{./img/Bottomonium_data_latt_prop/Normalized_0_wave_function_0.dat}{\zwfz}
  \pgfplotstableread{./img/Bottomonium_data_latt_prop/Normalized_0_wave_function_1.dat}{\zwfo}
  \pgfplotstableread{./img/Bottomonium_data_latt_prop/Normalized_1_wave_function_0.dat}{\owfz}
  \pgfplotstableread{./img/Bottomonium_data_latt_prop/Normalized_1_wave_function_1.dat}{\owfo}
  
  \pgfplotstableread{./img/Bottomonium_data_cornell_pot/Normalized_0_wave_function_0.dat}{\cpzwfz}
  \pgfplotstableread{./img/Bottomonium_data_cornell_pot/Normalized_0_wave_function_1.dat}{\cpzwfo}
  \pgfplotstableread{./img/Bottomonium_data_cornell_pot/Normalized_1_wave_function_0.dat}{\cpowfz}
  \pgfplotstableread{./img/Bottomonium_data_cornell_pot/Normalized_1_wave_function_1.dat}{\cpowfo}
  \centering
\subfigure
{
  \centering
  \label{fig:wf_1s_2s_bbar}
\begin{tikzpicture}[scale=0.8]
  \begin{axis}
     [xlabel=$r$,
      ylabel=$f(r)$,
      axis lines = left,
      axis x line=center,
      grid=both,
      minor y tick num = {2},
      minor x tick num = {1},
			x unit=\si{\femto \metre},
			y unit=\si{\femto \metre}^{-1/2},
			legend pos = {north east},
			legend cell align = left,
			legend style = {align=justify,font=\scriptsize},
			legend entries = {LGP Pot.\ 1S,Cornell Pot.\ 1S,LGP Pot.\ 2S,Cornell Pot.\ 2S},
		]
    \addplot [blue,very thick] table {\zwfz};
    \addplot [blue,very thick,style=dashed] table {\zwfo};
    \addplot [red,thick] table {\cpzwfz};
    \addplot [red,thick,style=dashed] table {\cpzwfo};
  \end{axis}
\end{tikzpicture}}\;
 \subfigure
{
   \centering
   \label{fig:wf_1p_2p_bbar}
 \begin{tikzpicture}[scale=0.8]
  \begin{axis}
     [xlabel=$r$,
      ylabel=$f(r)$,
      axis lines = left,
      axis x line=center,
      grid=both,
      minor y tick num = {1},
      minor x tick num = {1},
			x unit=\si{\femto \metre},
			y unit=\si{\femto \metre}^{-1/2},
			legend pos = {north east},
			legend cell align = left,
			legend style = {align=justify,font=\scriptsize},
			legend entries = {LGP Pot.\ 1P,Cornell Pot.\ 1P,LGP Pot.\ 2P,Cornell Pot.\ 2P},]
    \addplot [blue,very thick] table {\owfz};
    \addplot [blue,very thick,style=dashed] table {\cpowfo};
    \addplot [red,thick] table {\cpowfz};
    \addplot [red,thick,style=dashed] table {\cpowfo};
  \end{axis}
 \end{tikzpicture}}
\sourcebytheauthour
\end{figure}

For comparison we plot the wave function of some states using the lattice gluon propagator potential 
in Fig.\ \ref{fig:wf_plots_Bottomonium}. In the left panel we show S states and in the right panel P states.
We use solid lines for states 1S and 1P and dashed lines for 2S and 2P.

We obtain once again a similar behavior for both potentials (see Fig.\ \ref{fig:wf_plots_Bottomonium} 
and Table \ref{table:Results_bottom}), with the potential from the gluon propagator having a slightly 
better result, as an be noticed by a smaller value of $R$. The quark mass agrees less with the value
of $m_b$(1S) from Ref.\ \cite{Beringer:1900zz}. However it still nearer to the value of $m_b$(1S) than
from $m_b$($\overline{MS}$).

As was done previously, we use the values of the obtained parameters for computing the mass of the states 
not used as input in the fit, present in Table \ref{tb:bottomonium_experimental} (gray lines in Fig.\ \ref{fig:Bottomonium_Spectrum}).
We keep the search for $l = 0$, $1$, $2$, $3$ and for six states with each $l$. Again, for $l = 3$, we do not find any eigenstate.

We note that the prediction for the states $3S$ is somewhat more distant from the experimental results than the 
results that were used for the fit. Despite the increased error, it is still less than $1 \%$ and we consider that the models are able to
describe this state. We miss the right value for the $3P$ state by a little more than $1 \%$, making it harder to say that we
have a reasonable description of this state. The same does not happen to the state $4S$, which disagrees with the experiment
by more than $2 \%$ and is nearer to the $\Upsilon(10860)$ than to $\Upsilon(4S)$. The value calculated for the state $3D$ using the lattice potential
agrees with the value of mass of the $\Upsilon(10860)$, being inside the error we estimated for it. But the fact that there is 
no indication of a $2D$ state remains a question. In Figure \ref{fig:Bottomonium_Spectrum},
it seems that there is agreement between the $\Upsilon(11020)$ and the state ($4P$), but the quantum numbers for
the $\Upsilon$ forbid it to have an odd angular momentum and therefore the fact that they ``agree'' must be a coincidence.
Considering the amount of error in the description of the state $4S$, it seems far more probable that the $\Upsilon(11020)$
is the state $5S$ that our models were not able to describe accurately.

Our final conclusion is that our model has trouble describing the states that are not used to perform the fit. This is not
a surprise, considering that we neglected large relativistic corrections and used perturbation theory, making it necessary to
add the linear potential to obtain a confining potential. Using these approximations could imply that even if we 
use all states to perform the fit, we would not have a good description. We showed that this is not true and we can describe 
surprisingly well the known spectrum when using all states for the fit.

\section{Charmonium}
There is a rich literature about the application of the potential-model approach in the case of charmonium
\cite{PhysRevD.17.3090, Chaichian:1978th, Eichten:1979ms, PhysRevD.12.1999, Eichten:2002qv}.
This motivates us to apply the method to this case as well.

The charm mass according Ref.\ \cite{Beringer:1900zz}
is of \SI{1.275}{\giga \electronvolt}. We repeat the procedure of estimating the 
value of $\gamma$ as a check of the validity of the nonrelativistic approximation 
for this system. For the ground state, which is where our approximations hold better, we have
\begin{equation}
\gamma \;=\; \frac{E}{2 m_c} \;\approx\; \frac{3.09}{2.55} \;\approx\; 1.21\,,
\end{equation}
which is far larger than in the bottomonium case. If this were a classical particle, it would 
be traveling at approximately 32\% of the speed of light. In the worst-case scenario, the 
state of highest energy, we have
\begin{equation}
\gamma \;=\; \frac{E}{2 m_c} \;\approx\; \frac{4.42}{2.55} \;\approx\; 1.73\,,
\end{equation}
which is again far worse than in the bottomonium case and is the analogue of a classical particle 
traveling at 67\% of the speed of light. In this way, it is expected that our approach will be 
less suitable for the charmonium.

The nomenclature of charmonium particles follows a similar scheme to the one for bottomonium, and
is shown in Table \ref{tb:naming_scheme_charmonium}.
\begin{table}[h]%
\centering
\caption{Naming scheme for charmonium states.}
\label{tb:naming_scheme_charmonium}
\begin{tabu}{||c|c|c||}
\hline
Name		& S & L 	\\
\hline
$\eta_c$	& 0 & Even	\\
$h_c$		& 0 & Odd	\\
$\psi$		& 1 & Even	\\
$\chi_c$	& 1 & Odd	\\
\hline
\end{tabu}\vspace*{10pt}\\Source: BERINGER et al.\ \cite{Beringer:1900zz}.
\end{table}
An exception is made for the state $\psi(1S)$, which is called $J/\psi(1S)$ for historical reasons. The observed spectrum for
the charmonium can be found in Table \ref{tb:charmonium_experimental}. The procedure for generating input mass values is the
same as described in the previous section for the bottomonium.
\begin{table}[h!]%
\centering
\caption{Experimental spectrum for the charmonium and its its preparation as input for our calculations (see text).}
\label{tb:charmonium_experimental}
\resizebox{16cm}{!}{
\begin{tabu}{||c|c|c|c|c|c||}
\hline
Particle Name 		& Mass (\si{\giga \electronvolt}) & Decay width (\si{\giga \electronvolt})	& $l$ 				& Mass Input (\si{\giga \electronvolt}) & Mass Error (\si{\giga \electronvolt})\\ \hline
$\eta_c(1S)$		& 2.9837			    & \num{3.2E-2}				& \multirow{2}{*}{0}		& \multirow{2}{*}{3.0403 (used to fit)}		& \multirow{2}{*}{0.0026} \\
$J/\psi(1S)$		& 3.096916			  & \num{5.55E-6}				& & & \\ \hline
$\eta_c(2S)$		& 3.6394		  	  & \num{1.13E-2}				& \multirow{2}{*}{0}		& \multirow{2}{*}{3.6628 (used to fit)}		& \multirow{2}{*}{0.0080} \\
$\psi(2S)$		  & 3.686109			  & \num{3.03E-4}				& & & \\ \hline
$\chi_{c0}(1P)$ 	& 3.41475			  & \num{1.03E-2}				&	\multirow{4}{*}{1}	& \multirow{4}{*}{3.5017 (used to fit)}		& \multirow{4}{*}{0.0053} \\
$\chi_{c1}(1P)$		& 3.51066			  & \num{8.6E-4}				& & & \\
$h_c(1P)$		& 3.52538			  & \num{7E-4}					& & & \\
$\chi_{c2}(1P)$		& 3.5562			  & \num{1.97E-3}				& & & \\ \hline
$\chi_{c0}(2P)$ 	& 3.9184			  & \num{2.05E-2}				& \multirow{2}{*}{1}		& \multirow{2}{*}{3.9228 (used to fit)}		& \multirow{2}{*}{0.0045} \\
$\chi_{c2}(2P)$		& 3.9272			  & \num{2.4E-2}				& & & \\ \hline
$\psi(3770)$		& 3.77315			  & \num{2.72E-2}				& Even				& 3.77315				& 0.0272\\ \hline
$X(3872)$		& 3.87168			  & $< \num{1.2E-3}$				& Odd				& 3.87168				& 0.0012\\ \hline
$X(3940)$		& 3.942				  & \num{3.7E-2}				& Unknown			& 3.942					& 0.037\\ \hline
$\psi(4040)$		& 4.039				  & \num{8.0E-2}				& Even				& 4.039					& 0.08\\ \hline
$\psi(4160)$		& 4.153				  & \num{1.03E-1}				& Even				& 4.153					& 0.103\\ \hline
$X(4160)$		& 4.156				  & \num{1.39E-1}				& Unknown			& 4.156					& 0.139\\ \hline
$X(4260)$		& 4.250				  & \num{1.08E-1}				& Even				& 4.250					& 0.108\\ \hline
$X(4350)$		& 4.3506			  & \num{1.3E-2}				& Unknown			& 4.3506				& 0.013\\ \hline
$X(4360)$		& 4.361				  & \num{7.4E-2}				& Even				& 4.361					& 0.074\\ \hline
$\psi(4415)$		& 4.421				  & \num{6.2E-2}				& Even				& 4.421					& 0.062\\ \hline
$X(4660)$		& 4.664				  & \num{4.8E-2}				& Even				& 4.664					& 0.048\\ \hline
\end{tabu}}
\vspace*{10pt}\\Source: Adapted from BERINGER et al.\ \cite{Beringer:1900zz}.
\end{table}

As can be seen in Table \ref{tb:charmonium_experimental}, charmonium has more observed states, although most
of them do not have $l$ determined yet. For the fits, we only use states with $l$ already determined, so we will be fitting only
four states (namely $1S$, $2S$, $1P$ and $2P$).

The results from our computations are shown in Table \ref{tb:Results_Charmonium}. For the computations, we
kept the step used in the wave function integration at $dr = \SI{5e-3}{\per \giga \electronvolt}$. We searched the
eigenenergies in the range from \SI{3}{\giga \electronvolt} to \SI{5}{\giga \electronvolt} in steps of \SI{1E-4}{\giga \electronvolt}.
Ref.\ \cite{Beringer:1900zz} gives us just one value for the charm quark mass [$m_c = \SI{1.275(25)}{\giga \electronvolt}$] 
and we use this value as a fixed parameter in our computations. We search for the string force $F_0$ in the range from 
\SI{0.0}{\giga \electronvolt \squared} to \SI{0.2}{\giga \electronvolt \squared} in steps of \SI{0.01}{\giga \electronvolt \squared}.
\begin{table}[!h]
\centering
\caption{Comparison between the results for the charmonium obtained for the potential extracted 
using the lattice propagator and the usual Cornell potential.}
\label{tb:Results_Charmonium}
\resizebox{16cm}{!}{
\begin{tabu}{||c|c||c|c||c|c||}
\hline
\multicolumn{ 2}{||c||}{ $m_c = \SI{1.275(25)}{\giga\electronvolt}$}	& \multicolumn{ 1}{c}{Potential from}	 & $F_0 = \SI{0.17(2)}{\giga \electronvolt \squared}$	& \multicolumn{ 1}{c}{Cornell}		&  $F_0 = \SI{0.17(2)}{\giga \electronvolt \squared}$\\

\multicolumn{ 2}{||c||}{See Ref.\ \cite{Beringer:1900zz}}		& \multicolumn{ 1}{c}{Lattice Propagator}& $R = 0.0145$						& \multicolumn{ 1}{c}{Potential}		& $R  = 0.0115$\\

\hline
{\bf Particle}	& \multicolumn{ 1}{c||}{{\bf Experimental}} 		& {\bf Calculated Mass} 			& \multicolumn{ 1}{c||}{{\bf Deviation from}}	& {\bf Calculated Mass} 				& \multicolumn{ 1}{c||}{{\bf Deviation from}} \\

{\bf State}	& \multicolumn{ 1}{c||}{{\bf Mass (\si{\giga \electronvolt})}}		& {\bf ($\SI{+-1.e-4}{\giga \electronvolt}$)}	& \multicolumn{ 1}{c||}{{\bf Experiment (\si{\giga \electronvolt})}\footnotemark}	& {\bf ($\SI{+-1.e-4}{\giga \electronvolt}$)}	& \multicolumn{ 1}{c||}{{\bf  Experiment (\si{\giga \electronvolt})}\footnotemark[\value{footnote}]} \\
\hline\rowfont{\large}
1S\footnotemark		& 3.0403(26)						&    3.1583 					&     0.1180 (3.88\%)				&     3.1295 						&     0.0892 (2.93\%)\\
\hline\rowfont{\large}
2S\footnotemark[\value{footnote}] 		& 3.6628(80)						&    3.6779 					&     0.0151 (0.41\%)	 			&     3.6507 						&     0.0121 (0.33\%)\\
\hline\rowfont{\large}
1P\footnotemark[\value{footnote}]		& 3.5017(53) 						&    3.4841 					&     0.0176 (0.50\%)				&     3.4519						&     0.0498 (1.42\%)\\
\hline\rowfont{\large}
2P\footnotemark[\value{footnote}]		& 3.9228(45) 						&    3.9191 					&     0.0037 (0.09\%)				&     3.8929						&     0.0299 (0.76\%)\\
\hline\rowfont{\large}
3S	& ---				&    4.0905					&     ---					&     4.0671						&     --- \\
\hline\rowfont{\large}
4S	& ---				&    4.4525					&     ---					&     4.4319						&     --- \\
\hline\rowfont{\large}
5S	& ---				&    4.7825					&     ---					&     4.7637						&     --- \\
\hline\rowfont{\large}
3P	& ---				&    4.2955					&     ---					&     4.2731						&     --- \\
\hline\rowfont{\large}
4P	& ---				&    4.6357					&     ---					&     4.6159						&     --- \\
\hline\rowfont{\large}
5P	& ---				&    4.9507					&     ---					&     4.9327						&     --- \\
\hline\rowfont{\large}
1D	& ---				&    3.7459					&     ---					&     3.7151						&     --- \\
\hline\rowfont{\large}
2D	& ---				&    4.1369					&     ---					&     4.1117						&     --- \\
\hline\rowfont{\large}
3D	& ---				&    4.4879					&     ---					&     4.4659						&     --- \\
\hline\rowfont{\large}
4D	& ---				&    4.8111					&     ---					&     4.7915						&     --- \\
\hline
\end{tabu}
}\sourcebytheauthour
\end{table}

\begin{figure}[!h]
\centering
\caption{The experimental mass spectrum of the charmonium, with the computed spectrum using the two different potentials
(using data from Table \ref{tb:Results_Charmonium}). See explanation in the text}
\label{fig:Charmonium_Mass_Spectrum}
 \def\Ei{3.0403}
 \def\Eii{3.6628}
 \def\Eiii{3.5017}
 \def\Eiv{3.9228}
 \def\Ev{3.77315}
 \def\Evi{3.87168}
 \def\Evii{3.942}
 \def\Eviii{4.039}
 \def\Eix{4.051}
 \def\Ex{4.153}
 \def\Exi{4.156}
 \def\Exii{4.248}
 \def\Exiii{4.250}
 \def\Exiv{4.3506}
 \def\Exv{4.361}
 \def\Exvi{4.421}
 \def\Exvii{4.443}
 \def\Exviii{4.664}
 
 \def\LEi{3.1583}
 \def\LEii{3.6779}
 \def\LEiii{3.4841}
 \def\LEiv{3.9191}
 \def\LEv{4.0905}
 \def\LEvi{4.4525}
 \def\LEvii{4.7825}
 \def\LEix{4.2955}
 \def\LEx{4.6357}
 \def\LExi{4.9507}
 \def\LExiii{3.7459}
 \def\LExiv{4.1369}
 \def\LExv{4.4879}
 \def\LExvi{4.8111}
 
 \def\CEi{3.1295}
 \def\CEii{3.6507}
 \def\CEiii{3.4519}
 \def\CEiv{3.8929}
 \def\CEv{4.0671}
 \def\CEvi{4.4319}
 \def\CEvii{4.7637}
 \def\CEix{4.2731}
 \def\CEx{4.6159}
 \def\CExi{4.9327}
 \def\CExiii{3.7151}
 \def\CExiv{4.1117}
 \def\CExv{4.4659}
 \def\CExvi{4.7915}
 \begin{tikzpicture}[scale=2.5]
 
  \draw (3.2,5.7) -- ++(2.6,0) -- ++(0,-.65) -- ++(-2.6,0) -- ++(0,.65);
	\draw[very thick] (3.25,5.6) -- +(0.5,0) node[right] {\scriptsize Fit input};
	\draw[very thick,lightgray] (3.25,5.45) -- +(0.5,0) node[right] {\scriptsize \textcolor{black}{Experimental data}};
	\draw[thick,dashed,red] (3.25,5.3) -- +(0.5,0) node[right] {\scriptsize \textcolor{black}{Lattice propagator potential results}};
	\draw[thick,dotted,blue] (3.25,5.15) -- +(0.5,0) node[right] {\scriptsize \textcolor{black}{Cornell potential results}};

	\draw[very thick,->,>=latex] (0,2.5)--(0,5.5) node[right] (0,11.5) {Energy};

  \foreach \y in {3,3.5,...,5}
  	\draw (1pt,\y) -- (-3pt,\y) node[anchor=east] {\y};
  	
  \draw[very thick] (0.5,\Ei) -- (1.5,\Ei);
  \draw[very thick] (0.5,\Eii) -- (1.5,\Eii);
  
  \draw[very thick] (2.0,\Eiii) -- (3.0,\Eiii);
  \draw[very thick] (2.0,\Eiv) -- (3.0,\Eiv);
  
  \draw[very thick,lightgray] (0.5,\Ev) -- (4.5,\Ev);
	\draw (4.525,\Ev) -- (5,\Ev-0.05) node[right] {\scriptsize $\psi(3770)$};
  \draw[very thick,lightgray] (0.5,\Eviii) -- (4.5,\Eviii);
	\draw (4.525,\Eviii) --  (5,\Eviii-0.05) node[right] {\scriptsize $\psi(4040)$};

  \draw[very thick,lightgray] (0.5,\Ex) -- (4.5,\Ex);
	\draw (4.525,\Ex) -- (5,\Ex+0.05) node[right] {\scriptsize $\psi(4160)$,$\,X(4160)$};
  \draw[very thick,lightgray] (0.5,\Exi) -- (4.5,\Exi);

  \draw[very thick,lightgray] (0.5,\Exvi) -- (4.5,\Exvi);
	\draw (4.525,\Exvi) -- (5,\Exvi+0.1) node[right] {\scriptsize $\psi(4415)$};

  \draw[very thick,lightgray] (0.5,\Exviii) -- (4.5,\Exviii);
	\draw (4.525,\Exviii) -- (5,\Exviii+0.05)  node[right] {\scriptsize $X(4660)$};

  \draw[thick,dashed,red] (0.5,\LEi) -- (1.5,\LEi);
  \draw[thick,dashed,red] (0.5,\LEii) -- (1.5,\LEii);
  \draw[thick,dashed,red] (2.0,\LEiii) -- (3.0,\LEiii);
  \draw[thick,dashed,red] (2.0,\LEiv) -- (3.0,\LEiv);

  \draw[thick,dashed,red] (0.5,\LEv) -- (1.5,\LEv);
  \draw[thick,dashed,red] (0.5,\LEvi) -- (1.5,\LEvi);
  \draw[thick,dashed,red] (0.5,\LEvii) -- (1.5,\LEvii);
  
  \draw[thick,dashed,red] (2.0,\LEix) -- (3.0,\LEix);
  \draw[thick,dashed,red] (2.0,\LEx) -- (3.0,\LEx);
  \draw[thick,dashed,red] (2.0,\LExi) -- (3.0,\LExi);
  
  \draw[thick,dashed,red] (3.5,\LExiii) -- (4.5,\LExiii);
  \draw[thick,dashed,red] (3.5,\LExiv) -- (4.5,\LExiv);
  \draw[thick,dashed,red] (3.5,\LExv) -- (4.5,\LExv);
  \draw[thick,dashed,red] (3.5,\LExvi) -- (4.5,\LExvi);

  \draw[thick,dotted,blue] (0.5,\CEi) -- (1.5,\CEi);
  \draw[thick,dotted,blue] (0.5,\CEii) -- (1.5,\CEii);
  \draw[thick,dotted,blue] (2.0,\CEiii) -- (3.0,\CEiii);
  \draw[thick,dotted,blue] (2.0,\CEiv) -- (3.0,\CEiv);

  \draw[thick,dotted,blue] (0.5,\CEv) -- (1.5,\CEv);
  \draw[thick,dotted,blue] (0.5,\CEvi) -- (1.5,\CEvi);
  \draw[thick,dotted,blue] (0.5,\CEvii) -- (1.5,\CEvii);
  
  \draw[thick,dotted,blue] (2.0,\CEix) -- (3.0,\CEix);
  \draw[thick,dotted,blue] (2.0,\CEx) -- (3.0,\CEx);
  \draw[thick,dotted,blue] (2.0,\CExi) -- (3.0,\CExi);
  
  \draw[thick,dotted,blue] (3.5,\CExiii) -- (4.5,\CExiii);
  \draw[thick,dotted,blue] (3.5,\CExiv) -- (4.5,\CExiv);
  \draw[thick,dotted,blue] (3.5,\CExv) -- (4.5,\CExv);
  \draw[thick,dotted,blue] (3.5,\CExvi) -- (4.5,\CExvi);

  \node at (1.0,2.75) {S states};
  \node at (2.5,2.75) {P states};
  \node at (4.0,2.75) {D states};
 \end{tikzpicture}
\sourcebytheauthour
\end{figure}
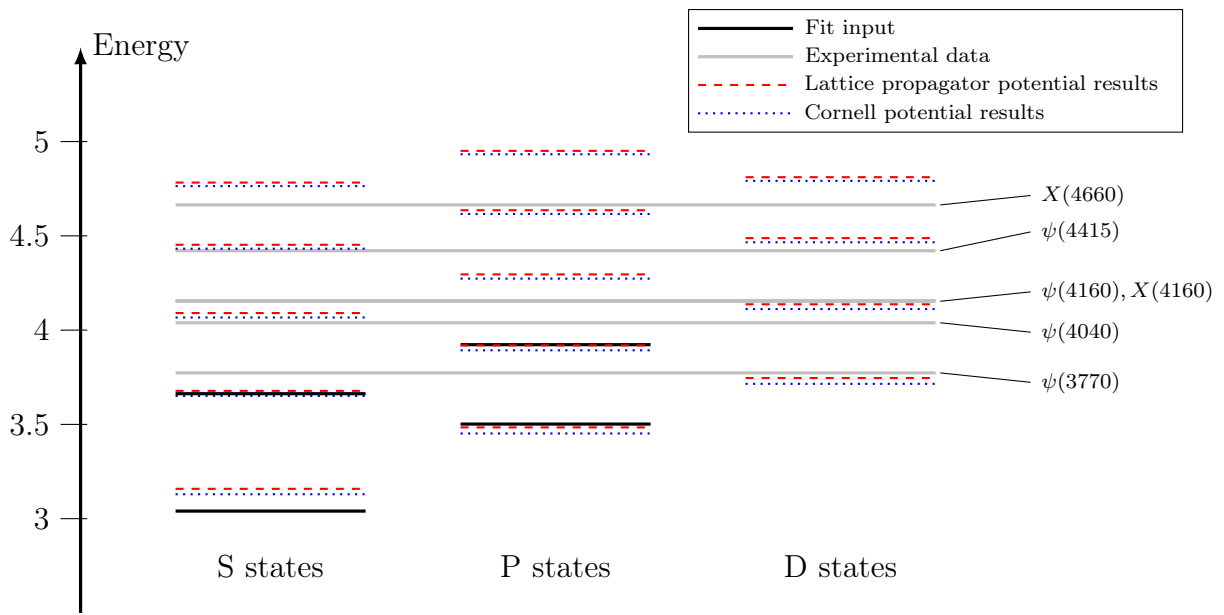

To compare the error of the three fits we performed, notice that $R$ is proportional to $\chi^2$. To obtain the $\chi^2$, we must divide by the square of our numerical error
(which is the same for all cases) and by the number of states used in the fit subtracted from the number of degree of freedoms, i.e.\ the number of free parameters
\begin{equation}
\chi^2 = \frac{R}{\sigma^2 (N-d.o.f)}\,.
\label{eq:}
\end{equation}

When we calculate the $\chi^2$, we obtain that for the bottomonium case, using eight states as input, $\chi^2 = 49.5$. If we use just four states as input, we get $\chi^2 = 31.6$
and for the charmonium, $\chi^2 = 48.3$. This way, the best fit happens for the bottomonium, when using fewer states as input. This is expected, since this system is less relativistic than
the charmonium and fewer states used as input usually means a fit that agrees more with the experimental data used.

\addtocounter{footnote}{-1}
\addtocounter{Hfootnote}{-1}
\footnotetext[\value{footnote}]{Relative errors are in parentheses.}
\addtocounter{footnote}{1}
\addtocounter{Hfootnote}{1}
\footnotetext[\value{footnote}]{These states are used in the fit.}

We notice that, as in the bottomonium, the results for both potentials are similar. In particular, 
they both yielded the same value for the string force (although this value for the charmonium case is not compatible with the one for the bottomonium case).
Nevertheless, the effects of the short-range interaction do play a role, as the spectra themselves are slightly different in the two models (see Table
\ref{tb:Results_Charmonium} and Fig.\ \ref{fig:Charmonium_Mass_Spectrum}).

\begin{figure}[h]
\caption{Plot of charmonium  wave functions obtained through our computations, using the potential from the lattice propagator (in blue) and the Cornell potential (in red).
The left panel corresponds to S states and the right panel to P states. We use solid lines for states 1S and 1P and dashed lines for 2S and 2P.}
\label{fig:wf_plots_Charmonium}
  \pgfplotstableread{./img/Charmonium_data_latt_prop/Normalized_0_wave_function_0.dat}{\zwfz}
  \pgfplotstableread{./img/Charmonium_data_latt_prop/Normalized_0_wave_function_1.dat}{\zwfo}
  \pgfplotstableread{./img/Charmonium_data_latt_prop/Normalized_1_wave_function_0.dat}{\owfz}
  \pgfplotstableread{./img/Charmonium_data_latt_prop/Normalized_1_wave_function_1.dat}{\owfo}
  
  \pgfplotstableread{./img/Charmonium_data_cornell_pot/Normalized_0_wave_function_0.dat}{\cpzwfz}
  \pgfplotstableread{./img/Charmonium_data_cornell_pot/Normalized_0_wave_function_1.dat}{\cpzwfo}
  \pgfplotstableread{./img/Charmonium_data_cornell_pot/Normalized_1_wave_function_0.dat}{\cpowfz}
  \pgfplotstableread{./img/Charmonium_data_cornell_pot/Normalized_1_wave_function_1.dat}{\cpowfo}
  \centering
\subfigure[
]{
  \centering
  \label{fig:wf_1s_2s_char}
\begin{tikzpicture}[scale=.8]
  \begin{axis}
     [xlabel=$r$,
      ylabel=$f(r)$,
      axis lines = left,
      axis x line=center,
      grid=both,
      minor y tick num = {2},
      minor x tick num = {1},
			x unit=\si{\femto \metre},
			y unit=\si{\femto \metre}^{-1/2},
			legend pos = {north east},    
			legend cell align = left,
			legend style = {align=justify,font=\scriptsize},
			legend entries = {LGP Pot.\ 1S,Cornell Pot.\ 1S,LGP Pot.\ 2S,Cornell Pot.\ 2S},]
    \addplot [blue,very thick] table {\zwfz};
    \addplot [blue,very thick,style=dashed] table {\zwfo};
    \addplot [red,thick] table {\cpzwfz};
    \addplot [red,thick,style=dashed] table {\cpzwfo};
  \end{axis}
\end{tikzpicture}}\;
 \subfigure[
]{
   \centering
   \label{fig:wf_1p_2p_char}
 \begin{tikzpicture}[scale=.8]
  \begin{axis}
     [xlabel=$r$,
      ylabel=$f(r)$
      axis lines = left,
      axis x line=center,
      grid=both,
      minor y tick num = {1},
      minor x tick num = {1},
			x unit=\si{\femto \metre},
			y unit=\si{\femto \metre}^{-1/2},
			legend pos = {north east},    
			legend cell align = left,
			legend style = {align=justify,font=\scriptsize},
			legend entries = {LGP Pot.\ 1P,Cornell Pot.\ 1P,LGP Pot.\ 2P,Cornell Pot.\ 2P},]
    \addplot [blue,very thick] table {\owfz};
    \addplot [blue,very thick,style=dashed] table {\cpowfo};
    \addplot [red,thick] table {\cpowfz};
    \addplot [red,thick,style=dashed] table {\cpowfo};
  \end{axis}
 \end{tikzpicture}}
\sourcebytheauthour
\end{figure}
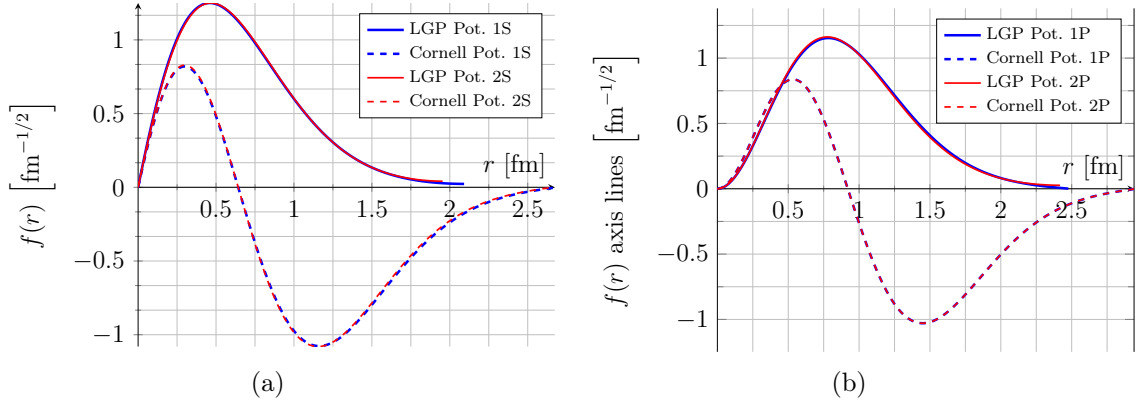

For comparison, we plot as well some obtained wave functions in Fig.\ \ref{fig:wf_plots_Charmonium}. As expected,
given the similarity of both potentials and of the obtained spectrum, the two wave functions
agree. The interesting fact is that for the charmonium the obtained string parameter is
much weaker, which results in a wave function that spreads to large values of $r$.

As was done for the bottomonium, we perform a search for states more excited than the ones used to perform the fit, and compare with the
observed spectrum in Table \ref{tb:charmonium_experimental}. The search is made for $l$ from $0$ to $3$, but no state was found 
for $l = 3$. We searched for six states for each $l$ in the range of energies from $\SI{3}{\giga \electronvolt}$ to $\SI{5}{\giga \electronvolt}$.
However, we found just five states in this range of energy for $l = 0,1$ and just four states for $l = 2$.

The states that clearly do not match any of the states found with the calculations are omitted from Fig.\ \ref{fig:Charmonium_Mass_Spectrum} for clarity\footnote{These 
states are the $X(3872)$, $\,X(3940)$,$\,X(4250)$,$\,X(4260)$,$\,X(4350)$ and $X(4360)$.}. In this figure it is possible to identify
$\psi(3770)$ as the state $1D$ (error of $0.72 \% - 1.54 \%$), $\psi(4040)$ as $3S$ (error of $0.70 \%- 1.28 \%$) and $\psi(4160)$ as $2D$ (error of $0.39 \% - 0.99 \%$).
If there are indeed particles in the state $2D$ with their quark constituents in the 
triplet configuration $s=1$, then it is expected that there are particles with $j = 1$, $2$ and $3$. The state $\psi(4160)$ has $j =1$ and $X(4160)$ has an unknown $j$. Therefore 
is tempting to consider $X(4160)$ as a state $\psi_{j=2,\,3}(2D)$. Also, we have a good match for $\psi(4415)$ with the state $4S$ (error of $0.25 \% - 0.71 \%$). 
Lastly, $X(4660)$ may appear initially to have a good agreement with the state $2P$, however its negative parity requires $l$ even and therefore our model is unable to describe it.

\section{Summary}
\label{sec:summary}
In this chapter, we applied the potential-model approach to compute the bottomonium and charmonium spectra. Our aim was not to reproduce
these spectra with high accuracy, but rather to study the effects of a modification of the simple Cornell potential to include a nonperturbative 
gluon propagator. To improve our results, one could introduce relativistic
corrections as well as spin-orbit and spin-spin interactions as done  in potential models, e.g.\ in Ref.\ \cite{Radford:2007vd}. The heavy-quarkonium spectrum is also obtained 
by using other techniques, such as lattice QCD calculations \cite{Liu:2012ze}, perturbative QCD \cite{Brambilla:2001qk} and effective field theories (EFTs) such as nonrelativistic QCD (NRQCD)
and potential nonrelativistic QCD (pNRQCD). EFTs are used both in analytical and in numerical studies, in combination with lattice formulations \cite{Pineda:2011dg, Brambilla:2014aaa}. Such calculations have achieved great precision for the description of a variety of states. Some of the challenges for a complete description of the spectra are: limitations on lattice simulations of fully
relativistic heavy quarks and difficulty to get information on the radius of quarkonia systems, which makes the distinction between perturbative and nonperturbative regimes
in pNRQCD ambiguous for some of the lowest bottomonium and charmonium states \cite{Brambilla:2010cs}. Also, obtaining the full spectrum presents a challenge for these techniques, as
there are obstacles to accurately get some properties of higher excited states. For example, in the lattice approach, higher excited states are more subject to statistical fluctuations in
their correlation functions, inducing errors in the mass determination. On the other hand, these methods account for splitting due to the different spins of the particles, allowing to
compute the masses of the $J/\psi(1S)$ and $\eta_c(1S)$ separately, for example. In several instances, they allow one also compute the quarks masses. With respect to the bottom quark mass, 
the results are often situated around \SI{4.2}{\giga \electronvolt} \cite{Pineda:2011dg, Brambilla:2010cs}, with some lattice results putting it 
at \SI{4.4}{\giga \electronvolt} \cite{Brambilla:2010cs}. This also agrees with values obtained for the bottom quark mass through QCD sum rules \cite{Brambilla:2010cs, Corcella:2002uu}.

 \cleardoublepage
\chapter{Conclusion}
\label{sec:conclusion}
\thispagestyle{capitulo}
\epigraph{``\textit{\ldots all good things must come to an end\ldots}''}{Q\\Star Trek -- The New Generation: All the good things\ldots} 

In this dissertation we initially reviewed some aspects of gauge theories. In particular,
we showed the equivalence of the path-integral formalism
with the conventional operator approach in Quantum Mechanics. The path integral
uses the classical action to describe the quantum mechanical system. We extended
this concept to be applied not just to spinless particles but to general quantum fields as well
(both gauge fields and matter fields). After introducing the $U(1)$ 
symmetry in electromagnetism, we showed how to introduce gauge symmetries given by the 
$SU(N)$ group. This step is of general interest, since all the forces of the 
standard model are gauge theories. Next we discussed the lattice formulation of $SU(N)$
gauge theories and presented a numerical application in the $SU(2)$ case.

We then turned to the study of heavy quarkonia states. The force 
binding these particles is the strong force, which has a nonperturbative nature. 
We showed, using the lattice formulation, that the 
interaction potential between two particles
rises linearly in the strong-coupling limit, which is compatible with the confinement of quarks.

The confining potential is predominant at long range, while the description of short-range 
interactions can be done using perturbative methods. More specifically, in the one-gluon-exchange approximation, the 
potential will be proportional to the Fourier transform of the gluon propagator. 
We obtained the short-range-interaction potential in two cases. In the first case we 
considered the gluon propagator to be inversely proportional to the square of the
transferred momentum. This case yields a Coulomb-like potential. This potential,
when added to the linearly rising term, is usually called Cornell potential. The 
second case considered came from using the gluon propagator obtained from 
a lattice simulation of $SU(2)$ pure gauge theory.

When we compare both potentials, we see only small differences between 
them, as Fig.\ \ref{fig:Comparison_Potentials} shows. However, the functional 
form of the second potential is quite interesting. In addition to the linear term, we see a clear indication that we are
in fact interpolating two limits, since the short range term is a Coulomb potential that is exponentially
damped (reminiscent of the Yukawa potential) and modulated by a trigonometric function. This
happens due to the pole structure of the propagator and comes from the non-Abelian structure 
of the theory, which determines gluon confinement \cite{Cucchieri:2011ig}.

Despite the fact that the potential from the lattice propagator presumably carries nonperturbative information, it 
does not yield significantly better results. In the bottomonium case, we obtain a slightly
better agreement with the experimental masses using this potential than when using the Cornell potential. 
We notice as well that the value obtained for $F_0$
in the bottomonium case does not agree with the one obtained in the charmonium case, meaning that
the potential obtained by fixing this parameter is not a general one.

One thing to note is that, especially in the charmonium case, we managed to describe reasonably well several states using the parameters 
determined by fitting the lowest states.
There were several states that our 
model could not explain, but these were mostly X-states. In Ref.\ \cite{Beringer:1900zz}, several of these states do not have
well determined isospin, which opens the possibility of them not being actually charmonium. If this is the case,
they should not be described by the computed spectrum anyway.

We remark that we use the potential model without any relativistic correction and therefore we do not expect our results 
to be very precise in the description of the quarkonia spectrum\footnote{See Section \ref{sec:summary}.} when compared with other works 
including relativistic corrections, as in Ref.\ \cite{Radford:2007vd}. Our aim here was to study how using a nonperturbative 
propagator would modify the Cornell potential. The results obtained showed that it introduces just minor corrections, far smaller 
than relativistic corrections. This can also be seen in the right panel of Fig.\ \ref{fig:Potential}, which shows that this potential differs very little
from Cornell potential.

Preliminary results of our study were presented at the XII Hadrons Physics Conference [see proceedings
in Ref.\ \cite{Serenone:2012yta}] and in the 31st International Symposium on Lattice 
Field Theory [see proceedings in Ref.\ \cite{Serenone:2014ota}]. We are also preparing a paper containing 
a more complete presentation of our study.

We are currently extending our analysis to the case of nonzero temperature.

\cleardoublepage
\phantomsection
\bibliographystyle{ifsc_abnt}
\addcontentsline{toc}{chapter}{References}
\bibliography{biblio}
\cleardoublepage

\appendix
\cleardoublepage
\chapter{Used Notation}
\label{sec:notation}
\thispagestyle{capitulo}

We use the standard notation present in the literature throughout this dissertation. However,
since the definition of what is commonly used can be ambiguous, we list our notation in this section
to avoid confusion.

We adopt natural units in all our calculations, i.e.\ $\hbar = c = 1$, meaning that all quantities are
given in terms of powers of the energy. Also, we set the permittivity of free space $\epsilon_0$ and the
permeability of free space $\mu_0$ equal to one. It is possible to show that this is equivalent 
to redefining the electric charge. Firstly, notice that in natural units, with electric charge in Coulombs, we have
\begin{equation}
 \mu_0 \;=\; \SI{4 \pi e-7}{\newton\per\ampere\squared} \;=\;
 \num{4 \pi e-7}\, \SI[parse-numbers = false]{\frac{\hbar}{c}}{\per\coulomb\squared}\,.
\end{equation}
By imposing $\mu_0 = 1$, we get
\begin{equation}
 \SI{1}{\coulomb}\;=\; \sqrt{\num{4 \pi e-7}{}\, \frac{\hbar}{c}}\;.
\end{equation}
We will generally adopt \si{\giga \electronvolt} for energy units. Conversion to 
other units can be achieved through multiplication by the appropriate factors of $\hbar$, $c$ and $\hbar c$.

We use Einstein notation for sums, i.e.\ repeated indices mean a sum over all possible values for them.
Also, when in Minkwoski space, we use lower-case Greek letters to indicate the components of a four-vector 
(e.g.\ $x^\mu\; |\; \mu = 0, 1, 2, 3$) and lower-case Latin letters of the middle of the alphabet 
(starting at $i$) to indicate the spatial components only of the four-vector (e.g.\ $x^i \;|\; i = 1, 2, 3$).
In Euclidean space we keep using the same notation to differentiate the four-vectors from spatial vectors.
However, we take Euclidean four-vector as $x^\mu\; |\; \mu = 1, 2, 3, 4$ and keeping indexing of $i$, i.e.\ time
will correspond to the fourth position instead of the zeroth. We use bold-face characters for three-vectors and normal
characters for four-vectors. However, in some circumstances we use normal characters to denote the norm of a three-vector,
i.e. $\abs{\vect{k}}\to k$.

We use two metric tensors. When in Minkowski space, we denote the metric tensor by
\begin{equation}
 g_{\mu \nu} = \diag({-1,1,1,1})\,.
\end{equation}
Components of contravariant vectors carry upper indices and do not carry any signs in its definition, e.g.\ they are
given by
\begin{equation}
 x = \begin{pmatrix} t \\ x^1 \\ x^2 \\ x^3 \end{pmatrix}\,,
\end{equation}
where $x^0 \equiv t$. Components of covariant vectors are indicated by a lower index are
related by the metric with contravariant vectors as $x^\mu = g^{\mu \nu} x_\nu$.

If we are in Euclidean space, the metric tensor coincides with the 
identity matrix and we denote it using the Kronecker delta $\delta_{\mu \nu}$.
Therefore there will be no difference between contravariant and covariant vectors and
all vectors components will be denoted by a lower indice for simplicity.

We use the Dirac representation for the gamma matrices
\begin{align}
     \gamma^0 \;&=\; \begin{pmatrix} 1 & 0 & 0 & 0 \\ 0 & 1 & 0 & 0 \\ 0 & 0 & -1 & 0 \\ 0 & 0 & 0 & -1 \end{pmatrix},\quad \gamma^1 \;=\; \begin{pmatrix} 0 & 0 & 0 & 1 \\ 0 & 0 & 1 & 0 \\ 0 & -1 & 0 & 0 \\ -1 & 0 & 0 & 0 \end{pmatrix} \nonumber \\[3mm]
    \gamma^2 \;&=\; \begin{pmatrix} 0 & 0 & 0 & -i \\ 0 & 0 & i & 0 \\ 0 & i & 0 & 0 \\ -i & 0 & 0 & 0 \end{pmatrix},\quad \gamma^3 \;=\; \begin{pmatrix} 0 & 0 & 1 & 0 \\ 0 & 0 & 0 & -1 \\ -1 & 0 & 0 & 0 \\ 0 & 1 & 0 & 0 \end{pmatrix}\,.
\label{eq:gamma_matrix}
\end{align}
To represent a four-vector $A_\mu$ contracted with these gamma matrices, we use the Feynman slash notation $\gamma^\mu A_\mu = \slashed{A}$.

We denote a group element of an arbitrary group $G$ by $g$. However, if we have a unitary group, such as $SU(N)$,
we denote its group elements by an upper-case Latin letter of the end of the alphabet (starting at $U$). If we want 
to specify a matrix element of this group element we use lower-case Greek letters of the beginning of the alphabet
(e.g.\ $U_{\alpha \beta}\; |\; \alpha, \beta = 1,2, \dots N$).

When associating group elements to links between points on a four-dimensional lattice we denote the group elements by
$U_\mu (x)$. This means that this group element is associated to the link between the points $x$ and $x+a e_\mu$, where $a$
stands for the lattice spacing and $e_\mu$ is a versor to indicate the direction of the link. The notation we adopt 
for a plaquette (the oriented multiplication of the matrices in a minimal square of the lattice) is $U_{\mu \nu}(x)$,
meaning a plaquette described by a closed path starting at the link $U_\mu(x)$ and rotating on the plane $\mu \nu$ 
in the counterclockwise direction.

We index the group generators by lower-case Latin letters from the beginning of the alphabet (e.g.\ $\lambda^a \;|\; 
a=1 \dots N^2-1$). For the $SU(2)$ case these are proportional to the Pauli matrices, i.e.\ $\lambda^a = \sigma_a/2$, where
\begin{equation}
 \sigma_1 \;=\;
\begin{pmatrix}
0&1\\
1&0
\end{pmatrix}\quad
\sigma_2  \;=\;
\begin{pmatrix}
0&-i\\
i&0
\end{pmatrix}\quad
\sigma_3  \;=\;
\begin{pmatrix}
1&0\\
0&-1
\end{pmatrix}\,.
\end{equation}

For the $SU(3)$, the generators are the Gell-Mann matrices
\begin{align}
\lambda_1 \;&=\; \frac{1}{2}\begin{pmatrix} 0 & 1 & 0 \\ 1 & 0 & 0 \\ 0 & 0 & 0 \end{pmatrix} &\qquad 
\lambda_2 \;&=\; \frac{1}{2}\begin{pmatrix} 0 & -i & 0 \\ i & 0 & 0 \\ 0 & 0 & 0 \end{pmatrix}&\quad 
\lambda_3 \;&=\; \frac{1}{2}\begin{pmatrix} 1 & 0 & 0 \\ 0 & -1 & 0 \\ 0 & 0 & 0 \end{pmatrix}&\nonumber \\[2mm]
\lambda_4 \;&=\; \frac{1}{2}\begin{pmatrix} 0 & 0 & 1 \\ 0 & 0 & 0 \\ 1 & 0 & 0 \end{pmatrix} &\qquad
\lambda_5 \;&=\; \frac{1}{2}\begin{pmatrix} 0 & 0 & -i \\ 0 & 0 & 0 \\ i & 0 & 0 \end{pmatrix} &\quad
\lambda_6 \;&=\; \frac{1}{2}\begin{pmatrix} 0 & 0 & 0 \\ 0 & 0 & 1 \\ 0 & 1 & 0 \end{pmatrix} &\nonumber \\[2mm]
\lambda_7 \;&=\; \frac{1}{2}\begin{pmatrix} 0 & 0 & 0 \\ 0 & 0 & -i \\ 0 & i & 0 \end{pmatrix} &\qquad
\lambda_8 \;&=\; \frac{1}{2\sqrt{3}} \begin{pmatrix} 1 & 0 & 0 \\ 0 & 1 & 0 \\ 0 & 0 & -2 \end{pmatrix}\,.& \,
\end{align}

We will find, in some instances, vectors and tensors carrying the group-generator (or color) indices. In this case 
we distinguish the indices using a comma, e.g.\ $F^{\mu \nu,\,a}$.

\cleardoublepage
\chapter{Group Theory and Group Integration}
\label{sec:group_theory_review}
\thispagestyle{capitulo}

This appendix aims at briefly reviewing some key concepts of group theory. We focus in the study of Lie Groups and Lie Algebras.
We will follow mostly Ref.\ \cite{Agostinho}. We also review group integrations using the Haar measure following
Refs.\ \cite[Chap.\ 8]{creutz1983quarks} and \cite[Chap. 3]{gattringer2009quantum}. Our approach will be more 
informal and in some cases we will state a result without proofing it. In those cases, the proof can be found in the already mentioned references.

\section{Definition of a Group and Useful Concepts}
\label{sec:group_def}

Consider a set of elements $G$ with a binary operation between them (a product). If this set of 
elements and this operation satisfy the properties
\begin{enumerate}
 \item If $g_1$, $g_2 \in G$ then  $g_1 g_2 = g \in G$.
 \item If $g_1$, $g_2$, $g_3 \in G$, $ (g_1 g_2) g_3 = g_1 (g_2 g_3)$.
 \item For any $g_1$, $g_2 \in G$ there exists one unique $g \in G$ and one unique $g' \in G$ such
 that $g g_1 = g_2\,$ and $\,g_1 g' = g_2$,
\end{enumerate}
then $G$ is a group under this binary operation.

Notice that the last axiom implies the existence of the identity element (just take $g_1 = g_2$). It is also possible to 
prove from these axioms the uniqueness of identity, i.e.\ the identity from the right side ($g e = g$), is the same as the 
identity from the left side ($e g = g$). And, since there is an identity that belongs to $G$, it is also 
possible to prove the existence of the inverse of an element (just take $g_2 = e$, where $e$ is the identity).

There are several examples of groups. As a trivial example, take the integer numbers as the set elements 
and the sum as the operation (product) between the elements. All the properties above are satisfied and 
we say that the integer numbers form a group under the sum operation. We can change the set from the integer
numbers to real numbers and we will obtain another group.

Let us see a second example. We consider all the rotations of an equilateral triangle $ABC$ that leave it invariant,
i.e.\ that just change the labels of the vertices but do not change the geometrical figure. We can easily enumerate
how many rotations there are: rotations by $g_{-2} = \SI{-240}{\degree}$, $g_{-1} = \SI{-120}{\degree}$,
$g_0 = \SI{0}{\degree}$, $g_1 = \SI{120}{\degree}$ and $g_2 = \SI{240}{\degree}$, with 
$g_3 = \SI{360}{\degree} = g_{-3} = \SI{-360}{\degree} = g_0$. The operation defined by the product of two elements 
of the set consists in applying the rotations consecutively in the order they appear. This set with this operation 
forms a group as well.

As a last example, and a far less trivial one, consider the set of $N \times N$ complex matrices that have determinant 
one and are unitary, i.e.\ $U U^\dagger = \dblone$. They form a group under the operation of matrix multiplication.
This kind of group is the one used in this dissertation. It is called the \emph{special unitary group of degree 
$N$}, abbreviated as $SU(N)$.
\vskip 3mm

Let us fix a group element $\bar{g}$ and consider any two group elements $g$ and $g'$. If they obey the relation
\begin{equation}
g = \bar{g}g'{\bar{g}}^{-1}\,,
\end{equation}
we say that \emph{$g$ is conjugate to $g'$}. This establishes an equivalence relation between these elements and
allows the creation of equivalence classes inside the group, i.e.\ we say that the set of elements that are 
conjugate to each other forms a \emph{conjugacy class inside} $G$. Notice that there is no intersection between two
distinct conjugacy classes.

Let us consider a group $G$ that has a subset $H$, which forms a group under the same operation
as the group $G$. This $H$ is called a subgroup of $G$.
This is nicely illustrated in the first example given above. Since the integers are a subset of the reals, and
both form a group under the sum operation, then the integers under the sum operation constitute a subgroup of the reals
under the sum.

The second example given above showed us that we can create a group that does not necessarily have a mathematical representation.
It may be defined as a physical construct as, for example, the rotations of a triangle.
However, we want to perform calculations involving the group elements and therefore we need a
\emph{representation}.

A representation is composed of mathematical objects $D$ that are functions of the
group elements $g$ and respect the group structure, i.e.\
\begin{align}
 D(g)D(g') = \,& D(g g')  \nonumber \\
 D(g^{-1}) =\, & D^{-1}(g) \\
 D(e) =\, & \dblone \nonumber\,.
\end{align}

Notice that a group $G$ can have several different representations. In the example of the triangle, the transformations could
be represented by
\begin{equation}
D(g_m)= 120\,m\,,\;m\,=\,-2,-1,0,1,2\,. 
\end{equation}
The group operation for these elements is the sum, represented as $D(g_m) D(g_n) = D(g_{m+n})$. Note that, 
to keep the closure of the group, every time that $D(g_{m+n} ) \geq 360\,$ or $\,D(g_{m+n} ) \leq 360$, we need 
to subtract or add $360$. Also, the group structure tells us that $D(g_3) = D(g_{-3}) = D(g_0)$.

However, we could represent it as well by a 2x2 matrix
\begin{equation}
D(g_m)= \begin{pmatrix}
   \cos \theta & -\sin \theta\\
   \sin \theta & \cos \theta
  \end{pmatrix}\,,\qquad \theta\,=\, \frac{2}{3} \pi m\,,
\end{equation}
where again $m = -2,-1,0,1,2$. These are two different representations of the same group. Also,
notice that there is nothing obliging the representations to have a one-to-one correspondence with the 
group elements. We could associate all the group elements to the unity matrix and that would be a representation
of the group as well (although a trivial and boring one). The representations that have a one-to-one correspondence
with the group elements are said to be faithful representations.

Note that in this second representation the triangle could be represented by a dimension-two column vector, which encodes 
the position of one of the vertices. If we apply $D(g_m)$ to this vector, it corresponds to performing the rotation of the 
triangle.

We say that two representations are equivalent if the relation $\,D'(g)=C D(g) C^{-1}\,$ holds for any $g \in G$, where $C$ is 
an operator that does not depend on $g$. As an example, we can create two equivalent representations by changing the 
basis of the vector space in which $D(g)$ acts.

The trace $\Tr\left[D(g)\right]$ gives us information about the representation and the group. If $D'(g)$ is an equivalent 
representation of $D(g)$, then the cyclic property of the trace says that they possess the same trace. Also, notice that
elements of the same conjugacy class in a given representation will have the same trace. The trace $\Tr\left[D(g)\right]$
is also called the character of the group element $g$ in the representation $D$.

\section{Lie Groups and Lie Algebras}

Notice that, in our example above, we have parametrized the group elements in the process of constructing the representations.
The parametrization of group elements by a set of parameters is a general step in constructing representations and
has an importance in the classification of groups. The groups where the parameters assume discrete values and have a 
finite number of elements, such as in the example of the triangle rotations, are said to be \emph{finite discrete groups}.
However, they can have infinite elements, as in the group of the integers under the sum. Such groups are called
\emph{infinite discrete groups}.

Another case happens if the parameters can assume continuous values, as happens in the case of the reals under the
sum operation. These groups are called \emph{continuous groups}. They can be divided into two subcategories.
If the parameters are restricted to a compact domain they are called \emph{compact continuous groups}. Otherwise, 
they are called \emph{non-compact continuous groups}. We will focus on the case of continuous groups because 
they are the ones used in this dissertation.

In all cases, it is possible to map these parameters onto a space, such as the points of a line in the case 
of real numbers. However, there is nothing restricting us to map it just to a flat (Euclidean) space. In general the mapping
will be associated to a \emph{manifold}. A manifold $M$ is a generalization of surfaces embedded in $\mathbb{R}^3$. 
Its definition is of a space that is locally Euclidean. This idea is better understood through the following illustration:\ 
imagine that the manifold is inhabited by ants. These ants would perceive the space surrounding them as a Euclidean
space and would believe to live in a flat land, the same way mankind believed that the Earth was flat in the past. They would
only perceive the true nature of their world when traveling large distances or flying to high altitude. 

A manifold is usually described by a parametrization. As an example, consider the surface of a sphere. A small ant 
on its surface would perceive its surrounding neighborhood as a Euclidean space. Also, we can associate each one 
of its points to points in a domain of $\mathbb{R}^2$. Notice that this surface can be described by two parameters 
$\theta$ and $\varphi$, which can be arranged in $\mathbb{R}^2$ to form a rectangle of sides $2 \pi$ by $\pi$.

If the manifold has a tangent space at each of its points, then it is called a \emph{differentiable manifold} \cite{choquet1982analysis}.

\emph{Lie groups} are the groups whose elements can be mapped onto a differentiable manifold. They can be related 
to a \emph{Lie algebra}. To define a Lie algebra, we start by the definition of \emph{field}\footnote{This concept 
should not be confused with the gauge fields or fermion fields studied in Chap. \ref{sec:Overview_Gauge_Field}.}.
A field $k$ is a set of elements with a sum operation and a multiplication operation, having the properties
\begin{itemize}
 \item The elements of $k$ commute under addition.
 \item The set $k$, without the neutral element of addition, is a group under multiplication.
 \item The multiplication is distributive with respect to the addition, i.e.\ $\,a(b+c) = ab + ac\,$.
\end{itemize}
Two usual examples of fields are the real numbers and the complex numbers.

A \textit{Lie algebra} $\mathcal{G}$ is a vector space over a field $k$ and is closed under a bilinear composition
law that respects the same properties as the commutator, namely
\begin{enumerate}
 \item $[x,a y + b z] = a [x,y] + b [x,z]$, where $a$ and $b$ are scalars.
 \item $[x,x] = 0\,$.
 \item $[x,[y,z]]+[z,[x,y]]+[y,[z,x]] = 0$ (Jacobi Identity).
\end{enumerate}

The relation between a Lie algebra $\mathcal{G}$ and a Lie group $G$ is given through the calculation of tangent spaces
to the manifold $M_G$ associated to the group. Let us consider a curve in the manifold defined by the collection of 
points $p$ identified by the set of coordinates $x^i(t)\, | i = 1 \dots \dim M$ and parametrized by $t$. The coordinates 
$x^i$ are the parameters used to describe the manifold. We consider as well a test function $f$, defined and differentiable 
in the neighborhood of the point $p$ where we wish to find the tangent space. Then, a tangent vector $V_p$ to the manifold at 
this point is
\begin{equation}
 V_p f \;=\; \left. \frac{d x^i (t)}{dt}\right|_{t=p} \frac{\partial }{\partial x^i}f\,.
 \label{eq:manifold_tangent_vector}
\end{equation}
Notice that, since the function $f$ is a test function, $V_p$ is only a function of the point $p$. We can generate as many 
tangent vectors as we want by choosing different curves in the manifold and then build a 
basis for the tangent space at the point $p$. This
is an unusual way to represent a vector. However, notice that the operators $\frac{\partial }{\partial x^i}$ are
linearly independent and can be used as a basis to write these vectors [see Ref.\ \cite{Agostinho}].

We can find in this manner vectors $V_p$ for all points on the manifold. Furthermore, if we can relate these vectors
in a continuous and differentiable way, we say that the set of vectors $V_p$ forms a \emph{vector field}
\begin{equation}
 V \;=\; V^i(x) \frac{\partial }{\partial x^i}\,.
 \label{eq:general_form_vec_field}
\end{equation}

If we consider
two vector fields $V$ and $W$, it is possible to show that they have the closure property under the commutator operation, 
i.e.\ $[V,W]$ is a vector field \cite{Agostinho}.
\begin{align}
 [V,W] \;&=\; V^i(x) \frac{\partial }{\partial x^i} \left[W^j(x) \frac{\partial }{\partial x^j}\right]
            -W^j(x) \frac{\partial }{\partial x^j} \left[V^i(x) \frac{\partial }{\partial x^i}\right] \nonumber \\[2mm]
       \;&=\; V^i(x) \frac{\partial W^j(x)}{\partial x^i}\frac{\partial }{\partial x^j}
            +V^i(x) W^j(x) \frac{\partial^2}{\partial x^i \partial x^j} \nonumber \\[2mm]
\;&\phantom{=}\;-W^j(x) \frac{\partial V^i(x)}{\partial x^j}\frac{\partial }{\partial x^i}
            -W^j(x) V^i(x) \frac{\partial^2}{\partial x^j \partial x^i} \nonumber \\[2mm]
       \;&=\; V^i(x) \frac{\partial W^j(x)}{\partial x^i}\frac{\partial }{\partial x^j}
            -W^i(x) \frac{\partial V^j(x)}{\partial x^i}\frac{\partial }{\partial x^j}
       \;=\; Y^j(x) \frac{\partial }{\partial x^j} = Y\,.
\end{align}
Using an analogous reasoning, it is possible to show that the set of vector fields obeys all the necessary
operations (listed above) to be considered a Lie algebra.

There is a set of vector fields that is special, called the Lie algebra of the Lie group. 
It forms a sub-algebra of the Lie algebra of the vector fields and its elements $\lambda^a$ obey the relation
\begin{equation}
 \label{eq:structure_constants_general}
 [\lambda^a,\lambda^b] = f^{a b c} \lambda^c\,,
\end{equation}
where $f^{a b c}$ is called the \textit{structure constant} of the group $G$. The reason for this name is that
$f^{a b c}$ does not depend on the point of the manifold where the vector field is being evaluated. Therefore
it carries global information about this algebra and the group. Also, this independence of the point gives us 
the freedom to choose the point of the manifold associated with the identity element $\,e\,$ to evaluate the tangent 
vectors $\lambda^a$. Moreover, there will be enough vectors to form a base for the tangent space at the evaluation point.

These vector fields are called the \textit{generators} of the group $G$. This name comes from the fact that
we can generate the group elements near $e$ by
\begin{equation}
 g \;=\; \exp(i \omega^a \lambda^a)\,,
\label{eq:exponential_map}
\end{equation}
where $\omega^a$ is a set of parameters of the group $G$. In some cases,
when some conditions are met, this can be extended to the whole group.

Moreover, if the generators are 
Hermitian, it is more useful to write Eq.\ \ref{eq:structure_constants_general} as
\begin{equation}
\label{eq:structure_constants_hermitian}
 [\lambda^a,\lambda^b] \;=\; i f^{a b c} \lambda^c\,, 
\end{equation}
since now the structure constants will be real.

There is one useful property of the structure constant that we shall demonstrate. Consider Eq.\ \ref{eq:structure_constants_hermitian}.
We multiply both sides by another generator and take the trace. We assume that the normalization
\begin{equation}
 \Tr (\lambda^a \lambda^b) \;=\; \frac{1}{2} \delta^{a b}
\end{equation}
is valid. We obtain
\begin{equation}
 \Tr\left([\lambda^a ,\lambda^b] \lambda^d\right) \;=\; i f^{a b c} \Tr(\lambda^c \lambda^d) \;=\; f^{a b d}\; \frac{i}{2}\,.
\end{equation}

We then isolate the structure constant, expand the commutator and use the cyclic property of the trace to obtain
\begin{equation}
 f^{a b d} = -2 i\left[ \Tr(\lambda^a \lambda^b \lambda^d) - \Tr(\lambda^b \lambda^a \lambda^d) \right] =
-2 i\left[ \Tr(\lambda^d \lambda^a \lambda^b) - \Tr(\lambda^a \lambda^d \lambda^b) \right] = f^{d a b}\,.
\end{equation}

Therefore, the structure constant remains invariant if we make a cyclic permutation.

\section{Example of Lie Group: The SO(3) Group}

In the previous section we presented several abstract concepts, some of which may seem disconnected
from physics at first glance. We will try to connect the mathematics of the previous section with 
well known properties of the angular momentum in quantum mechanics, as an example of connection between
Lie groups, Lie algebras and physics. In this example, we follow Ref.\ \cite{claudequantum}.

It is easy to convince ourselves that the set of rotations in the space $\mathbb{R}^3$ forms
a group. Each possible rotation will be a group element. It is a continuous compact group that can be represented
by $3 \times 3$ real matrices. These matrices $O$ must be orthogonal, i.e.\ $O O^T = \dblone$ and must satisfy $\det O = 1$. This group is
called the \emph{special orthogonal group of degree 3} or $SO(3)$. The orientation of a classical object can be 
represented by a column vector $\vect{v}$ of dimension 3. The operation of rotating this object is performed by $O$
acting in $\vect{v}$.

Let us consider now a quantum system in the state $| \psi \rangle$. After the spatial rotation
is performed it will be described by the state $| \psi' \rangle$, which is related to the previous
state by
\begin{equation}
 | \psi' \rangle \;\equiv\; R | \psi \rangle\,.
\end{equation}
The operator $R$ represents the rotation performed. It possesses the same group properties of $O$, being
an element of $SO(3)$ as well, although in a different representation.

The wave function is defined by  $\,\psi(\vect{r}) = \langle \vect{r}\, | \psi \rangle$. Also, it is natural
to assume that after the rotation, the value of $\,\psi(\vect{r}\,'_0)\,$ will be the same of the wave function
before the rotation $\psi(\vect{r}_0)$, where $\vect{r}\,'_0 = O \vect{r}_0$. Therefore we have
\begin{equation}
 \psi' (\vect{r}\,') \;=\; \langle \vect{r}\,' | \psi' \rangle \;=\; \langle \vect{r} O^T| R | \psi \rangle 
                  \;=\; \langle \vect{r}\, | \psi \rangle \;=\; \psi(\vect{r})
\end{equation}
Therefore we obtain
\begin{equation}
 R | \vect{r}\, \rangle \;=\; | O \vect{r}\, \rangle\,.
\end{equation}
This relates the two representations.

We will now seek to discover the Lie algebra of $SO(3)$. For this, let us consider an infinitesimal rotation
around an axis $\hat{u}$ by an angle $d \alpha$. It can be written as
\begin{equation}
O_{\hat{u}}(d \alpha)\, \vect{v} \;=\; \vect{v} + \hat{u} \times \vect{v}\, d \alpha\,.
\label{eq:infinitesimal_rotation}
\end{equation}

Let us apply this rotation to a wave function $\psi(r)$
\begin{equation}
 \langle \vect{r}\, | R_{\hat{u}}(d \alpha) | \psi \rangle \;=\; \langle \vect{r} O_{\hat{u}}^T (d \alpha) | \psi \rangle 
                                              \;=\; \psi\left[O_{\hat{u}}^{-1} (d \alpha) \vect{r}\,\right]\,.
 \label{eq:wave_func_infinites_rotate}
\end{equation}

We then expand $\psi\left[O_{\hat{u}}^{-1} (d \alpha) \vect{r}\,\right]$ around $d \alpha = 0$
\begin{align}
 \psi\left[O_{\hat{u}}^{-1} (d \alpha) \vect{r}\,\right] \;&=\; \psi \left[ \vect{r} - \hat{u} \times \vect{r}\, d \alpha \right]
          \;=\; \psi(\vect{r}\,) - \frac{\partial \psi(\vect{r}\,)}{\partial x^i} (\hat{u} \times \vect{r}\, d \alpha) \cdot \hat{x}_i \nonumber \\[2mm]
          \;&=\;\psi(\vect{r}\,) - \frac{\partial \psi(\vect{r}\,)}{\partial x^i} \epsilon_{j k l} u_k r_l \hat{x}_j \cdot \hat{x}_i d \alpha
          \;=\; \psi(\vect{r}\,) - \frac{\partial \psi(\vect{r}\,)}{\partial x^i} \epsilon_{k l i} u_k r_l d \alpha \nonumber \\[2mm]
          \;&=\;\psi(\vect{r}\,) - \hat{u} \cdot \vect{r} \times \left[\frac{\partial \psi(\vect{r}\,)}{\partial x^i} \hat{x}_i\right]\,.
\end{align}

Remembering that the linear-momentum operator can be written as $p_i = -i \partial/\partial x^i$, we obtain
\begin{equation}
 \psi\left[O_{\hat{u}}^{-1}(d \alpha) \vect{r}\,\right] \;=\; \psi(\vect{r}\,) - i \hat{u} \cdot \vect{r} \times \vect{p} \,\psi(\vect{r}\,) d \alpha \;=\; 
 \left[1 -i \hat{u} \cdot \vect{L} \,d \alpha \right] \psi(\vect{r}\,)\,,
\end{equation}
where $\vect{L}$ is the angular momentum components $\vect{L} \;=\; \vect{r} \times \vect{p}$. Therefore, we get
\begin{equation}
R_{\hat{u}} (d \alpha) \;=\; 1 -i \hat{u} \cdot \vect{L} \,d \alpha
\label{eq:infinitesimal_rotation_op}
\end{equation}

We identify the angular momentum operators $\hat{u}\cdot \vect{L}$ with the Lie algebra of the group $SO(3)$. Firstly, notice
that they have the form of Eq.\ \ref{eq:general_form_vec_field}. Also, it is well known that
\begin{equation}
[L_i,L_j] \;=\; i \varepsilon_{i j k} L_k\,,
\end{equation}
where $\varepsilon_{i j k}$ is the totally anti-symmetric Levi-Civita tensor, which is identified with the structure 
constant of the Algebra. This relation can be extended by defining $L_u = \hat{u} \cdot \vect{L}$ and considering 
two unit vectors $\hat{u}$ and $\hat{v}$
\begin{equation}
[L_u,L_v] \;=\; [L_i,L_j]\, u_i v_j \;=\; i u_i v_j\, \varepsilon_{i j k}\, L_k \;=\; i \vect{L} \cdot \hat{u} \times \hat{v} \;=\; i L_{\hat{u} \times \hat{v}} 
\end{equation}

Notice that $\hat{u}, \hat{v}$ and $\hat{u} \times \hat{v}$ form a basis. Therefore, using each one the $L_u$'s, 
it is possible to build a basis of a space. This space is the tangent space of the manifold of the $SO(3)$ group near the identity. In this case,
the manifold is a sphere of radius $\pi$ embebed in $\mathbb{R}^4$. Each point inside the sphere can be associated 
with a rotation. The axis of the rotation is specified by the direction of the vector connecting the point and the center of 
the sphere. The rotation angle is given by the norm of this vector. The identity element, where the tangent space is calculated,
is the center of the sphere.

To show that the angular-momentum operators form the Lie algebra of $SO(3)$, it remains to show that it is possible to
generate finite rotations from the infinitesimal rotations and therefore generate group elements. To this end, consider 
two rotations around the same axis $\hat{u}$. They are related by
\begin{equation}
R_{\hat{u}}(\beta) R_{\hat{u}}(\alpha) \;=\; R_{\hat{u}}(\alpha + \beta)
\end{equation}

Now, let us suppose that the second rotation is an infinitesimal one, i.e.\ $\beta = d \alpha$. Using Eq.\ \ref{eq:infinitesimal_rotation_op}
we get
\begin{equation}
R_{\hat{u}}(\alpha + d \alpha) \;=\; R_{\hat{u}}(d \alpha) R_{\hat{u}}(\alpha) \;=\; \left[1 - i \hat{u} \cdot \vect{L} \,d \alpha \right]R_{\hat{u}}(\alpha)\,.
\end{equation}

Therefore we have the following differential equation in the limit of $d \alpha \to 0$
\begin{equation}
\frac{R_{\hat{u}}(\alpha + d \alpha) - R_{\hat{u}}(\alpha)}{d \alpha} \;=\; -i \hat{u} \cdot \vect{L}\, R_{\hat{u}}(\alpha) 
\;\to\; \frac{d R_{\hat{u}}(\alpha)}{d \alpha} \;=\; -i \hat{u} \cdot \vect{L} \,R_{\hat{u}}(\alpha)\,.
\end{equation}

The solution of this differential equation is
\begin{equation}
R_{\hat{u}}(\alpha) \;=\; e^{-i \hat{u} \cdot \vect{L}\, d\alpha}\,.
\end{equation}

The above expression shows us that it is possible to generate any element $R_{\hat{u}}(\alpha)$ of the $SO(3)$ group through 
the angular-momentum operators, i.e.\ its Lie Algebra. Also, notice that this expression is the same as Eq.\ \ref{eq:exponential_map}.

The lesson learned through this example is that the operators in quantum mechanics are the Lie algebra of a group. Often the group in 
question is the the symmetry group of the system being studied. The wave function is an element of the field in which the Lie algebra
acts. In our study in Chapter \ref{sec:Overview_Gauge_Field}, the symmetry group of the system was the group $SU(N)$ and the fields $F_{\mu \nu}$
and $A_\mu$ where Lie-algebra elements. The field over which they acted was the column vector $c$.

\section{Group Integration}
\label{sec:group_integration}
In this section we aim at defining the meaning of the integral over group elements. Firstly, we want to
preserve the linearity of the integral operation, i.e.
\begin{equation}
 \int{ dg\,\left[ a f(g) + b h(g)\right]} \;=\; a \int{ dg\, f(g)} + b \int{dg\, h(g)}\,,
\end{equation}
where $a$ and $b$ are complex numbers, $g$ represents a dummy group element and  $f$, $h$ are functions
of $g$. We impose the additional constraint of left-invariance
\begin{equation}
 \int dg\, f(g) \;=\; \int dg\, f(g' g)\,,
 \label{eq:haar_measure_def}
\end{equation}
where $g'$ is a fixed group element. The motivation for this comes from the fact that if $g$ and $g' \in G$
then $g' g \in G$. Since we are integrating over the whole group, $g' g$ will run over all group elements 
and therefore the integral result must not change. This is analogous in ordinary integration to shifting
the integral variable by a constant when the integration runs over the whole domain of real numbers. The 
measure $dg$ obeying this condition are called \emph{Haar measure}.

The above conditions will allow us to find the measure $dg$. We will
restrict the discussion to compact Lie groups, which can be parametrized by a set of $N$ parameters
$\alpha_i$. We will denote this set of parameters simply by $\alpha$ to avoid cluttering the notation. The
same applies to $d\alpha$, i.e.\ $d\alpha = d\alpha_1 d\alpha_2 \dots d\alpha_N$. We start by assuming that it is 
possible to write
\begin{equation}
 \int{dg(\alpha)\, f[g(\alpha)]} \;=\; \int{ d\alpha\, J(\alpha) f[g(\alpha)]}\,.
\end{equation}

We use the left invariance, i.e.\ $g[\beta(\gamma,\alpha)] = g(\gamma)g(\alpha)$, to obtain
\begin{equation}
 \int{d\alpha J(\alpha) f[g(\alpha)]} \;=\; \int{d\beta J(\beta) \norm{\frac{\partial \beta(\gamma,\alpha)}{\partial \alpha}}^{-1} f[g(\beta)]}\,.
\end{equation}

Since the function $f$ is arbitrary, and setting $\gamma = e$ after evaluating the derivatives, we obtain
\begin{equation}
 J(\alpha) \;=\; \left. \left(J(\beta) \norm{\frac{\partial \beta(\gamma,\alpha)}{\partial \alpha}}^{-1}\right)\right|_{\gamma = e} = 
            K \norm{\frac{\partial \beta(\gamma,\alpha)}{\partial \alpha}}^{-1}_{\gamma = e}\,.
\end{equation}
The constant $K$ gives us freedom to choose a normalization. We will choose $K$ such that
\begin{equation}
\int{dg\, 1} \;=\; 1\,.
\label{eq:Haar_measure_normalization}
\end{equation}

This process is useful if one wishes to perform these calculations using a mathematical software such as Maple.
For manual calculation, this is a simple but tedious task. Fortunately, it is possible to solve most of the relevant
integrals using the defining property of the Haar measure (see Eq.\ \ref{eq:haar_measure_def}) \cite{gattringer2009quantum}.

Let us identify $f(U) = U^{a b}\,$ and apply this property. We obtain
\begin{equation}
\int{dU\, U^{a b} }\;=\; \int{dU\, (V U)^{a b}} \;=\; V^{a c} \int{dU\,U^{c b}}\,.
\label{eq:haar_integral_1}
\end{equation}
The above equation can be satisfied in only two ways. We can restrict $V$ to be $\dblone$ or the integral must be $0$. Since the
definition of the Haar measure requires $V$ to be arbitrary, we conclude that the integral must vanish.

Another useful integral comes from from setting $f(U) = U^{a b} U^{c d}$. We use again the definition of the Haar measure
\begin{equation}
\int{dU\, U^{a b} U^{c d}} = V^{a e} V^{f c}\int{dU\, U^{e b} U^{f d}}\,.
\label{eq:haar_integral_2}
\end{equation}
Once again we are not allowed to make any hypothesis about $V$, which leads us to the conclusion that the result of the integral
is zero\footnote{This is not always true. If we have a situation where $U^{c d} = (U^\dagger)^{a b}$, this integral
will not always vanish, since this will be the situation described in Eq.\ \ref{eq:haar_integral_3}. A situation like this
happens with integration of $SU(2)$ group elements.}. 

There is one other integral that is useful to be evaluated. Let us consider $f(U) = U^{a b} (U^\dagger)^{c d}$.
In the special case where $c = b$ and we sum over $b$, we arrive at $f(U) = \delta_{a d}$. Using the normalization of the metric
in Eq.\ \ref{eq:Haar_measure_normalization}, we obtain that
\begin{equation}
\int{dU  U^{a b} (U^\dagger)^{b d}} = \sum_{i=1}^N{\int{dU  U^{a i} (U^\dagger)^{i d}}}
\;=\; \delta_{a d}\,.
\end{equation}
 
Notice that the terms inside the sum are all equivalent and therefore contribute by the same amount. To justify this,
observe that if we swap the two rows and then swap the two columns of an SU(N) matrix, we obtain another element of SU(N). However 
in both cases we will sum over these elements. This shows us that it does not matter over which component we are integrating. 
The result will always be the same, meaning that the two integrals will contribute with a factor $1/N$.

The other case is $c \neq b$. This case is similar to the one in Eq.\ \ref{eq:haar_integral_2} and can be proven to give zero
using the same method. The obtained results can be summarized as
\begin{equation}
\int{dU\,  U^{a b} (U^\dagger)^{c d}} \;=\; \frac{1}{N}\,\delta_{b c}\, \delta_{a d}\,.
\label{eq:haar_integral_3}
\end{equation}

\section{Example: The SU(2) Group}
An element $U$ of the SU(2) group can be parametrized by
 \begin{equation}
  U \;=\; \{a_0+i \vect{a}\cdot \vect{\sigma} | a_0^2+\vect{a}\cdot\vect{a}\;=\;1 \}\,,
    \label{eq:SU2_parametrization_appendix}
 \end{equation}
where $\sigma_i$ are the Pauli matrices, given by
\begin{equation}
 \sigma_1 \;=\;
  \begin{pmatrix}
   0&1\\
   1&0
  \end{pmatrix} \quad
 \sigma_2 \;=\;
  \begin{pmatrix}
   0&-i\\
   i&0
  \end{pmatrix} \quad
 \sigma_3 \;=\;
  \begin{pmatrix}
   1&0\\
   0&-1
  \end{pmatrix}
\end{equation}

Let us now calculate $U_3[c(a,b)] = U_1(a) U_2(b)$. Using the property $\sigma_i \sigma_j = \delta_{i j} + i \varepsilon_{i j k} \sigma_k$
we obtain that
\begin{align}
 c_0 &= a_0 b_0 - a_i b_i\,, \\
 c_i &= a_0 b_i + b_0 a_i - \varepsilon_{i j k} a_j b_k\,,
\end{align}
which leads us to the Jacobian
\begin{equation}
  \frac{\partial c}{\partial b} \;=\;
	\begin{pmatrix}
	a_0 & -a_1 & -a_2 & -a_3 \\
	a_1 &  a_0 &  a_3 & -a_2 \\
	a_2 & -a_3 &  a_0 &  a_1 \\
	a_3 &  a_2 & -a_1 &  a_0
	\end{pmatrix}\,.
\end{equation}
Inverting this matrix and taking its determinant will lead us to
\begin{equation}
\norm{ \left(\frac{\partial c}{\partial b}\right)^{-1}}\;=\;\frac{1}{\left(a_0^2+a_1^2+a_2^2+a_3^2\right)^2} \;=\; 1\,.
\end{equation}
Therefore we conclude that
\begin{equation}
dU \;=\; K\, \delta(a^2-1)\, da_0 da_1 da_2 da_3\,.
\end{equation}
The Dirac delta function is introduced by hand to guarantee that in the integration we do no
leave the group by summing a set of parameters that do not correspond to our parametrization. However it can be easily
removed if we use (four-dimensional) spherical coordinates
\begin{align}
a_0 & = r \sin \theta \sin \varphi \sin \alpha \\
a_1 & = r \sin \theta \sin \varphi \cos \alpha \\
a_2 & = r \sin \theta \cos \varphi \\
a_3 & = r \cos \theta\,.
\end{align}
The range of the variables will be $r \in [0,\infty]$, $\,\theta \in [0,\pi]$, $\,\varphi \in [0,\pi]$ and $\,\alpha \in [0,2\pi]$.
The Jacobian will give us the relation $\,d^4 a = r^3 \sin^2 \theta \sin \varphi dr d\theta d\varphi d\alpha$. Notice also that
we obtain now $\,a_0^2+\vect{a}\cdot\vect{a} = r^2\,$ and the integral over $r$ will simply fix $\,r=1$. We can now set $f(U) = 1$
and obtain the normalization $\,K = (2 \pi^2)^{-1}$. We can use this parametrization to show that
\begin{equation}
\int{dU\, U^{a b}} = 0\,.
\end{equation}

We also have the results
\begin{align}
\int{dU\, U^{a a} U^{b b}} \;= &\;  \frac{1}{2}\,, \nonumber \\[2mm]
\int{dU\, U^{a b} U^{b a}} \;= &\; -\frac{1}{2}\,,
\end{align}
with $a \neq b$. If the components do not fit in one of these cases, the result will be 0.

Also, we verify that the result
\begin{equation}
\int{dU \, U^{a b} (U^\dagger)^{c d}} \;=\; \frac{1}{2}\,\delta_{b c}\, \delta_{a d}
\end{equation}
holds, as was expected.

\cleardoublepage
\chapter{Monte Carlo Simulation of the Harmonic Oscillator}
\label{sec:HO_results}
\thispagestyle{capitulo}
 
We aim at computing observables for the harmonic oscillator, which possesses the Lagrangian (in Euclidean space)
\begin{equation}
\mathcal{L} \;=\; \frac{m v^2}{2} + \frac{m \omega^2 x^2}{2}\,,
\label{eq:HO_Hamiltonian}
\end{equation}
where $m$ is the mass of the oscillating particle and $\omega$ is the angular frequency of the oscillator.
The purpose of doing so is to illustrate the use of the path-integral formalism (introduced in Section 
\ref{sec:path_integral}) in combination with Monte Carlo simulations (Section \ref{sec:MC_methods}). 
The fact that this problem has a well known solution makes it ideal as an illustration, 
since it enables the comparison of our results with the known solution. This appendix was partially
developed at the DESY Summer Student Proggramme of 2012 in collaboration with Aleksandra S\l{}apik. The
full work developed there can be found in Ref.\ \cite{DSSP:2012}.

\section{Analytic Solution with Path Integrals}
 
The harmonic oscillator is one of the few systems that have an analytic solution, which makes it an excellent
system for testing algorithms. We will compare our numerical results to the solution of the problem [see Ref.\
\cite{Creutz:1980gp}] for the case of a particle of unitary mass. In this section, we will state these results 
and generalize them.
 
The discrete Euclidean action used in Ref.\ \cite{Creutz:1980gp} is given by
\begin{equation}
 S \;=\; a \sum_{j=0}^{N-1} \left(\frac{x_{j+1}-x_j}{a}\right)^2 \,+\, \frac{\mu^2 x_j^2}{2}\,,
\end{equation}
where $\mu$ is a positive coefficient, related to the oscillator frequency $\omega$ and to its mass $m$ in Eq.\ \ref{eq:HO_Hamiltonian}.
By discretizing the action we mean that we divided the time in $N$ slices as in Section \ref{sec:path_integral}. It is
Euclidean because we already performed the Wick rotation $\,\tau = i t$.

The average of the operator $x^2$ is given by 
\begin{equation}
 \langle x^2 \rangle \;=\; \frac{1}{2\mu\sqrt{1+\frac{a^2 \mu^2}{4}}}\left(\frac{1+R^N}{1-R^N}\right)\,,
\end{equation}
where
\begin{equation}
 R \;=\; 1 \,+\, \frac{a^2 \mu^2}{2} \,-\, a \mu \sqrt{1+\frac{a^2 \mu^2}{4}}\,.
\end{equation}
 
Ref.\ \cite{Creutz:1980gp} also gives the expression for the correlation functions $\langle x_{i+j}\, x_i \rangle$
\begin{equation}
 \langle x_{i+j}\, x_i \rangle \;=\; \frac{R^j \,+\, R^{N-j}}{2(1-R^N)\,\mu\,\sqrt{1+\frac{a^2 \mu^2}{4}}}\,.
\end{equation}
 
We use the value of $\,\langle x^2 \rangle\,$ to calculate the ground-state energy of the system through the virial theorem
(see Eq.\ \ref{eq:Virial_Theorem})
\begin{equation}
 E_0 \;=\; \mu^2 \langle x^2 \rangle \;=\; \frac{\mu}{2\sqrt{1+\frac{a^2 \mu^2}{4}}}\left(\frac{1+R^N}{1-R^N}\right)\,.
\end{equation}
We can use as well the correlation functions to calculate $E_1$ (see Eq.\ \ref{eq:Calc_E1})
\begin{equation}
 \label{eq:E1}
 E_1 \;=\; E_0 -\frac{1}{a} \ln \left[\frac{R^{j+1}-R^{N-j-1}}{R^j - R^{N-j}}\right]\,.
\end{equation}

These calculations were all done with the mass set equal to unity. However we are interested in the action
\begin{equation}
  S \;=\; a \sum_{j=1}^N m\left( \frac{x_{j+1}-x_j}{a}\right)^2 + \frac{ \mu^2 x_j^2}{2}\,.
\end{equation}
To solve our problem, we rescale $x$ as $x/\sqrt{m}$. The action becomes then
\begin{equation}
  S \;=\; a \sum_{j=1}^N \left( \frac{x_{j+1}-x_j}{a}\right)^2 + \frac{\mu^2 x_j^2}{2 m}\,.
\end{equation}
This means that in every expression it will be necessary to replace $\mu$ by $\mu/\sqrt{m}$. We find then
\begin{align}
 \label{eq:x_squared}
 \langle x^2 \rangle  \;&=\; \frac{1}{2\mu\sqrt{m+\frac{a^2 \mu^2}{4}}}\left(\frac{1+R^N}{1-R^N}\right) \\[2mm]
 \label{eq:R}
 R \;&=\;  1+\frac{a^2 \mu^2}{2 m} - a \mu \sqrt{\frac{1}{m}+\frac{a^2 \mu^2}{4 m^2}} \\[2mm]
 \label{eq:correlators}
 \langle x_{i+j} x_i \rangle  \;&=\; \frac{R^j+R^{N-j}}{2(1-R^N)\mu\sqrt{m+\frac{a^2 \mu^2}{4}}} \\[2mm]
 \label{eq:E0}
 E_0 \;&=\; \frac{\mu}{2\sqrt{m+\frac{a^2 \mu^2}{4}}}\left(\frac{1+R^N}{1-R^N}\right) \;=\; \mu^2 \langle x^2 \rangle \,.
\end{align}
Notice that the expression for $E_1$ is the same as in Eq.\ \ref{eq:E1}, but with the definition of R given in Eq.
\ref{eq:R}.

\section{Numerical Results}
 
We aim to compute the observables for the harmonic oscillator. Our program calculates the expected value for the square of the
position $\langle x^2 \rangle$ since, through the virial theorem (Eq.\ \ref{eq:Virial_Theorem}), $E_0 = \mu^2 \langle x^2 \rangle$.
The parameters chosen for our simulations were
$m = 0.5$ and $\mu^2 = 2$. Also, we draw a new value for $x$ in a range between $x - \sqrt{a}$ and $x + \sqrt{a}$. Also, we
apply the Metropolis update at a given lattice site $\overline{n} = 10$ times. For these parameters we have $E_0 = 1.0$ in the continuum limit  
and that is the value we hope to find numerically.

In all cases we consider the equilibrium to be reached after $100$ Monte Carlo iterations. We also take measurements for $M = 5$,
that is every five Monte Carlo iterations. This is to try to enforce that the quantities measured be
completely decorrelated from each other. Since we are interested in the continuum limit, we run our simulation with increasingly 
smaller lattice spacing, but keeping the quantity $Na\,$ fixed to $25$. In each case we made $10^5$ Monte Carlo iterations to ensure 
a good precision. The results can be seen in Table \ref{tb:Results_HO_Na_25}.

\begin{table}[h]
\centering
\caption{Results for $E_0$ using $10^5$ Monte Carlo iterations, $\overline{n} = 10$ and measurements made every five Monte Carlo
iterations.}
\label{tb:Results_HO_Na_25}
\begin{tabu}{||c|c|c|c||}
\hline
 Lattice Spacing $a$		& $E_0$ Analytical 	& $E_0$ Computed	&Deviation from theory	\\ \hline
$\num{5.0E-1}$					& $0.89443$					& $0.89416(65)$		& $\num{ 2.7000E-4}$	\\
$\num{2.5E-1}$					& $0.97014$					& $0.97205(64)$		& $\num{-1.9100E-3}$	\\
$\num{5.0E-2}$					& $0.99875$					& $1.01907(63)$		& $\num{-2.0320E-2}$	\\
$\num{2.5E-2}$					& $0.99969$					& $1.02280(64)$		& $\num{-2.3110E-2}$	\\
$\num{5.0E-3}$					& $0.99999$					& $1.03073(54)$		& $\num{-3.0740E-2}$	\\
\hline
\end{tabu}
\vspace*{10pt}\\Source: S\L{}APIK; SERENONE; \cite{DSSP:2012}.
\end{table}

\vspace{0.5cm}

Our estimate for the error is given by
\begin{equation}
 \Delta E_0 = \frac{\sigma}{\sqrt{N}}\,,
\end{equation}
where $\sigma$ stands for the standard deviation from the data obtained from the Monte Carlo simulation and $N$ is the number of 
measurements.  
 
The main feature we see in this graph is that the smaller the $a$ we get, the more we deviate from the analytic value. One possible
reason for this is that our paths are much alike even after five complete Monte Carlo iterations when $a$ is small, which violates our work
hypothesis for applying the algorithm. To test if that is really the case, 
we increase the number of Monte Carlo iterations measurements to $M = 25$ and test the program with the same set of 
parameters. To keep the total number of measurements constant (which would affect our error) we increase the number of Monte Carlo
steps to 499600. The results are in Table \ref{tb:Results_HO_Na_25_M=25}.

\begin{table}[h]
\centering
\caption{Results for $E_0$ using 499600 Monte Carlo steps, $\overline{n} = 10$ and measurements made every 25 Monte Carlo
iterations.}
\label{tb:Results_HO_Na_25_M=25}
\begin{tabular}{||c|c|c|c||}
\hline
 Lattice Spacing $a$	& $E_0$ Analytical 	& $E_0$ Computed	&Deviation from theory	\\ \hline
$\num{5.0E-1}$				& $0.89443$					& $0.89242(29)$		& $\num{ 2.01E-3}$	\\
$\num{2.5E-1}$				& $0.97014$					& $0.96998(28)$ 	& $\num{ 1.60E-4}$	\\
$\num{5.0E-2}$				& $0.99875$					& $0.99963(28)$ 	& $\num{-8.80E-4}$	\\
$\num{2.5E-2}$				& $0.99969$					& $0.99546(27)$ 	& $\num{-4.23E-3}$	\\
$\num{5.0E-3}$				& $0.99999$					& $1.00418(31)$ 	& $\num{-4.19E-3}$  \\
\hline
\end{tabular}
\vspace*{10pt}\\Source: S\L{}APIK; SERENONE; \cite{DSSP:2012}.
\end{table}

\vspace{0.5cm}

We can see clearly a reduction of the error, although not enough to get our computed value and analytical value to agree. Ideally, we should
increase the number of simulations between measurements to $M = 50$ in order to obtain a more precise value. However this would require 
approximately one million Monte Carlo iterations to keep the number of measurements fixed. Because of this, it becomes too expensive to
make such a simulation.

\begin{figure}[h]
 \centering
 \caption{Continuum limit analysis for $E_0$.}
 \label{fg:fit_E0_T_25}
   \includegraphics[width=0.75\textwidth]{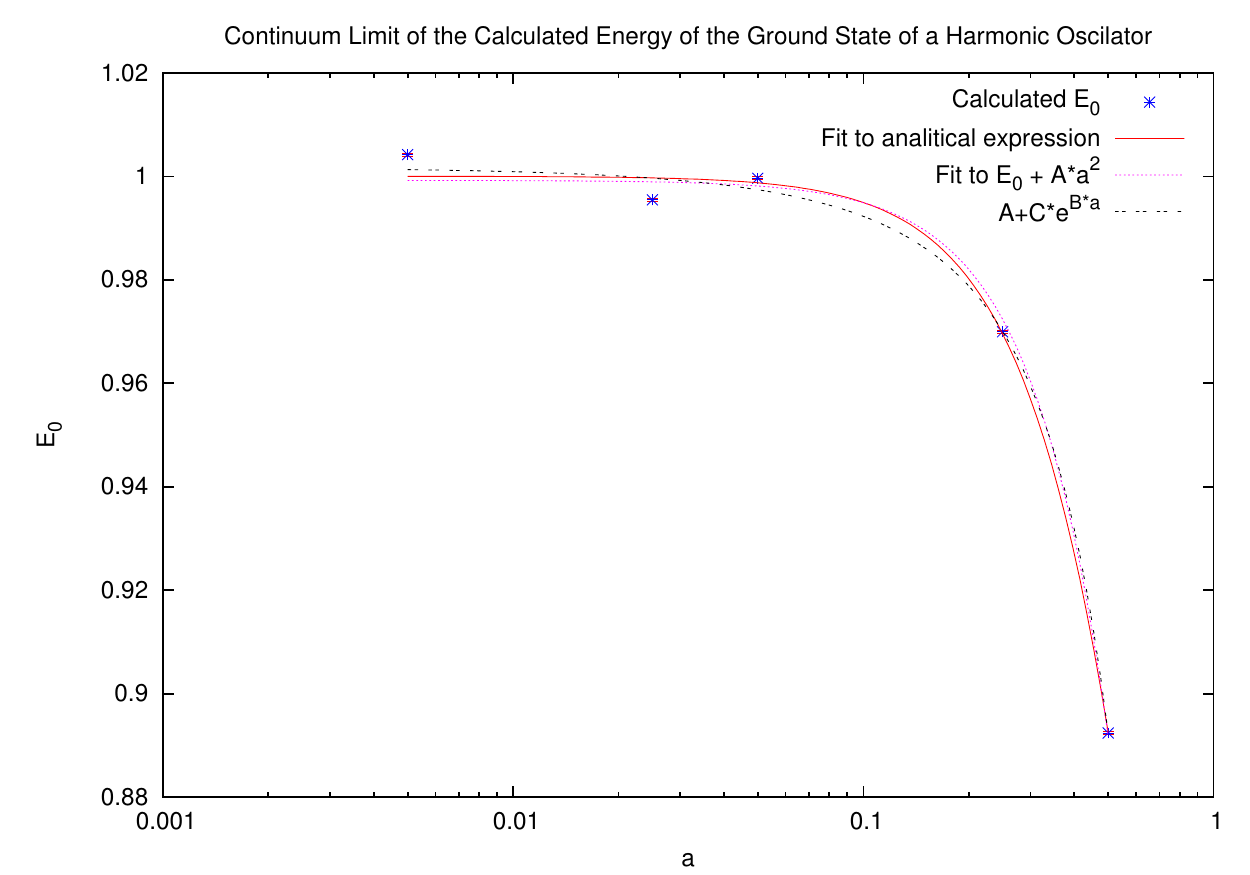}
\vspace*{10pt}\\Source: S\L{}APIK; SERENONE; \cite{DSSP:2012}.
 \end{figure}

We proceed to plotting the data from Table \ref{tb:Results_HO_Na_25_M=25}. The idea now is to fit these data to a function
and then take the limit $a \rightarrow 0$. Since this problem has an analytic solution given by Eq.\ \ref{eq:E0}, we use it,
treating the mass $m$ and the parameter $\mu$ as free parameters in our fit. Since this method is not general, we also try 
to fit to a polynomial. But from Fig.\ \ref{fg:fit_E0_T_25} we see that this function approaches the continuum value asymptotically.
This means that at least its first derivative will be zero at $a=0$. This hints to us an even function and therefore we will include
only even powers in the polynomial. Finally we try also an ad hoc function $f(a) = A + C e^{-B a}$. The fits are shown in Fig.\ \ref{fg:fit_E0_T_25} and
the resulting parameters in Table \ref{tb:Fit_parameters_E0}.

\begin{table}[h]
\centering
\caption{Fitting parameters for functions in Fig.\ \ref{fg:fit_E0_T_25}.}
\label{tb:Fit_parameters_E0}
\begin{tabular}{||c|c|c||}
\hline
$E_0/\sqrt{1+B x^2}$	& $E_0 + A x^2$ 		& $A+C e^{B x}$ Calculated 	\\ \hline
$E_0 = 1.000(2)$			& $E_0 = 0.999(2)$	& $A= 1.02(1)$		\\
$B = 1.02(2)$					& $A = -0.43(2)$		& $B= 3.6(8)$		\\
											&										& $C =-0.02(1)$		\\
\hline
\end{tabular}
\vspace*{10pt}\\Source: S\L{}APIK; SERENONE; \cite{DSSP:2012}.
\end{table} 

The first point to notice is that for larger lattices we have larger fluctuations than expected by our estimate of the error.
This can again be explained by the fact that the paths between measurements are not completely decorrelated for small $a$. But, since these fluctuations
are random, our fit converges to the expected value of $E_0$ (see Eq.\ \ref{eq:E0}). Also, we see that the best result is, as expected, the fit to the 
analytic expression. But since this is a special case and the fit to a second order polynomial yielded
a good result as well, this will be our preferred fitting method from now on.

We want also to predict the energy of the first excited state. As stated in Section \ref{sec:QM_Conn_w_SM}, we use two strategies.
One is the fit of the correlation function between lattice sites $\Gamma^{(2)}_c$ and the other is the direct application of Eq.\ \ref{eq:Calc_E1}. 
Our program does these calculations automatically and, for the same run in Table 
\ref{tb:Results_HO_Na_25_M=25}, we have the results of Table \ref{tb:Results_E1_HO_Na_25_M=25}.

\begin{table}[h]
\centering
\caption{Results for $E_1$ using $499600$ Monte Carlo iterations, $\overline{n} = 10$ and measurements made every each $25$ Monte Carlo iterations.}
\label{tb:Results_E1_HO_Na_25_M=25}
\begin{tabular}{||c|c|c|c||}
\hline
Lattice Spacing $a$		& $E_1$ Analytical 	& $E_1$ Calculated	& $E_1$ Calculated 				\\ 
											&										& Through Fit				& Through Eq.\ \ref{eq:Calc_E1}			\\ \hline
$\num{5.0E-1}$				& $2.81928$					& $2.687(22)$				& $ 1.94438(64)$	\\
$\num{2.5E-1}$				& $2.94987$					& $2.863(12)$				& $ 2.79011(46)$	\\
$\num{5.0E-2}$				& $2.99792$					& $2.889(37)$				& $ 2.96684(57)$	\\
$\num{2.5E-2}$				& $2.99948$					& $2.844(67)$				& $ 2.96918(56)$	\\
$\num{5.0E-3}$				& $2.99998$					& $2.228(400)$			& $ 2.94124(63)$  \\
\hline
\end{tabular}
\vspace*{10pt}\\Source: S\L{}APIK; SERENONE; \cite{DSSP:2012}.
\end{table}
\vspace{0.5cm}
 
We can see clearly that these results are not in agreement with the analytic formula. This means that the correlation effects influence more the calculation of $E_1$ than
the calculation of $E_0$. To get better results, we increase the number of Monte Carlo iterations to 999100 paths, but we choose to make a measurement only
every 50 Monte Carlo iterations. With the purpose of making the program run faster, we reduce the parameter $\overline{n}$ to 5. We summarize the results 
in Table \ref{tb:Results_E1_HO_Na_25_M=50}. 

\begin{table}[h]
\centering
\caption{Results for $E_1$ using $999100$ Monte Carlo iterations, $\overline{n} = 5$ and measurements made only every $50$ Monte Carlo iterations.}
\label{tb:Results_E1_HO_Na_25_M=50}
\begin{tabular}{||c|c|c|c||}
 \hline
Lattice Spacing $a$		& $E_1$ Analytical 	& $E_1$ Calculated 	& $E_1$ Calculated 				\\ 
											&										& Through Fit				& Through Eq.\ \ref{eq:Calc_E1}			\\ \hline
$\num{5.0E-1}$				& $2.81928$					& $2.8212(38)$			& $ 2.49800(52)$	\\
$\num{2.5E-1}$				& $2.94987$					& $2.94069(94)$			& $ 2.94307(44)$	\\
$\num{5.0E-2}$				& $2.99792$					& $2.99614(75)$			& $ 3.00084(41)$	\\
$\num{2.5E-2}$				& $2.99948$					& $2.99709(44)$			& $ 2.99732(40)$	\\
$\num{5.0E-3}$				& $2.99998$					& $2.98905(28)$			& $ 2.98908(44)$  \\
\hline
\end{tabular}
\vspace*{10pt}\\Source: S\L{}APIK; SERENONE; \cite{DSSP:2012}.
\end{table}
\vspace{0.5cm}
 
We can see that even though we have not achieved a total agreement with the analytic value, we have more reasonable results. We proceed in taking the continuum 
limit by the same procedure used for obtaining the value of $E_0$. Again we see an asymptotic behavior of the points and therefore we use a polynomial $E_1+B x^2$. 

\begin{table}[h]
\centering
\caption{Parameters used for the fit of the graph in Fig.\ \ref{fg:fit_E1_T_25}.}
\label{tb:Fitting_Parameters}
\begin{tabular}{||c|c|c||}
\hline
Lattice Spacing $a$	& $E_1$ Calculated Through Fit	& $E_1$ Calculated Through Eq.\ \ref{eq:Calc_E1}\\ \hline
$E_1$								& $  2.993(4)$									& $  3.01(2)$	\\
$A$									& $  -0.69(3)$									& $  -2.00(17)$	\\
\hline
\end{tabular}
\vspace*{10pt}\\Source: S\L{}APIK; SERENONE; \cite{DSSP:2012}.
\end{table}

\begin{figure}[h]
\centering
  \caption{Continuum-limit analysis for $E_1$ using the two methods described in Section \ref{sec:QM_Conn_w_SM}.}
  \label{fg:fit_E1_T_25}
  \includegraphics[width=0.75\textwidth]{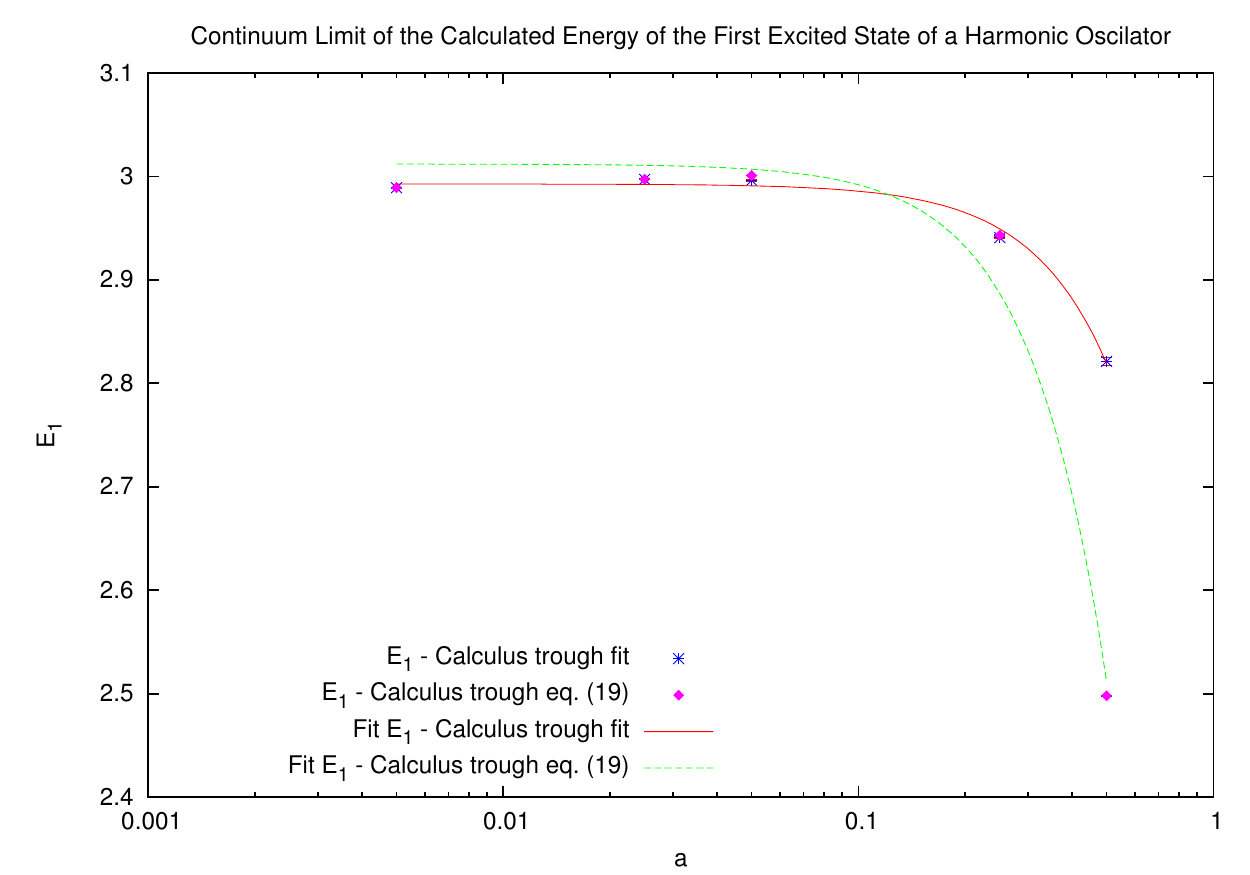}
\vspace*{10pt}\\Source: S\L{}APIK; SERENONE; \cite{DSSP:2012}.
\end{figure}

As we can see from the graph in Fig.\ \ref{fg:fit_E1_T_25}, the point calculated using Eq.\ \ref{eq:Calc_E1} for $a = 0.5$ 
completely disagrees with the analytic solution. There are two sources of systematic errors that could explain these errors.

\begin{figure}[h]
\centering
\caption{ $\langle x(0) x(\tau) \rangle$ as function of $\tau$ for $a=0.5$.}
\label{fg:corr_func_tau_a=0.5} 
\includegraphics[width=0.65\textwidth]{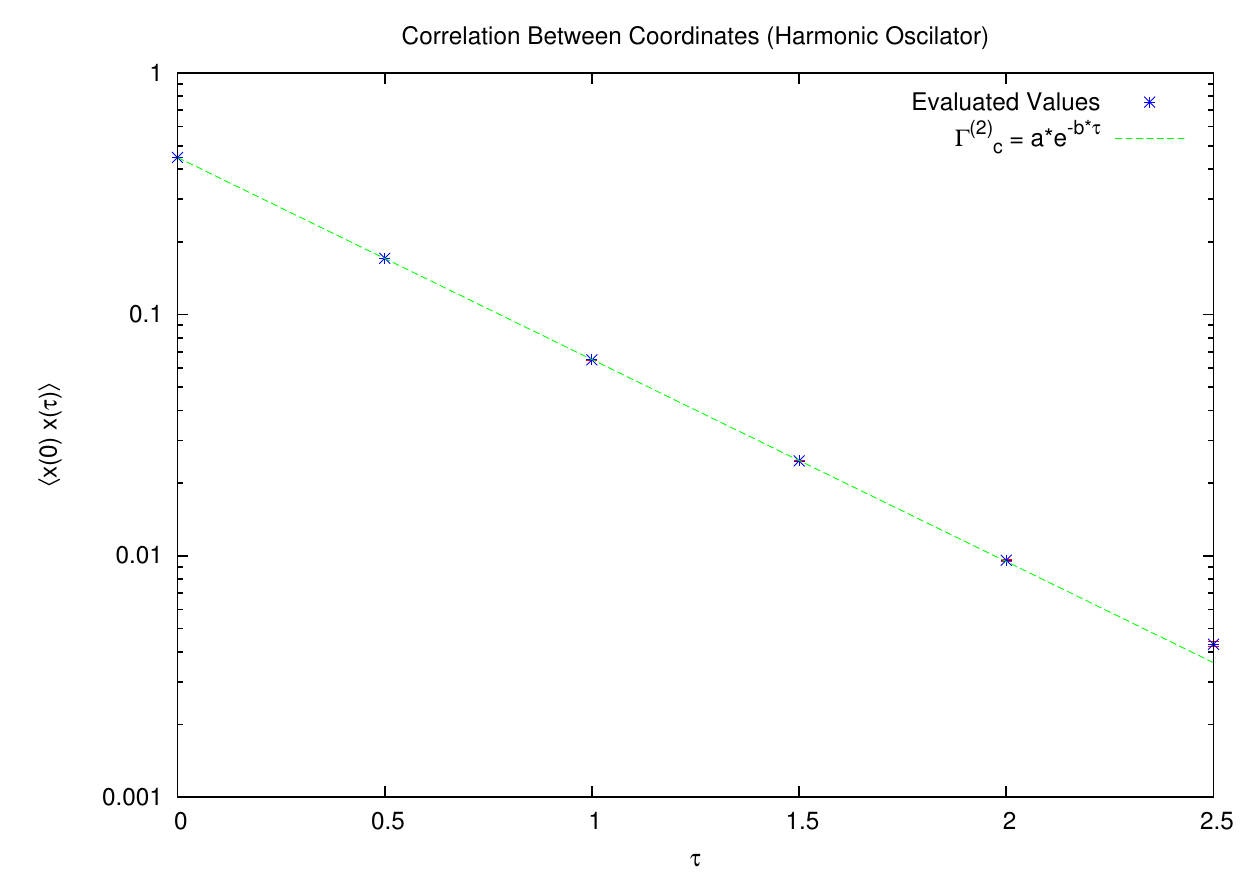}
\vspace*{10pt}\\Source: S\L{}APIK; SERENONE; \cite{DSSP:2012}.
\end{figure}

The first of them is that, according to the theory presented in Section \ref{sec:QM_Conn_w_SM}, the application of Eq.\ \ref{eq:E1}
can only be done for $\tau \rightarrow \infty$ (large $j$). We actually do not do this and our $j$ is always set to 4. The second source of 
error is to impose periodic boundary conditions. Since we impose that $\,x(\tau_{N+1}) =  x(\tau_0)$, it is expected that 
$\Gamma^{(2)}_c(\tau_{N+1}) = \Gamma^{(2)}_c(\tau_0)$. For the same reason, $\Gamma^{(2)}_c(\tau_N) = \Gamma^{(2)}_c(\tau_1)$ and so on.
The effect is that, in the plot of $\Gamma^{(2)}_c (\tau)$, instead of a single decaying exponential there is a contribution of a rising exponential
as well. The effect of this rising exponential is negligible near the origin, but becomes bigger at large $\tau$, originating deviations such as 
the ones that can be seen in Fig.\ \ref{fg:corr_func_tau_a=0.5} for $\tau=2.5$. This indicates the need to find a balance between these two sources of errors.
 
In our case, our program keeps the value of $j$ fixed at 4 in Eq.\ \ref{eq:E1}. Since $\tau_j= j a$, keeping $j$ fixed means that we are using smaller $\tau$ in the calculations
and therefore we are minimizing the effects of the errors due to the influence of boundary conditions. The cost of that is an increase in errors due to violating our working hypothesis.
However, since the errors coming from the use of Eq.\ \ref{eq:E1} usually get smaller for smaller $a$, we can say that the effects of the boundary conditions are far more relevant. We must 
highlight however that this analysis is just a qualitative approach. These errors should be carefully analyzed in a quantitative way. This analysis should also include
the propagation of the errors of $E_0$ in the calculation of $E_1$, since in all cases we calculate just the difference $E_1-E_0$ and then add $E_0$ to the result. We do not do this analysis here.

We turn our attention again to the analysis of Fig.\ \ref{fg:fit_E1_T_25} and the parameters of Table \ref{tb:Fitting_Parameters}. 
We can observe that the calculation made with Eq.\ \ref{eq:Calc_E1} do not agree with the analytic value of $E_1 = 3$, which is expected given our explanation
of why this method gives worse results. However, we reach agreement with the analytic solution using the method in which we fit our data to Eq.\ \ref{eq:gamma_n_c_E1_isolated}.

\end{document}